\documentclass[acmtog]{acmart}
\acmSubmissionID{431}

\usepackage{booktabs} %

\usepackage{enumitem}
\usepackage[capitalise]{cleveref}
\usepackage[normalem]{ulem}
\usepackage{physics}
\usepackage{bbm}
\usepackage{array}
\usepackage{multirow}
\usepackage{tikz}
\usepackage{xcolor}
\usepackage{stmaryrd}
\usetikzlibrary{shapes.misc}
\usetikzlibrary{spy}
\tikzset{cross/.style={cross out, draw=black, minimum size=2*(#1-\pgflinewidth), inner sep=0pt, outer sep=0pt},
cross/.default={1pt}}

\setlist{leftmargin=*}

\newcommand{\lingxiao}[1]{{\scriptsize\color{brown}\textbf{LL: #1}}}

\newcommand{\rev}[1]{{#1}}

\newcommand{\RR}{\mathbb{R}}
\newcommand{\defeq}{:=}
\newcommand{\vol}{\textup{vol}}
\newcommand{\avg}{\textup{avg}}
\newcommand{\area}{\textup{area}}
\newcommand{\proj}{\textup{proj}}
\newcommand{\pull}{\textup{pull}}

\newcommand{\sto}{\shortrightarrow}
\DeclareGraphicsExtensions{.pdf,.png,.jpg}

\usepackage[ruled]{algorithm2e} %

\SetAlFnt{\small}
\SetAlCapFnt{\small}
\SetAlCapNameFnt{\small}
\SetAlCapHSkip{0pt}

\setcopyright{rightsretained}
\acmJournal{TOG}
\acmYear{2021}
\acmVolume{40}
\acmNumber{6}
\acmArticle{256}
\acmMonth{12}
\acmDOI{10.1145/3478513.3480568}
\begin{document}

\title{Interactive All-Hex Meshing via Cuboid Decomposition}
\author{Lingxiao Li}
\email{lingxiao@mit.edu}
\author{Paul Zhang}
\email{pzpzpzp1@mit.edu}
\author{Dmitriy Smirnov}
\email{smirnov@mit.edu}
\author{S. Mazdak Abulnaga}
\email{abulnaga@mit.edu}
\author{Justin Solomon}
\email{jsolomon@mit.edu}
\affiliation{%
  \institution{
  \\
  Massachusetts Institute of Technology}
  \streetaddress{77 Massachusetts Avenue}
  \city{Cambridge}
  \state{MA}
  \postcode{02139}
  \country{USA}}

\authorsaddresses{Authors’ address: Lingxiao Li, lingxiao@mit.edu; Paul Zhang, pzpzpzp1@mit.edu; Dmitriy Smirnov, smirnov@mit.edu; Mazdak Abulnaga, abulnaga@mit.edu; Justin Solomon, jsolomon@mit.edu, Massachusetts Institute of Technology, 77 Massachusetts Avenue, Cambridge, MA, 02139, US}

\begin{abstract}
Standard PolyCube-based hexahedral (hex) meshing methods aim to deform the input domain into an axis-aligned PolyCube volume with integer corners; if this deformation is bijective, then applying the inverse map to the voxelized PolyCube yields a valid hex mesh. A key challenge in these methods is to maintain the bijectivity of the PolyCube deformation, thus reducing the robustness of these algorithms. In this work, we present an interactive pipeline for hex meshing that sidesteps this challenge by using a new representation of PolyCubes as unions of cuboids. We begin by deforming the input tetrahedral mesh into a near-PolyCube domain whose faces are loosely aligned to the major axis directions. We then build a PolyCube by optimizing the layout of a set of cuboids with user guidance to closely fit the deformed domain. Finally, we construct an inversion-free pullback map from the voxelized PolyCube to the input domain while optimizing for mesh quality metrics. We allow extensive user control over each stage, such as editing the voxelized PolyCube, positioning surface vertices, and exploring the trade-off among competing quality metrics, while also providing automatic alternatives. We validate our method on over one hundred shapes, including models that are challenging for past PolyCube-based and frame-field-based methods. Our pipeline reliably produces hex meshes with quality on par with or better than state-of-the-art. We additionally conduct a user study with 21 participants in which the majority prefer hex meshes they make using our tool to the ones from automatic state-of-the-art methods. This demonstrates the need for intuitive interactive hex meshing tools where the user can dictate the priorities of their mesh.
\end{abstract}



\begin{teaserfigure}
	\includegraphics*[width=\linewidth]{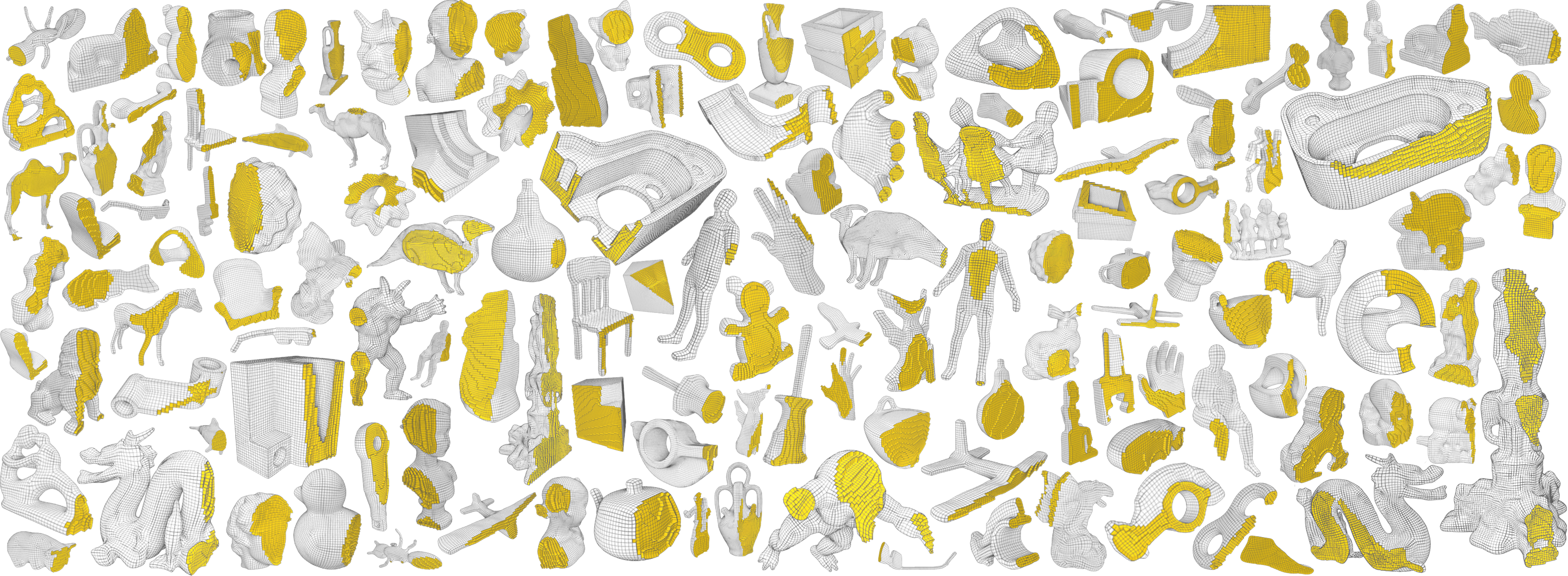}
	\caption{A sampling of hexahedral meshes produced by our method. Our user-in-the-loop interactive pipeline facilitates the creation of high-quality hex meshes, giving the user the option to control the final result at any desired level of granularity.}
	\label{fig:teaser}
\end{teaserfigure}

\maketitle

\section{Introduction}
Hexahedral (hex) meshes historically have been preferred over tetrahedral meshes in various graphics and simulation applications due to their reduced numerical error and usage of fewer elements \cite{shepherd2008hexahedral}. 
In particular, the regular structure of layers of hex elements in a hexahedral mesh enables natural support of tensor product function bases \cite{liu2015feature} and multilevel hierarchies of nested meshes for efficient PDE solvers \citep{nieser2011cubecover}.
However, few reliable techniques exist to generate a high-quality hex mesh that conforms to the input domain while having desirable properties such as low distortion and uniform edge lengths.

Among hex meshing methods, PolyCube-based algorithms stand out for their relative robustness. The PolyCube-based pipeline typically starts by deforming the input mesh into a \emph{near-PolyCube}, a mesh whose surface normals are roughly axis-aligned \citep{huang2014l1}. Various heuristics and repairs can be applied to transform the near-PolyCube into a \emph{PolyCube}, whose surface normals are exactly axis aligned \citep{sokolov2015fixing,zhao2019polycube}. 
If the deformation map from these two steps maintains bijectivity, and the PolyCube has integer corners, then one can voxelize the PolyCube and pull it back through the map to obtain a boundary-aligned hex mesh of the input domain \citep{gregson2011all,livesu2013polycut,fu2016efficient}.
The obtained hex mesh can then be improved using various techniques, from optimizing element quality \citep{livesu2015practical} to pushing boundary singularities inwards \citep{cherchi2019selective}.

However, obtaining a bijective deformation map from the input domain to a PolyCube while satisfying integer constraints remains unsolved \citep{sokolov2015fixing, protais2020robust}.
In this work, we circumvent this difficulty by decomposing a near-PolyCube deformed from the input into a union of axis-aligned cuboids, which represents the generated PolyCube.
Borrowing ideas from computer vision \cite{tulsiani2017learning, smirnov2020deep}, we optimize the PolyCube structure by optimizing over constituent cuboids' parameters.
This representation effectively controls the complexity of the PolyCube via the number of cuboids, resulting in few corners.
At the same time, it is compact and resolution-independent compared to methods based on voxelization \citep{yu2014optimizing, yang2019computing}.

Although our PolyCube generation using cuboids sidesteps the typical robustness issues in deformation-based methods \cite{gregson2011all, huang2014l1, sokolov2015fixing}, it comes at the cost of losing the map between the PolyCube and the near-PolyCube (and, by extension, the input mesh). 
Instead of trying to recover this missing link, we compute a locally injective map from a voxelized PolyCube directly to the input domain to get the final hex mesh.
This is made possible by three components: a smooth distortion energy \citep{garanzha2021foldover} that forces local injectivity of the map, an inversion-free pullback step that guides the hex mesh to progressively deform to the input domain, and a bi-directional proximity energy that encourages the recovery of the input surface.
A mesh quality optimization step then follows to further improve the final hex mesh.

Rather than making our pipeline fully automatic, we instead create an interactive system to give the user significant freedom on how their hex meshes are generated so as to accommodate application-dependent requirements.
As the resulting hex mesh largely depends on the PolyCube structure, we allow the user to build the PolyCube interactively by adding or modifying the constituent cuboids in an intuitive way, while also providing the automatic option to adjust the existing cuboids using our PolyCube optimization method.
In addition to PolyCube generation, the user can substantially affect other parts of the pipeline, e.g., by digging or extruding layers on the voxelized PolyCube, modifying surface vertex positions of the final hex mesh, and exploring the trade-off among competing metrics in the refinement step.

Compared to past automatic and interactive hex-meshing methods, our system reliably \rev{generates} all-hex meshes for a wide range of input domains with mesh quality on par \rev{with} or \rev{surpassing that of} prior work.
At the same time, our system allows intuitive and extensive user control across all stages of the pipeline.
We perform a user study with 21 participants; most participants are satisfied with their results over the ones from \url{www.hexalab.net} and enjoy the fact that they are able \rev{to make} fine-grained adjustments.
%
%
%
%
%
%
%
%
%

\paragraph*{Contributions.}
We present an end-to-end interactive pipeline for robust hexahedral meshing based on cuboid decomposition.
Our main contributions include:
\begin{itemize}
  \item
    a method for continuous optimization of PolyCube structure via cuboid decomposition and signed distance fields;
  \item
    a method for computing a low-distortion, inversion-free volumetric map from a voxelized PolyCube to the input mesh; and
  \item
    an integrated interactive system that gives the user extensive and intuitive controls over the proposed pipeline with automatic alternatives.
\end{itemize}

\section{Related Works}
\paragraph{PolyCube construction and hex meshing.}
\citet{tarini2004polycube} first suggest using PolyCube maps to create seamless texture mappings with manual PolyCube construction.
\citet{lin2008automatic} subsequently introduce the first automatic approach to construct a PolyCube based on segmentation using Reeb graphs but with a limited set of primitives. %

\citet{gregson2011all} propose a hex-meshing pipeline  that deforms the input domain into a PolyCube and then pulls back a voxelized PolyCube to obtain the hex mesh.
To obtain a PolyCube, they segment the surface into charts based on rotations and flatten the charts by solving a Poisson equation.
\citet{livesu2013polycut} cast the segmentation in PolyCube generation as a multi-label graph cut problem.
\citet{huang2014l1} suggest using the $\ell_1$-norm of surface normals to encourage cubeness during the deformation while constraining the total surface area to prevent degeneracy.
\citet{fang2016all} cut tunnel loops of the input surface to allow generation of hex meshes with a much larger class of singularity patterns using PolyCubes, but their method is expensive due to consistency constraints across the cuts.
\citet{fu2016efficient} alternate between smoothing the surface normals and deforming the surface to be axis-aligned until a valid PolyCube is generated, using the AMIPS energy \citep{fu2015computing} to enforce inversion-free deformation.
More recently, \citet{guo2020cut} cut PolyCube edges open to inject internal singularities connected with the boundary while preserving a set of prescribed feature curves.
None of these methods guarantees the generation of a bijective map that deforms the input domain to a valid PolyCube, \rev{a challenge that our approach sidesteps}, and non-exhaustive heuristic fixes are typically used.
Many \citep{gregson2011all, livesu2013polycut, fu2016efficient} also limit the type of PolyCubes to the ones where every corner is adjacent to three charts, a sufficient but not necessary condition \citep{eppstein2010steinitz}, whereas our proposed pipeline has no such limitation (\cref{fig:flex_charts}). 

\citet{yu2014optimizing} directly voxelize the near-PolyCube deformed from the input mesh, \rev{a procedure of obtaining a PolyCube} with guaranteed success similar to ours.
Morphological operations then simplify the voxelized PolyCube.
However, they separate computation of the surface and volume components of the backward mapping, while we formulate a cohesive optimization scheme that computes both simultaneously, avoiding situations where a fixed surface mapping does not admit a low-distortion mapping of the volume.
As a follow-up, \citet{yang2019computing} use erasing-and-filling operators to reduce the number of corners in the voxelized PolyCube, but they are only concerned with producing a surface PolyCube map.
In contrast to both voxelization-based methods, our pipeline builds the PolyCube in a top-down manner: cuboids are placed one by one to form the PolyCube whose complexity increases gradually.

\rev{
Aside from robustness issues, all these fully-automatic pipelines consist of multiple stages where the correctness of each successive stage relies heavily on the success of previous stages.
As the errors can accumulate unpredictably, it is strenuous to tweak the parameters to yield desirable results. In contrast, our interactive system allows the user to ensure the quality of each stage separately using intuitive controls before moving onto the next one.
}

\paragraph{Interactive hex meshing.}
Contrary to the extensive research in interactive quadrilateral meshing \citep{campen2014dual,jakob2015instant,ebke2016interactively}, few methods \rev{for} interactive hex meshing have been proposed.
\citet{takayama2019dual} introduce dual-sheet hexahedralization by asking the user to design sheets that are combinatorial duals of a hex layout followed by a primalization step that recovers the hex mesh.
The representation of dual sheets as zero isosurfaces of implicit functions allows intuitive user editing and simplifies computation.
Although their method can generate a large class of all-hex meshes with internal singularity patterns, the sheet configuration needs to satisfy a series of complicated conditions and requires a manual fix otherwise.
Moreover, these conditions are not sufficient to obtain a valid hex topology, and
generating hex meshes with uniform edge lengths is difficult in their framework.%

The industry standard for hex meshing is CUBIT, an interactive tool \citep{cubit}. 
CUBIT operates directly on CAD geometries and supports user-guided sweeping operations by which a quad mesh can be extruded into hexes.
Usage of CUBIT requires a significant amount of training, and their technical details are unpublished.

\paragraph{Other hex meshing techniques}
A promising line of work extends cross-field-based quadrilateral meshing \citep{bommes2009mixed} to 3D by computing a smooth boundary-aligned frame field on an input domain to guide the hex layout  \citep{huang2011boundary,solomon2017boundary,ray2016practical,drpmbo}. 
While frame-field-based methods do not limit their singularities to the surface and can generate high-quality meshes, they frequently fail even on fairly simple domains \citep{viertel2016analysis}.  
\citet{liu2018singularity,corman2019symmetric} add topological constraints to frame field generation, but these methods have not managed to increase robustness of end-to-end field-based hex mesh generation. 

\citet{gao2019feature} use an octree to adaptively generate hex elements that fill the input domain while preserving features. 
However, their method is slow and cannot generate hexes of uniform size.
\citet{livesu2019loopy} use a surface frame field to guide creation of cuts that partition the input mesh, generating hex-dominant meshes rather than all-hex meshes.
We compare against these methods in \cref{sec:results}.

Various postprocessing techniques \citep{livesu2015practical, fu2015computing, gao2015hexahedral, cherchi2019selective, marschner2020hexahedral} have been suggested to improve the quality of a hex mesh.
In our pipeline, mesh refinement is incorporated seamlessly in a final stage.
We further allow customization of the surface layout as well as exploration of competing quality metrics.

\section{System Overview}
\begin{figure*}
  \centering
  \newcolumntype{C}[1]{>{\centering\arraybackslash}m{#1}}
  \hspace*{-0.2in}
  \begin{tabular}{C{3cm} C{3cm} C{3cm} C{3cm} C{3cm}}
    \includegraphics[height=1.2in,trim=4.0in 1.in 2.2in 0, clip]{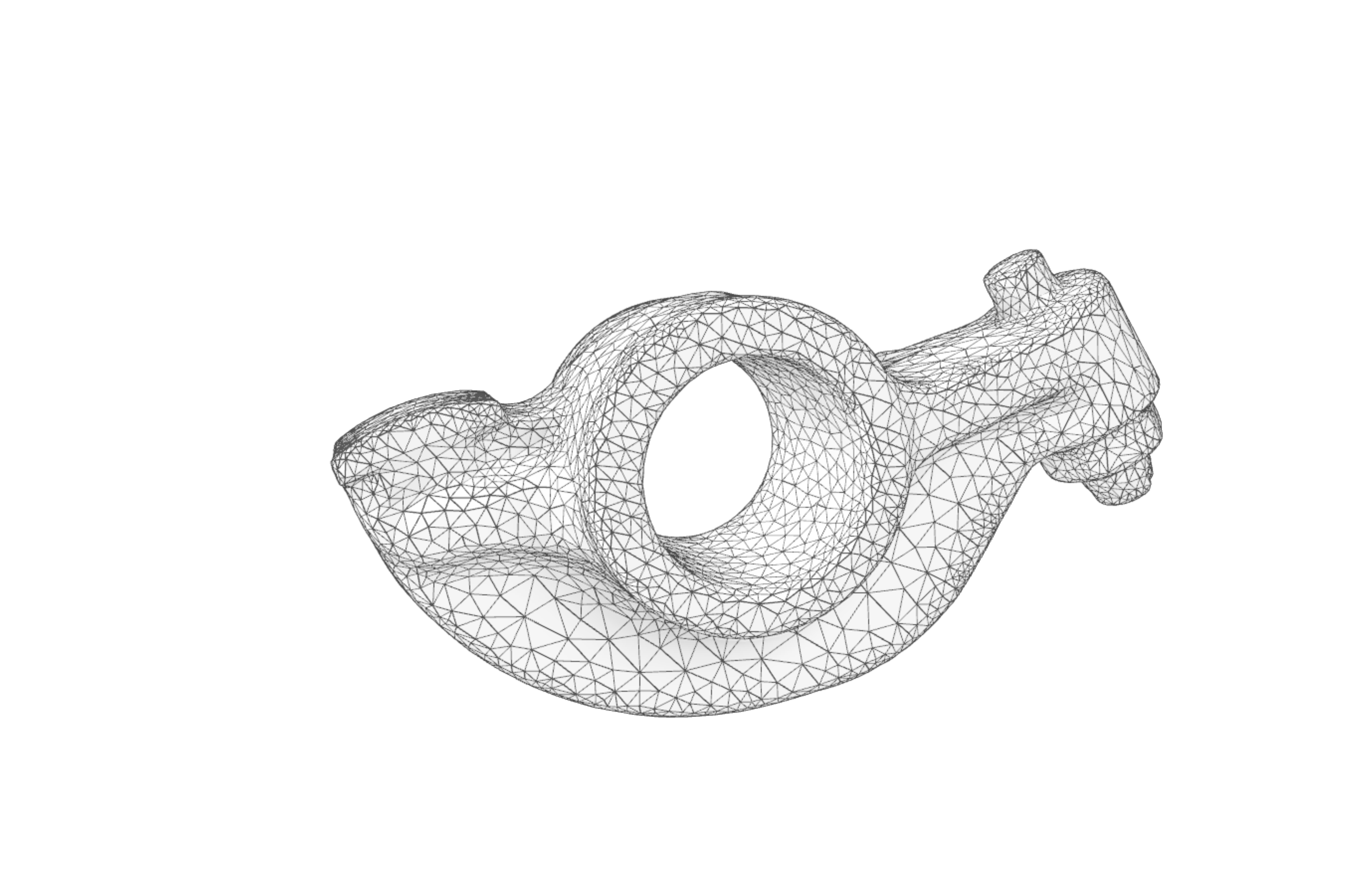} &
    \includegraphics[height=1.2in,trim=4.0in 1.in 2.2in 0, clip]{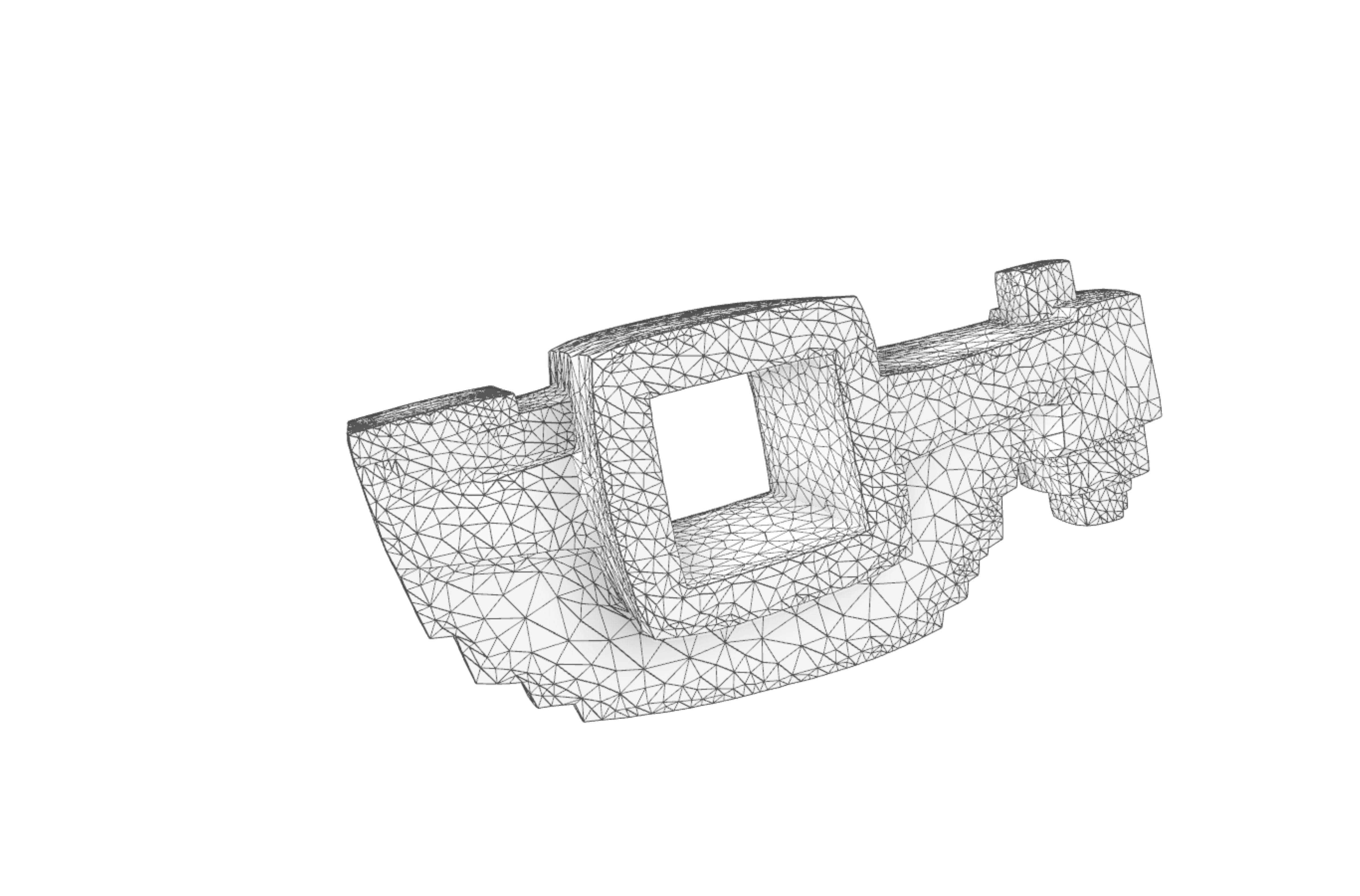} &
    \includegraphics[height=1.2in,trim=4.0in 1.in 2.2in 0, clip]{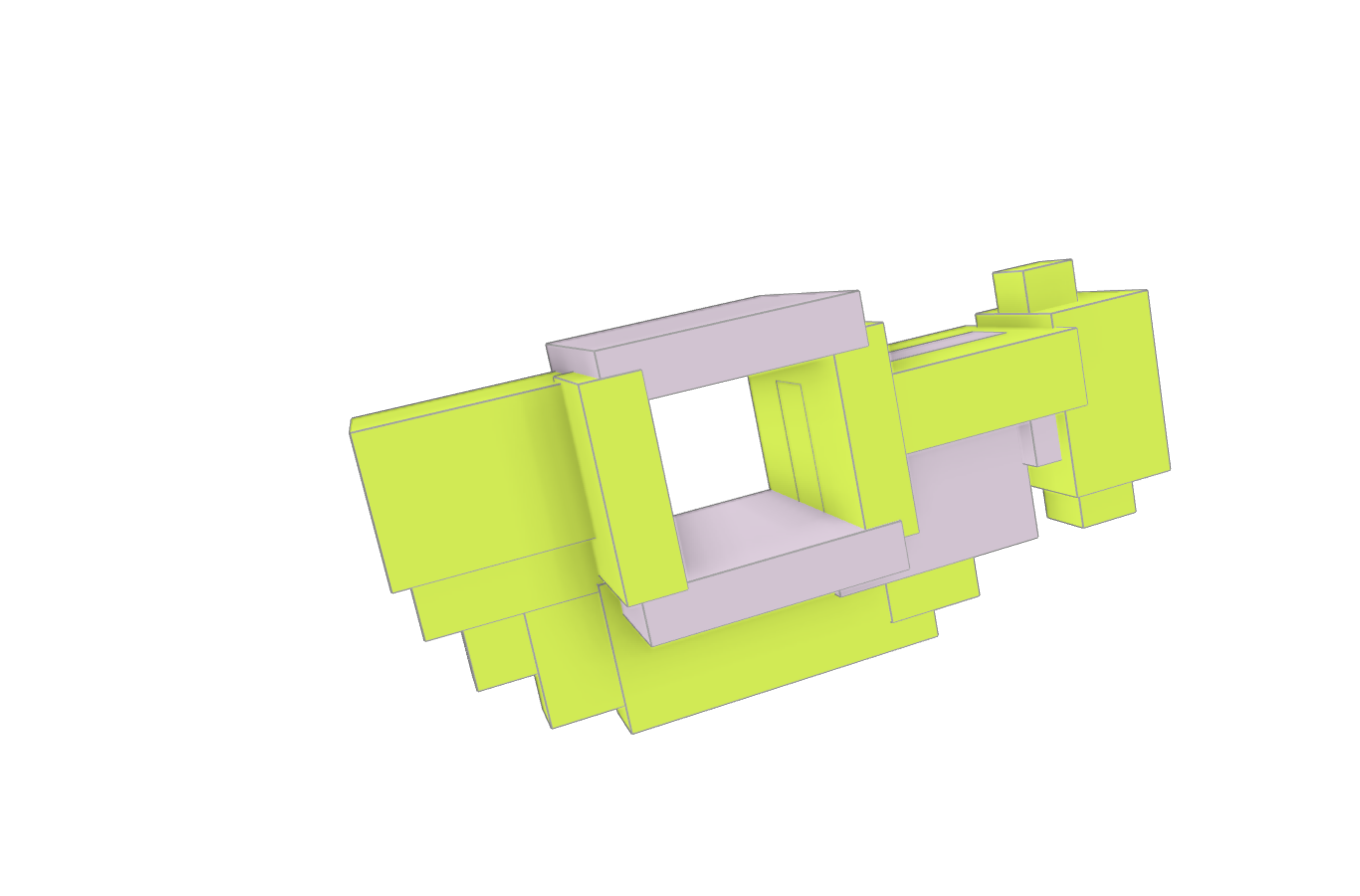} &
    \includegraphics[height=1.2in,trim=4.0in 1.in 2.2in 0, clip]{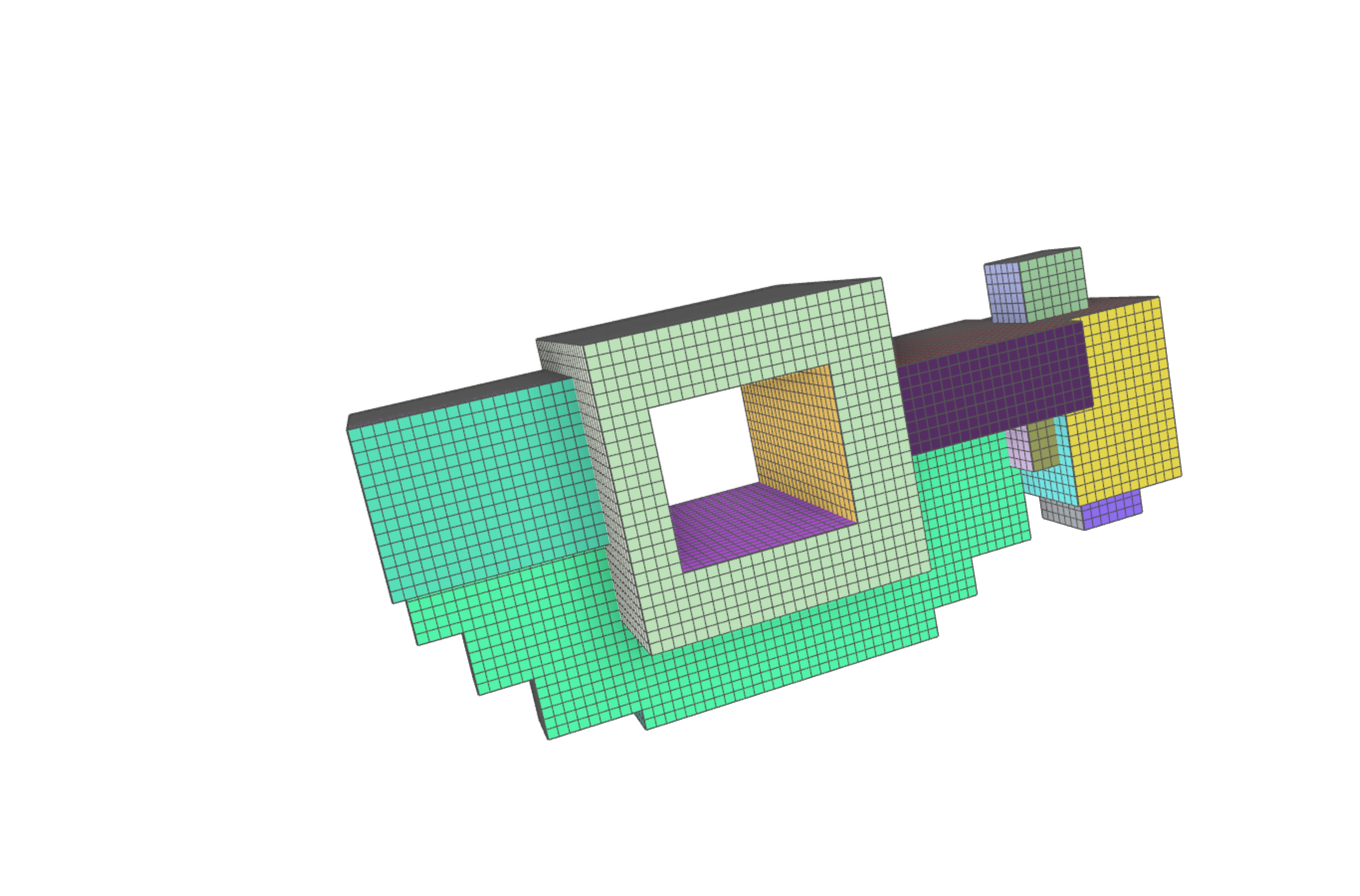} &
    \includegraphics[height=1.2in,trim=4.0in 1.in 2.2in 0, clip]{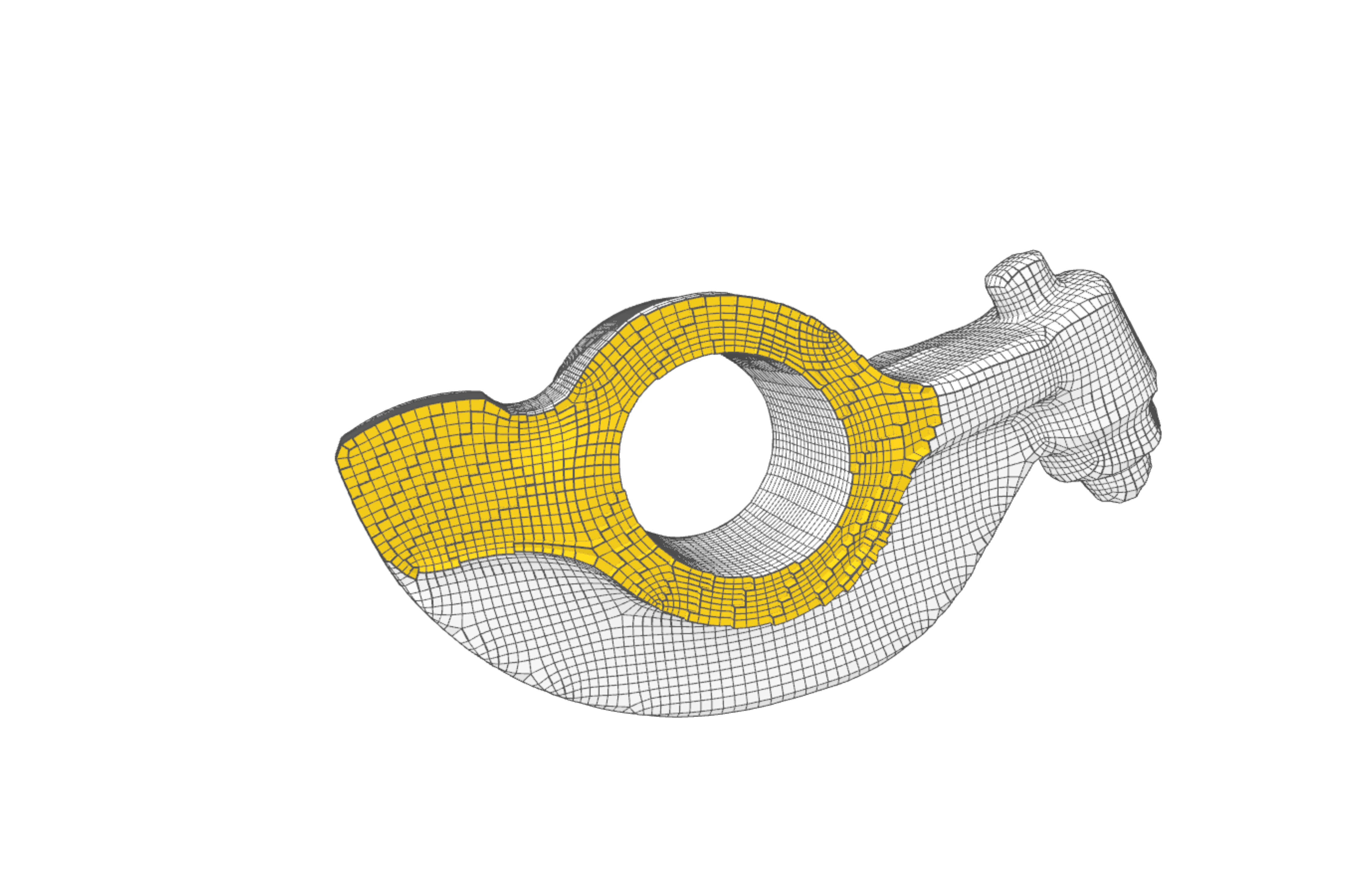}  \\
    {\footnotesize (a) Input tet mesh} & {\footnotesize (b) Deformed tet mesh} & {\footnotesize (c) Decomposed PolyCube} & {\footnotesize (d) Discretized PolyCube} & {\footnotesize (e) Output hex mesh}
  \end{tabular}
  \caption{Overview of our interactive hex meshing pipeline. (b)--(e) correspond to the output of the four stages of the pipeline. The output hex mesh (e) is sliced open, with yellow indicating interior quad faces for visualization.}
  \label{fig:pipeline_stages}
\end{figure*}

Our system inputs a tetrahedral mesh and, with user guidance, generates a hexahedral mesh whose surface matches that of the input. 
If the input is only a triangular surface mesh, we run \cite{hu2020fast} to tetrahedralize its volume as preprocessing. All input meshes are centered and \rev{rescaled} to fit within a unit bounding box.
Like past PolyCube-based methods, we limit the class of generated hex meshes to those with the topology of a voxelized PolyCube hex mesh; in particular, all singularities are on the surface (except for a layer of global padding).
\cref{fig:pipeline_stages} illustrates our pipeline.

Our system employs several user-in-the-loop stages that give the user fine-grained control over the pipeline. 
After deforming the input mesh into a near-PolyCube shape whose faces are almost axis-aligned (\cref{stage:deformation}, \cref{fig:pipeline_stages}(b)), the system generates a PolyCube composed of a collection of cuboids %
whose union approximates the deformed shape in a semi-automatic fashion (\cref{stage:decomposition}, \cref{fig:pipeline_stages}(c)).
The PolyCube is then voxelized to a hex mesh, which the user can further modify (\cref{stage:discretization}, \cref{fig:pipeline_stages}(d)).
Lastly, the voxelized PolyCube is mapped to the input mesh so that the mapped surface agrees with the input mesh surface (\cref{stage:hexahedralization}, \cref{fig:pipeline_stages}(e)). 

Please refer to the supplemental video for a demo of our system.

\subsection{Deformation stage}
\label{stage:deformation}
In the first stage of the pipeline, an input tetrahedral mesh is deformed into a near-PolyCube shape (see the second column of \cref{fig:pipeline_stages}). 
While the deformed shape is not a (strict) PolyCube (its face normals can deviate from the main axis directions), this deformation makes it easier to approximate the shape using cuboids in the following stage by, e.g., aligning it with the coordinate axes and reducing the numbers of corners and creases.

\begin{figure}
  \centering
  \newcolumntype{C}[1]{>{\centering\arraybackslash}m{#1}}
  \begin{tabular}{C{1.8cm} C{1.8cm} C{1.8cm} C{1.8cm}}
    \includegraphics[width=0.8in, trim=4.5in 1in 5in 1in, clip]{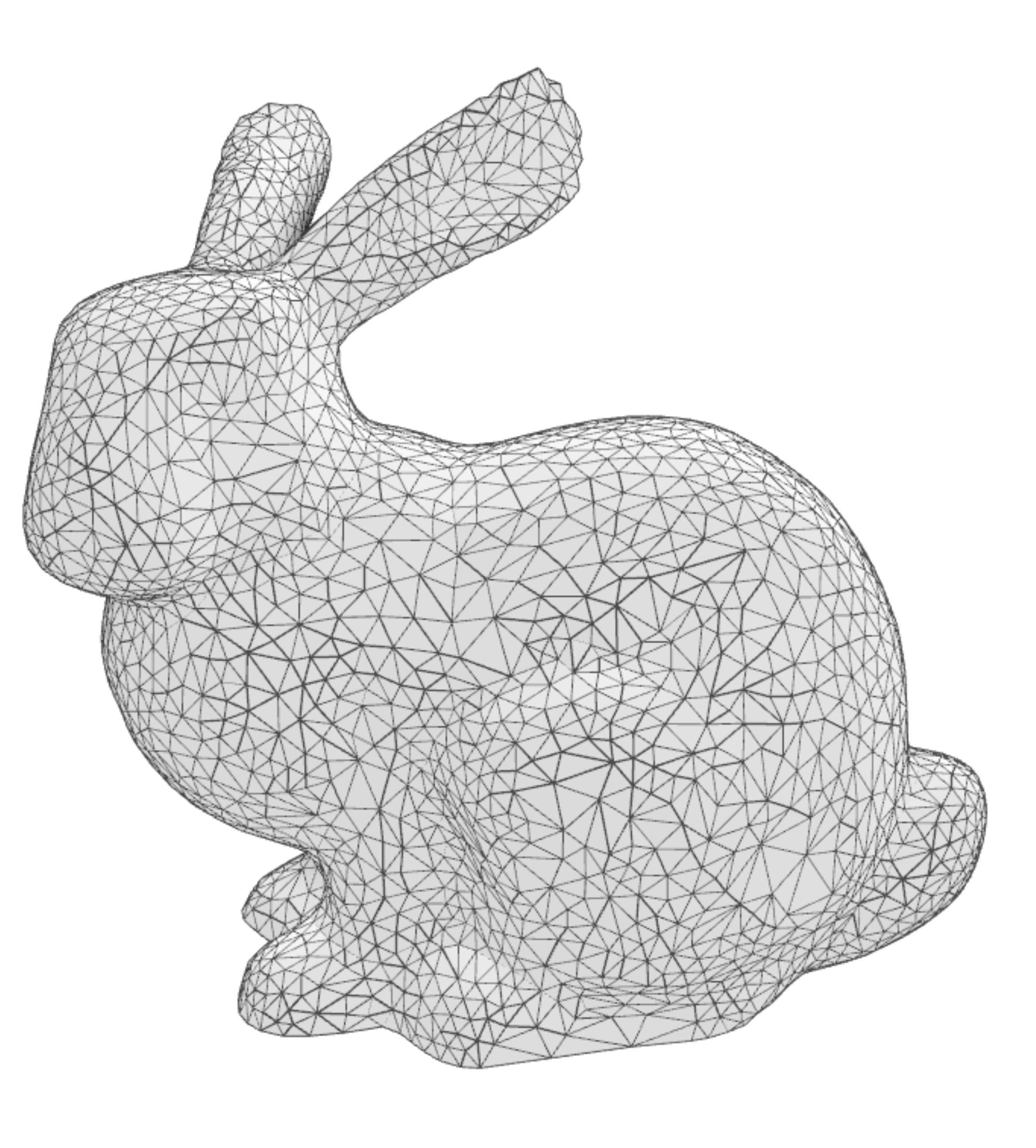} &
    \includegraphics[width=0.8in, trim=4.5in 1in 5in 1in, clip]{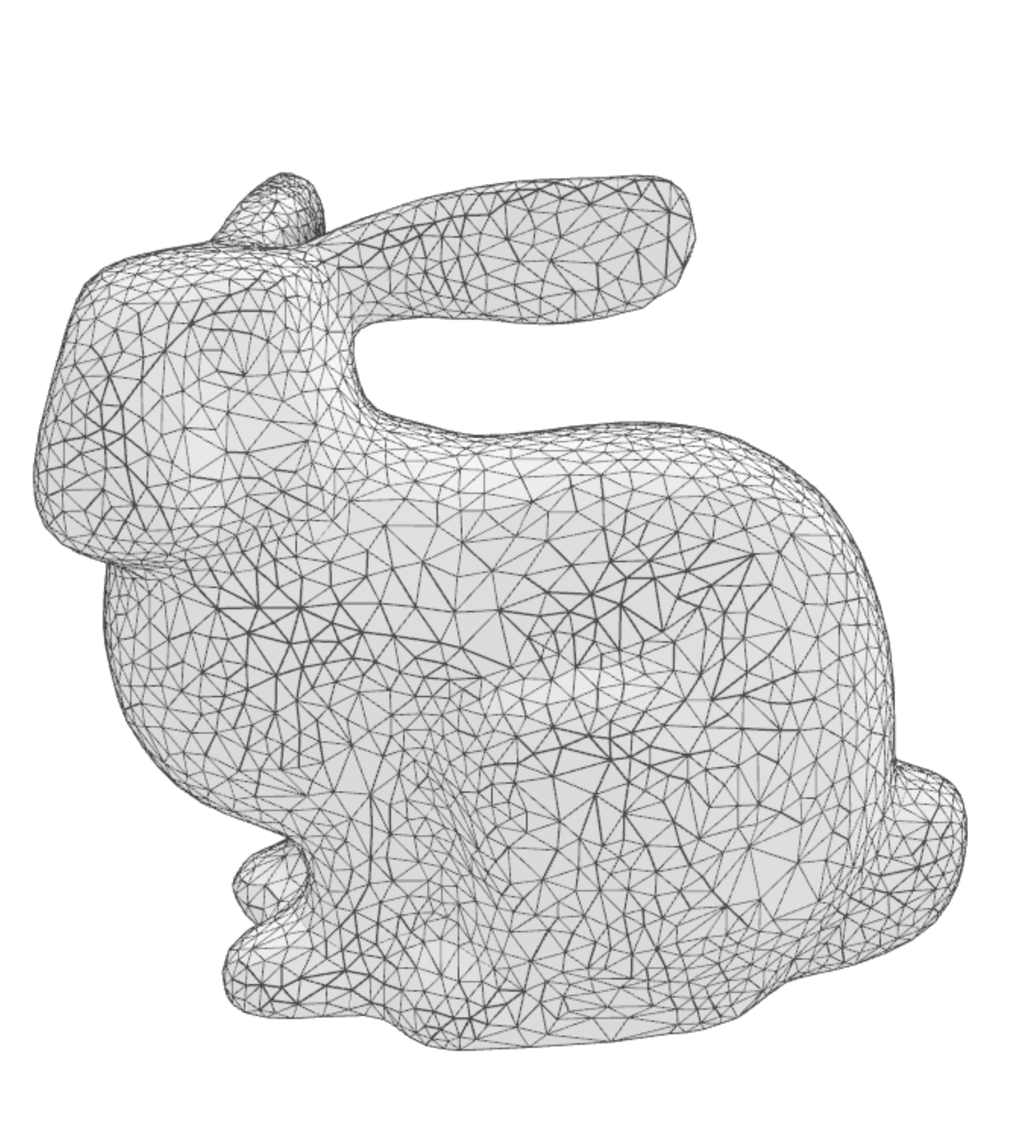} &
    \includegraphics[width=0.8in, trim=4.5in 1in 5in 1in, clip]{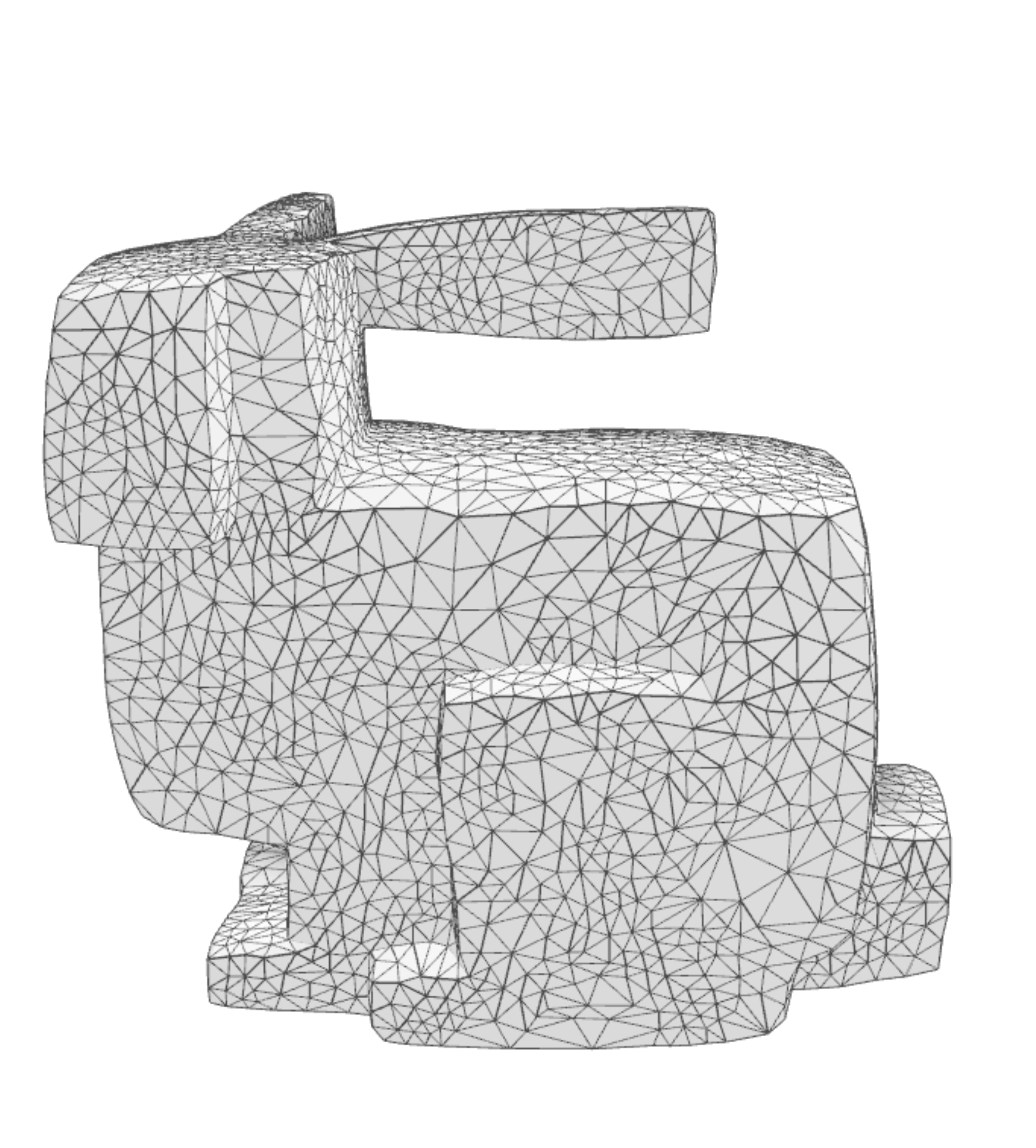} &
  \includegraphics[width=0.8in, trim=4.5in 1in 5in 1in, clip]{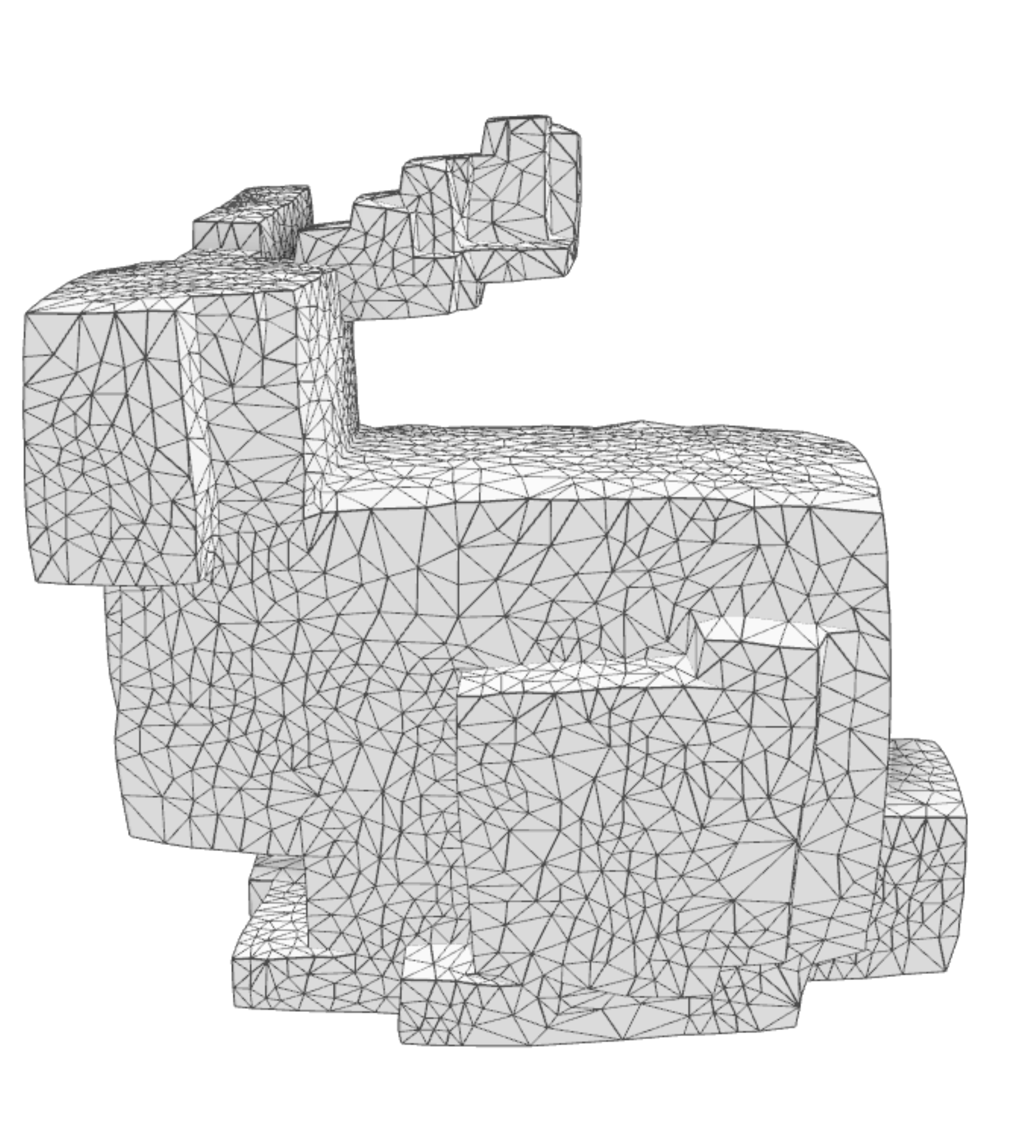} \\
    {\footnotesize (a) Input mesh} & {\footnotesize (b) Low cubeness} & {\footnotesize (c) Ramping up to high cubeness} & 
{\footnotesize (d) Directly using high cubeness}
\end{tabular}
  \caption{User-guided deformation. Starting with an input mesh (a), first using a low cubeness value ($=0.3$) orients the coarse features such as the bunny ears to be axis-aligned (b). Then, increasing the cubeness value ($=3.0$) deforms the shape closer to a PolyCube by creating sharp edges (c). 
  In contrast, directly using a high cubeness value creates unnecessary stairs on the ears of the bunny (d).}
  \label{fig:cubeness_adjustment}
\end{figure}

To achieve the goals above, a low-distortion deformation map is computed that encourages normals to be aligned with \rev{the} main axis directions (\cref{sec:opt_deformation}).
The user has the option to control the deformation by changing \emph{cubeness} and \emph{smoothness} parameters that control how PolyCube-like and smooth they want the surface to be.
In addition, they can modify the parameters before resuming the deformation process. 
A typical use case is to start with a low cubeness value so the shape is globally axis-aligned and then gradually increase the cubeness parameter to deform the shape closer to a PolyCube with few corners (\cref{fig:cubeness_adjustment}).

\subsection{Decomposition stage}
\label{stage:decomposition}
In this second stage, the user guides the creation of a PolyCube represented as a collection of axis-aligned cuboids whose union approximates the near-PolyCube from the previous stage.
The quality of the generated PolyCube is determined by how closely it approximates the deformed shape and its complexity, as more complex PolyCubes (e.g., with more corners) lead to more surface singularities in the resulting hex mesh.

\begin{figure}
  \centering
  \hspace*{-0.1in}
  \newcolumntype{C}[1]{>{\centering\arraybackslash}m{#1}}
  \begin{tabular}{C{1.7cm}  C{1.7cm}  C{1.7cm} C{1.7cm}}
    \includegraphics[height=0.7in,trim=3.5in 1.8in 0 1.5in, clip]{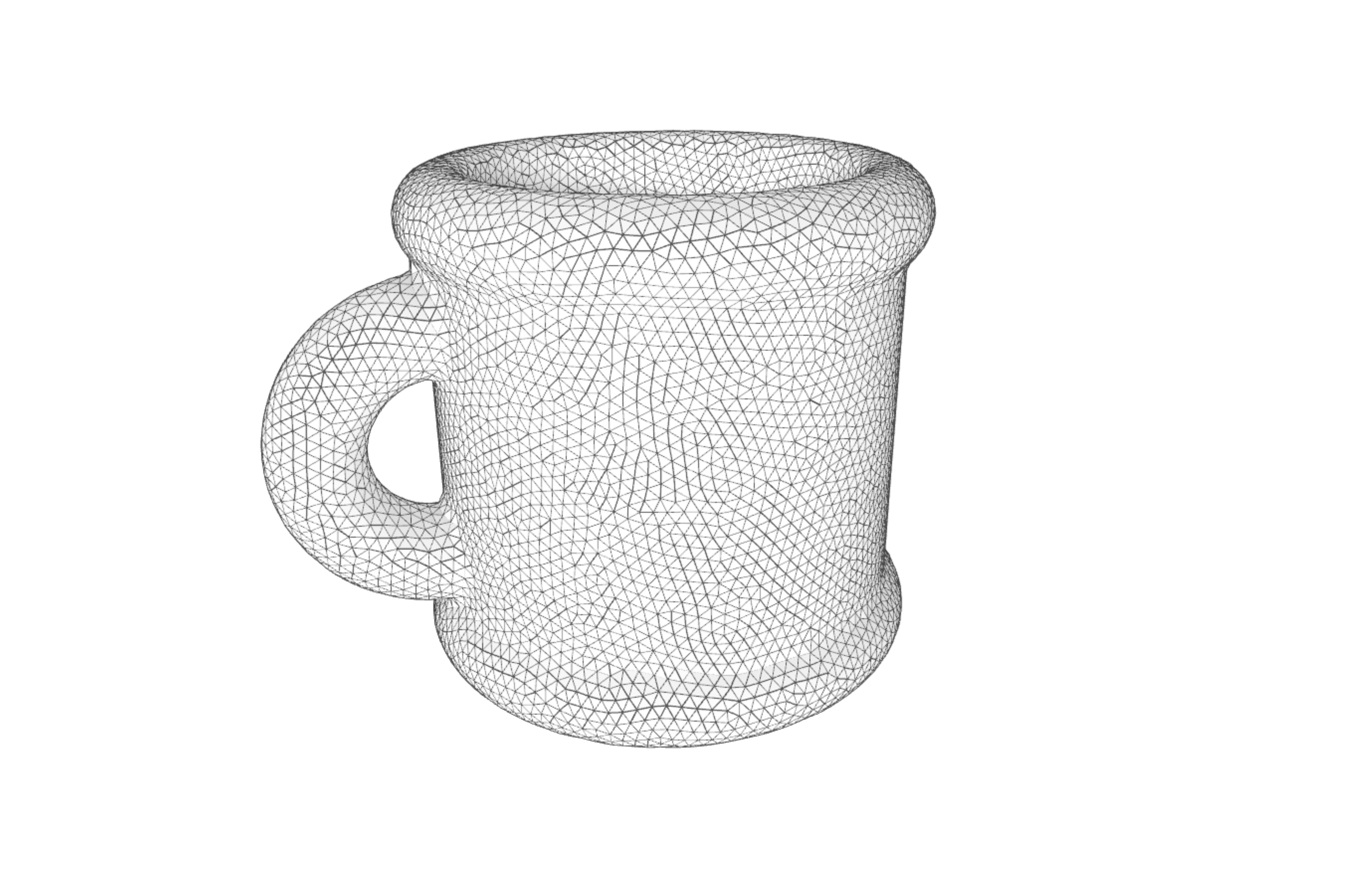} 
    \includegraphics[height=0.7in,trim=3.5in 1in 0 1.5in, clip]{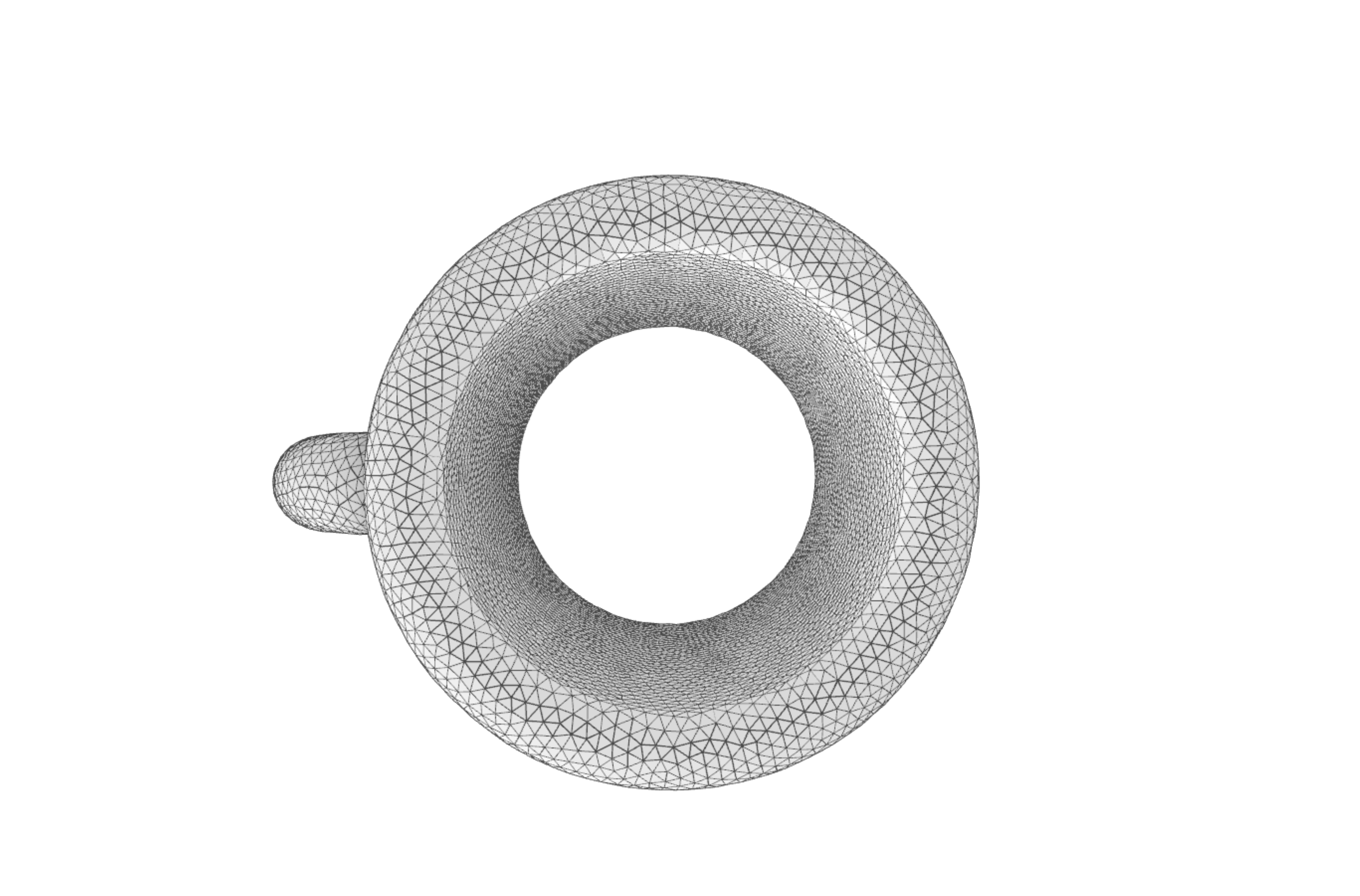}
    &
    \includegraphics[height=0.7in,trim=3.5in 1.8in 0 1.5in, clip]{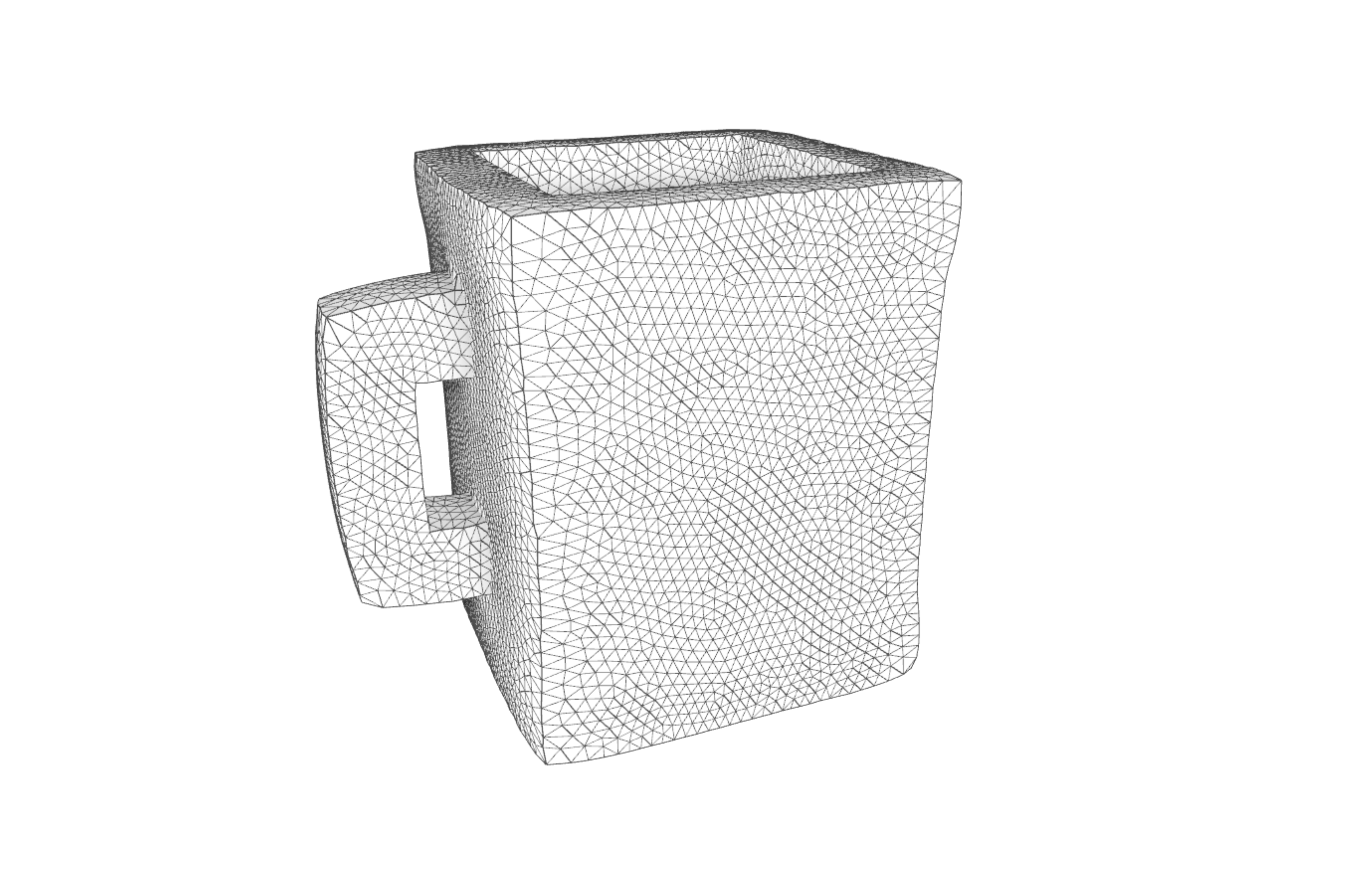} 
    \includegraphics[height=0.7in,trim=3.5in 1in 0 1.5in, clip]{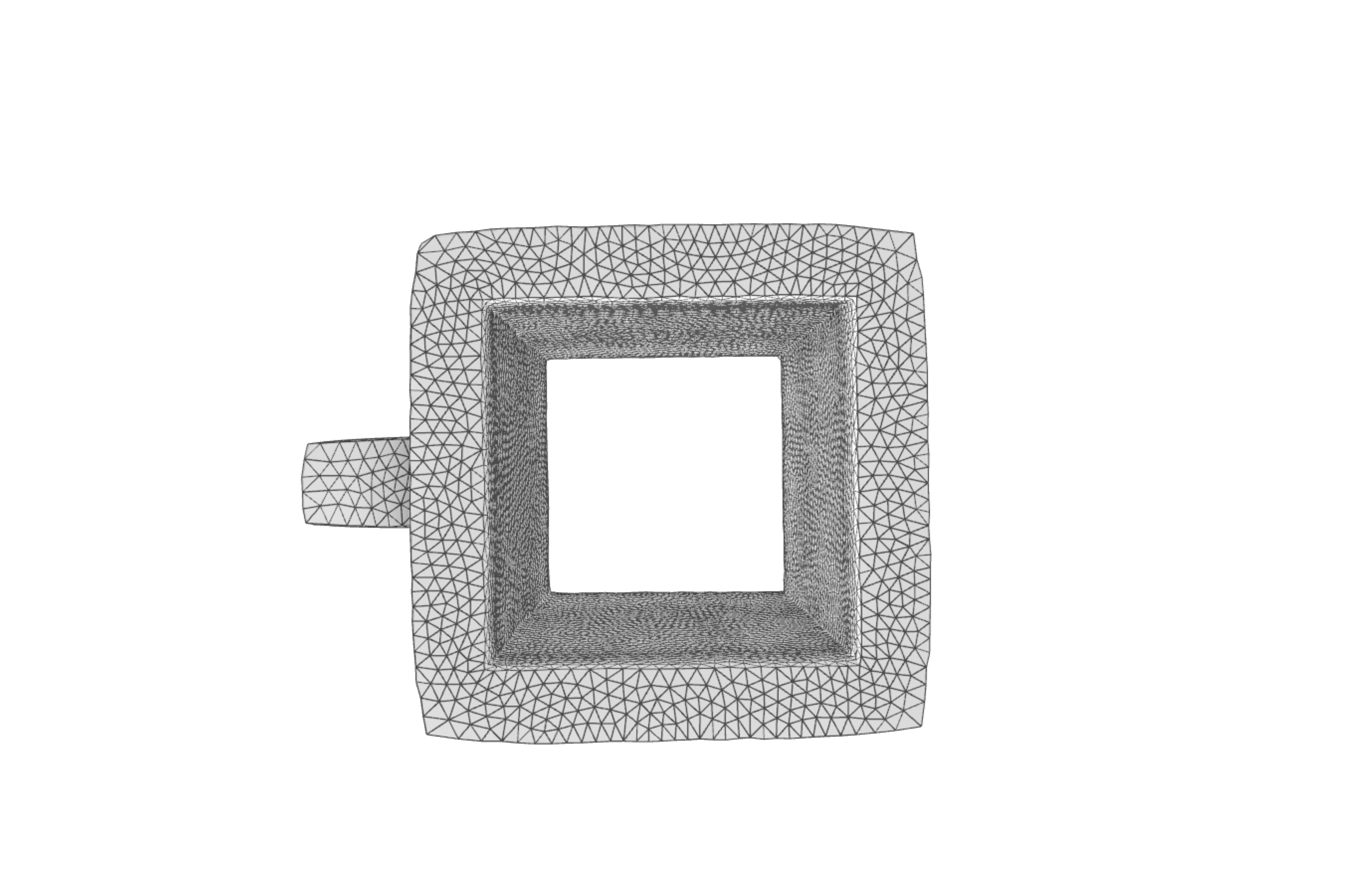}
     &
    \includegraphics[height=0.7in,trim=3.5in 1.8in 0 1.5in, clip]{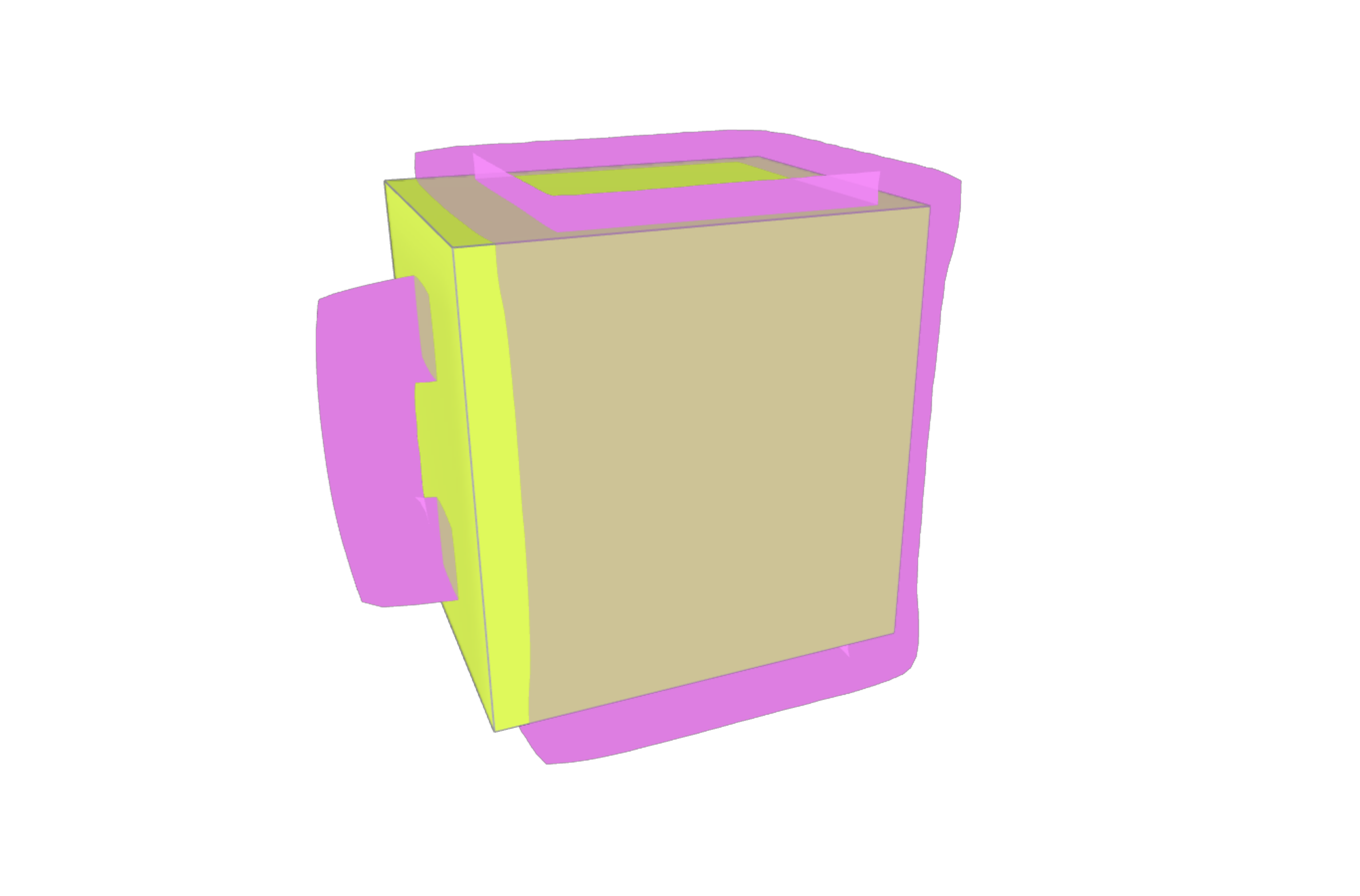} 
    \includegraphics[height=0.7in,trim=3.5in 1in 0 1.5in, clip]{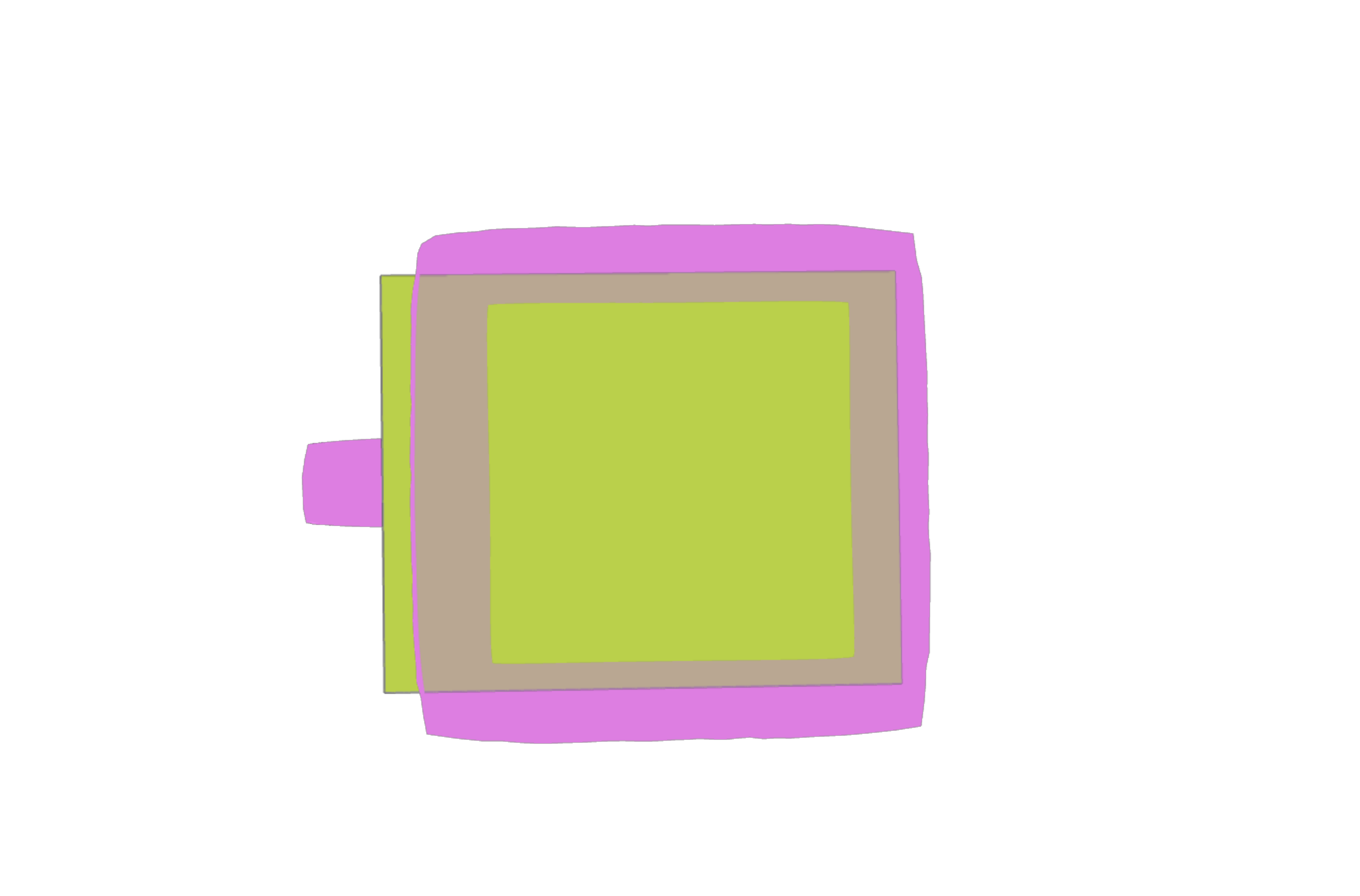}
     &
    \includegraphics[height=0.7in,trim=3.5in 1.8in 0 1.5in, clip]{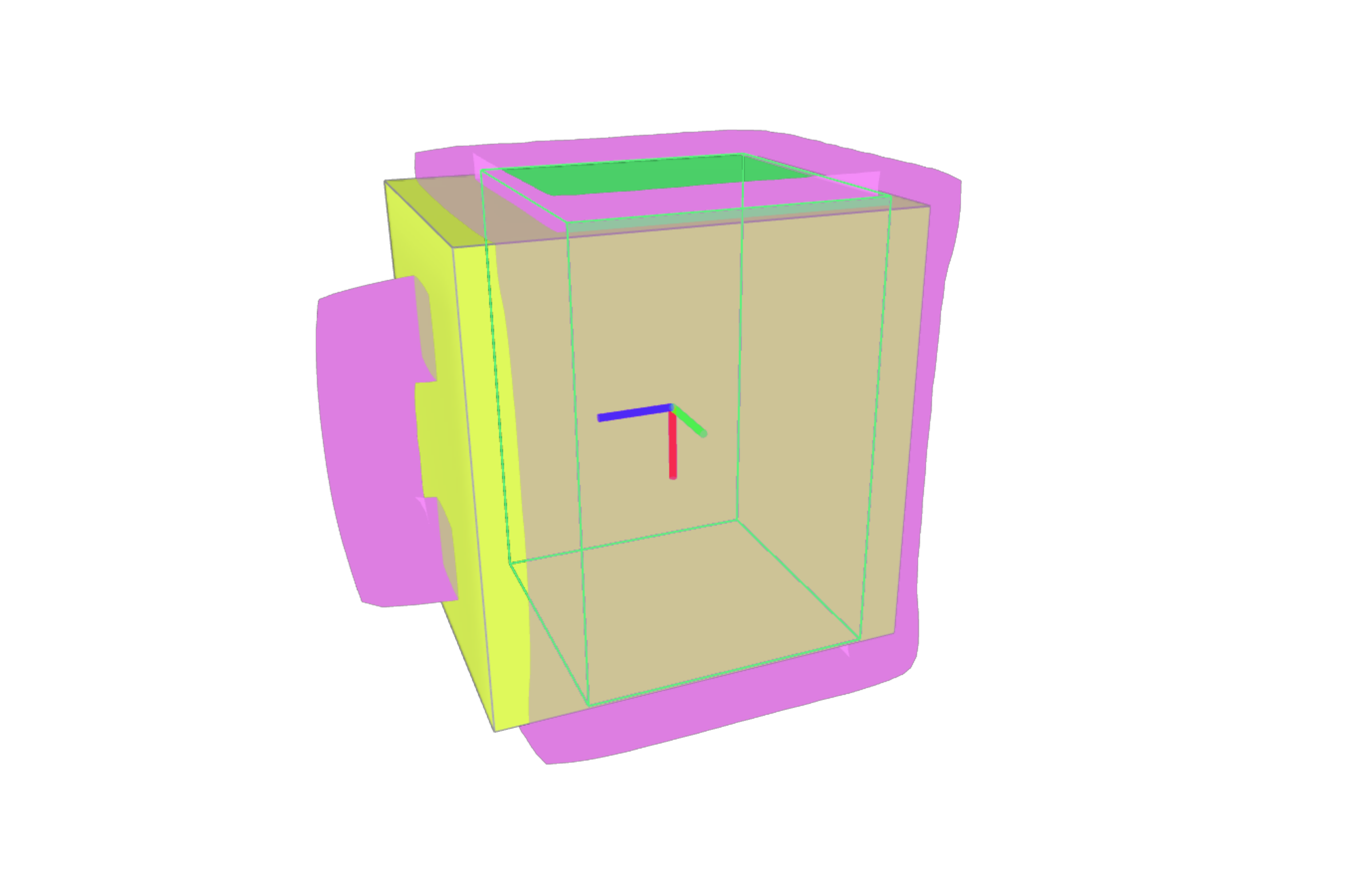} 
    \includegraphics[height=0.7in,trim=3.5in 1in 0 1.5in, clip]{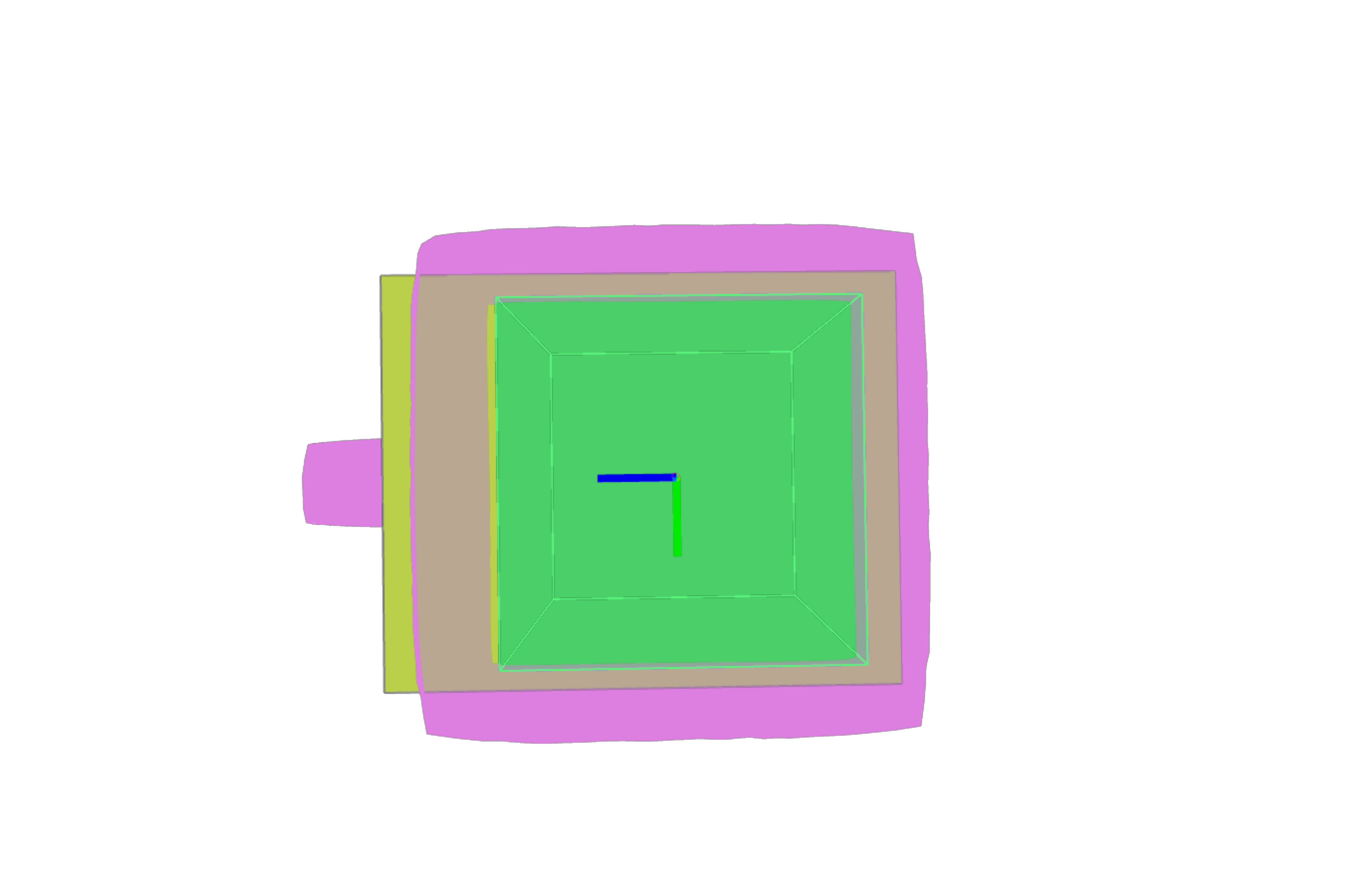}
    \\
    {\footnotesize (a) Input} & {\footnotesize (b) Deformed} & {\footnotesize (c) Optimized single cuboid} & {\footnotesize (d) Suggested \textit{Subtract} region}\\
    \includegraphics[height=0.7in,trim=3.5in 1.8in 0 1.5in, clip]{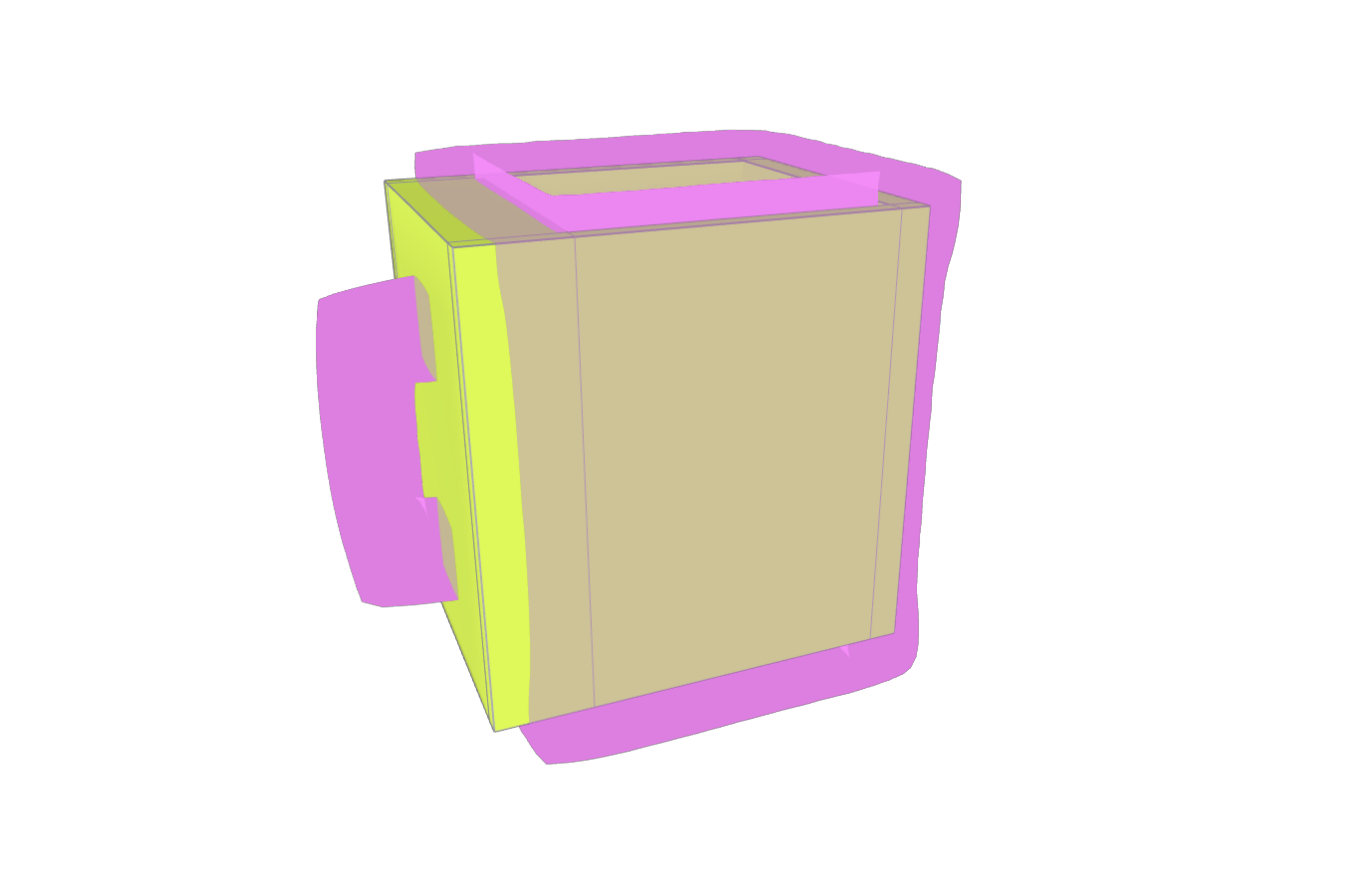} 
    \includegraphics[height=0.7in,trim=3.5in 1in 0 1.5in, clip]{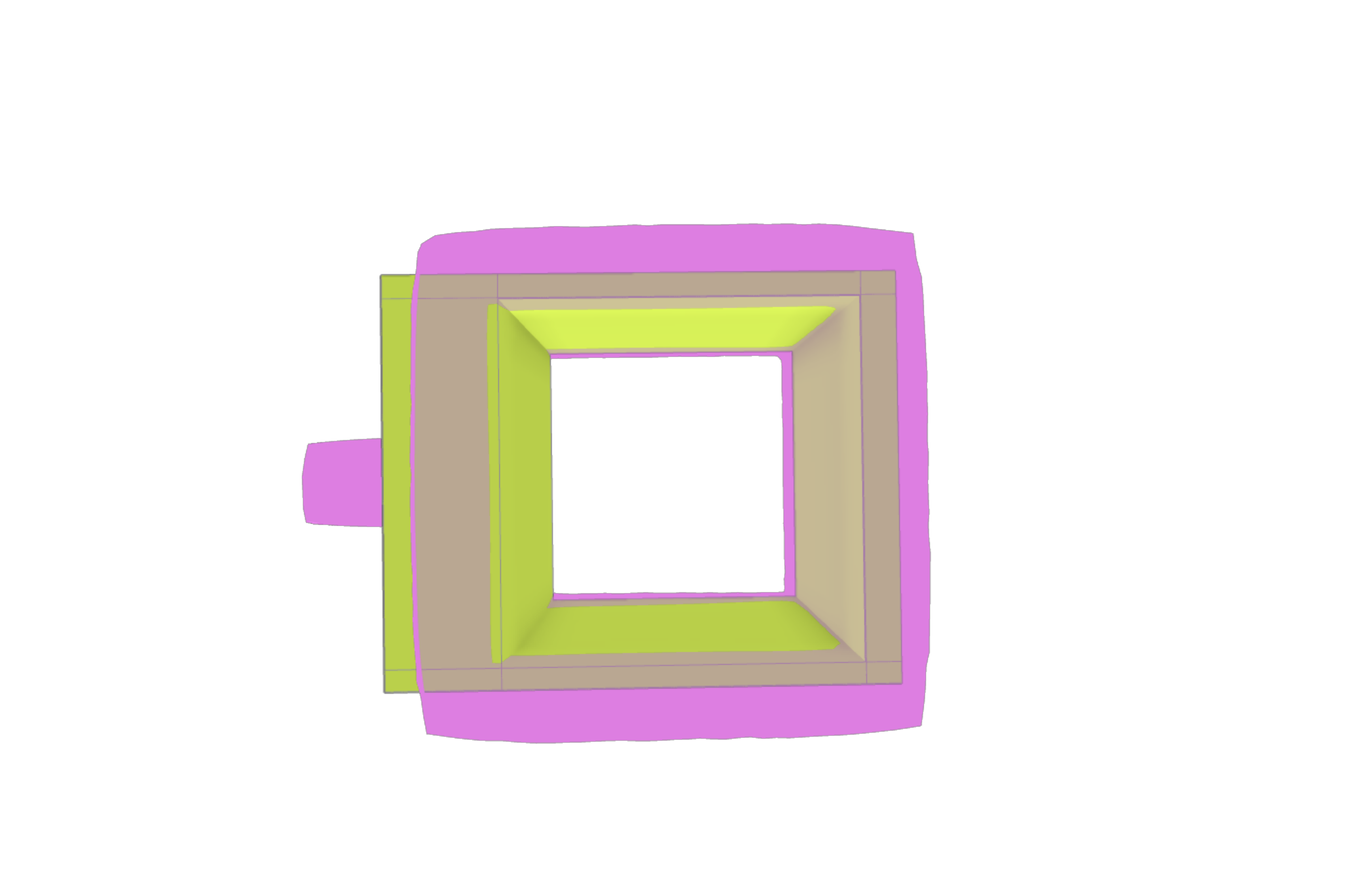}
    &
    \includegraphics[height=0.7in,trim=3.5in 1.8in 0 1.5in, clip]{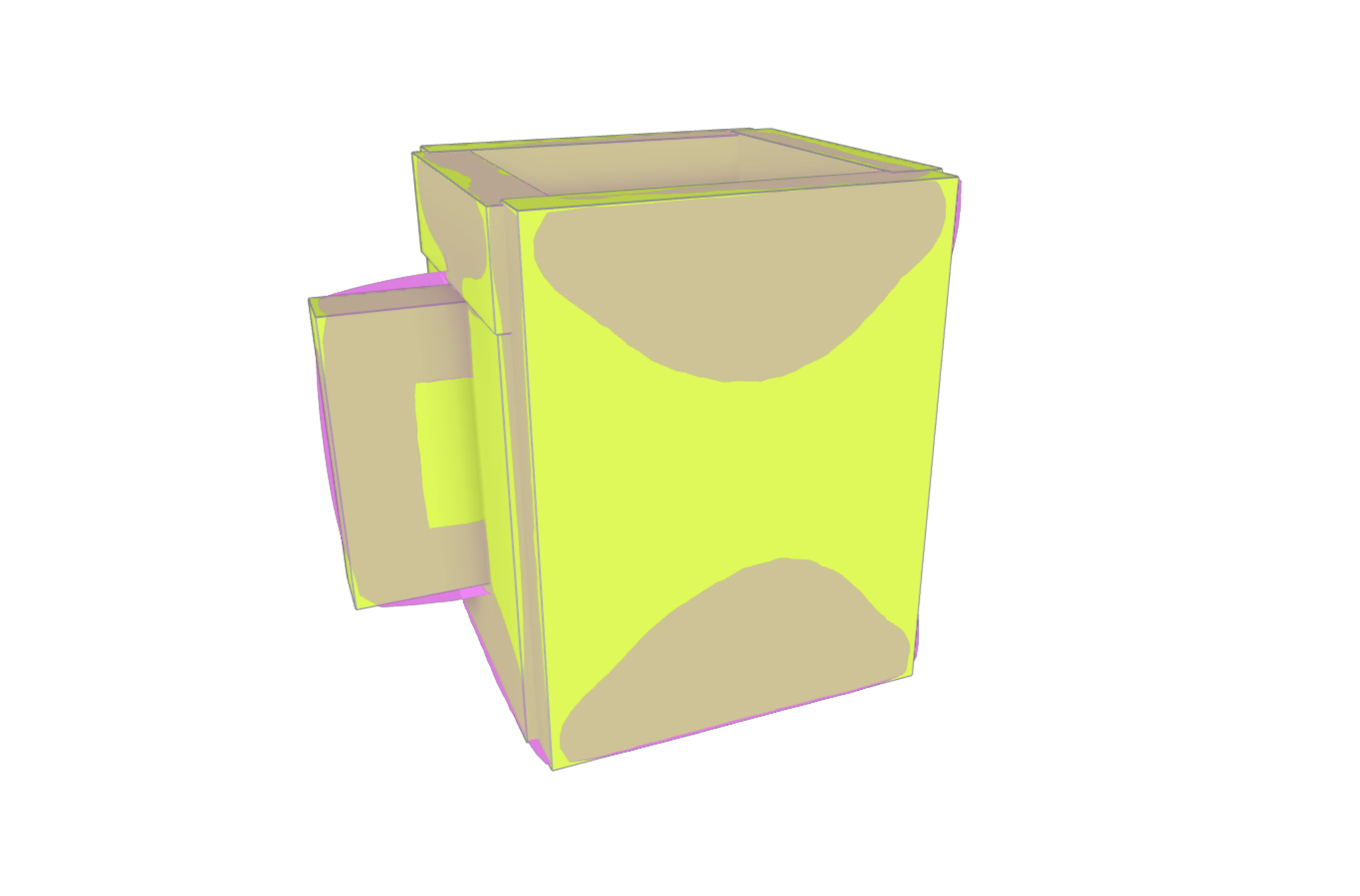} 
    \includegraphics[height=0.7in,trim=3.5in 1in 0 1.5in, clip]{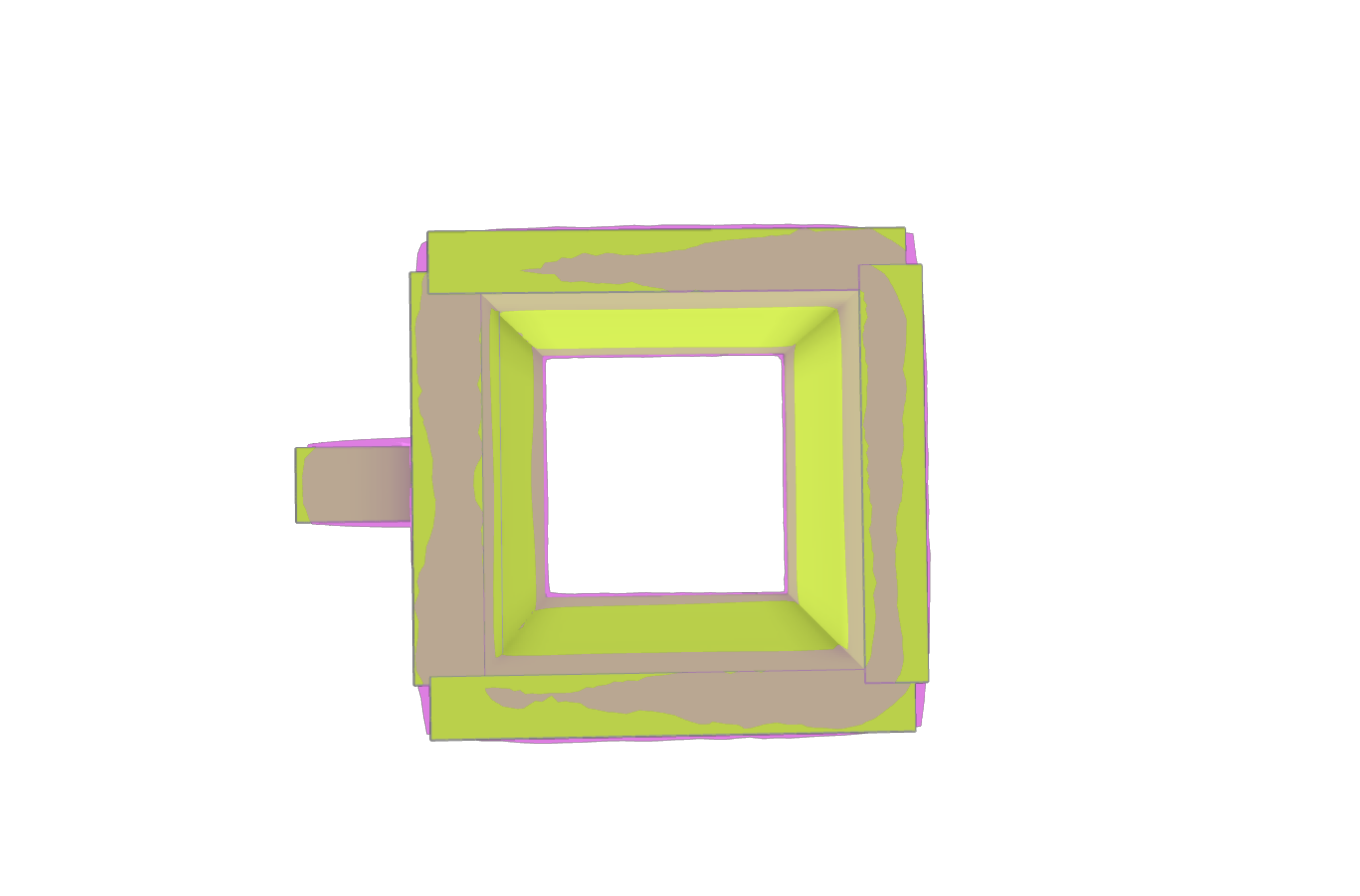}
    &
    \includegraphics[height=0.7in,trim=3.5in 1.8in 0 1.5in, clip]{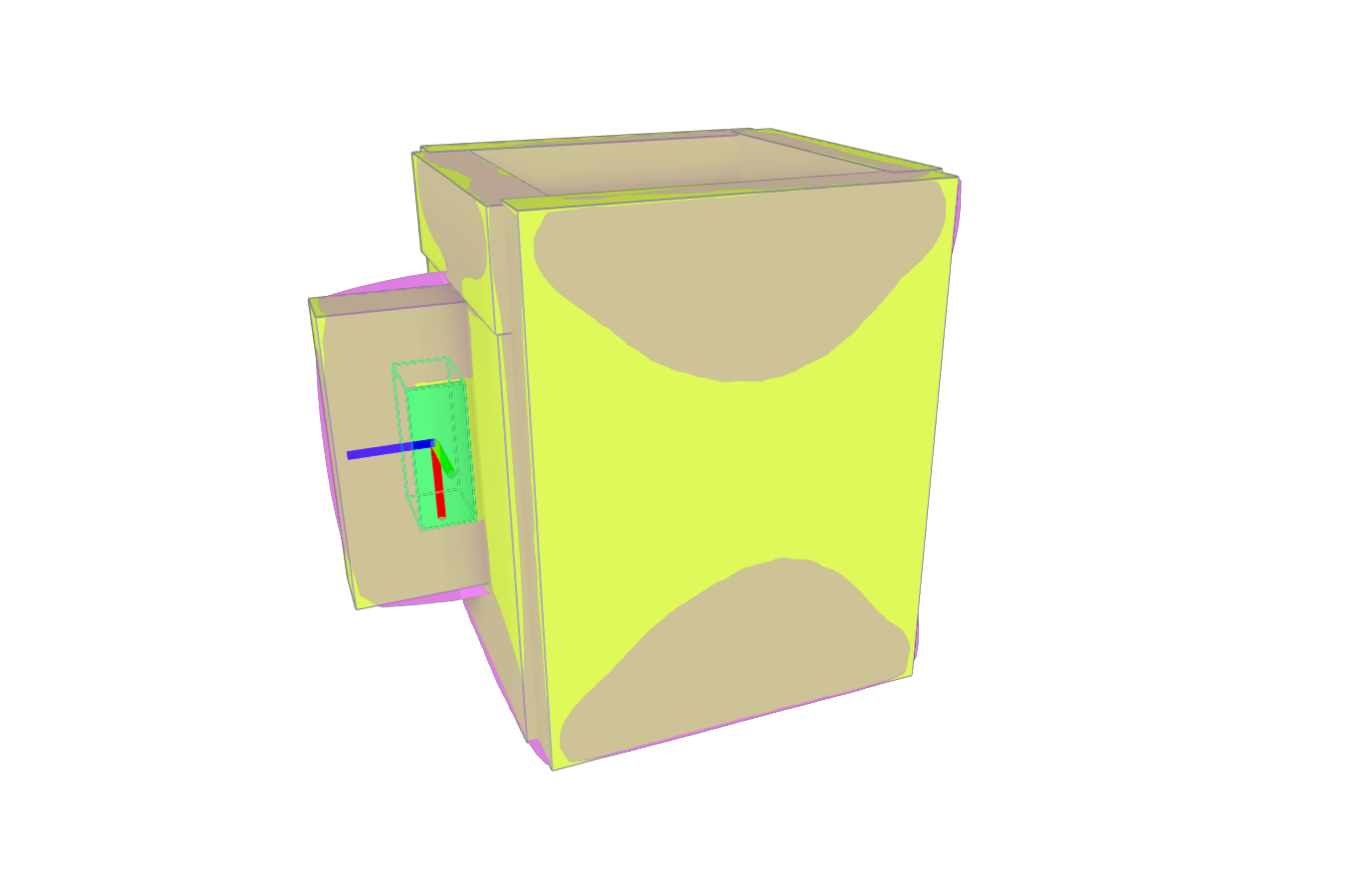} 
    \includegraphics[height=0.7in,trim=3.5in 1in 0 1.5in, clip]{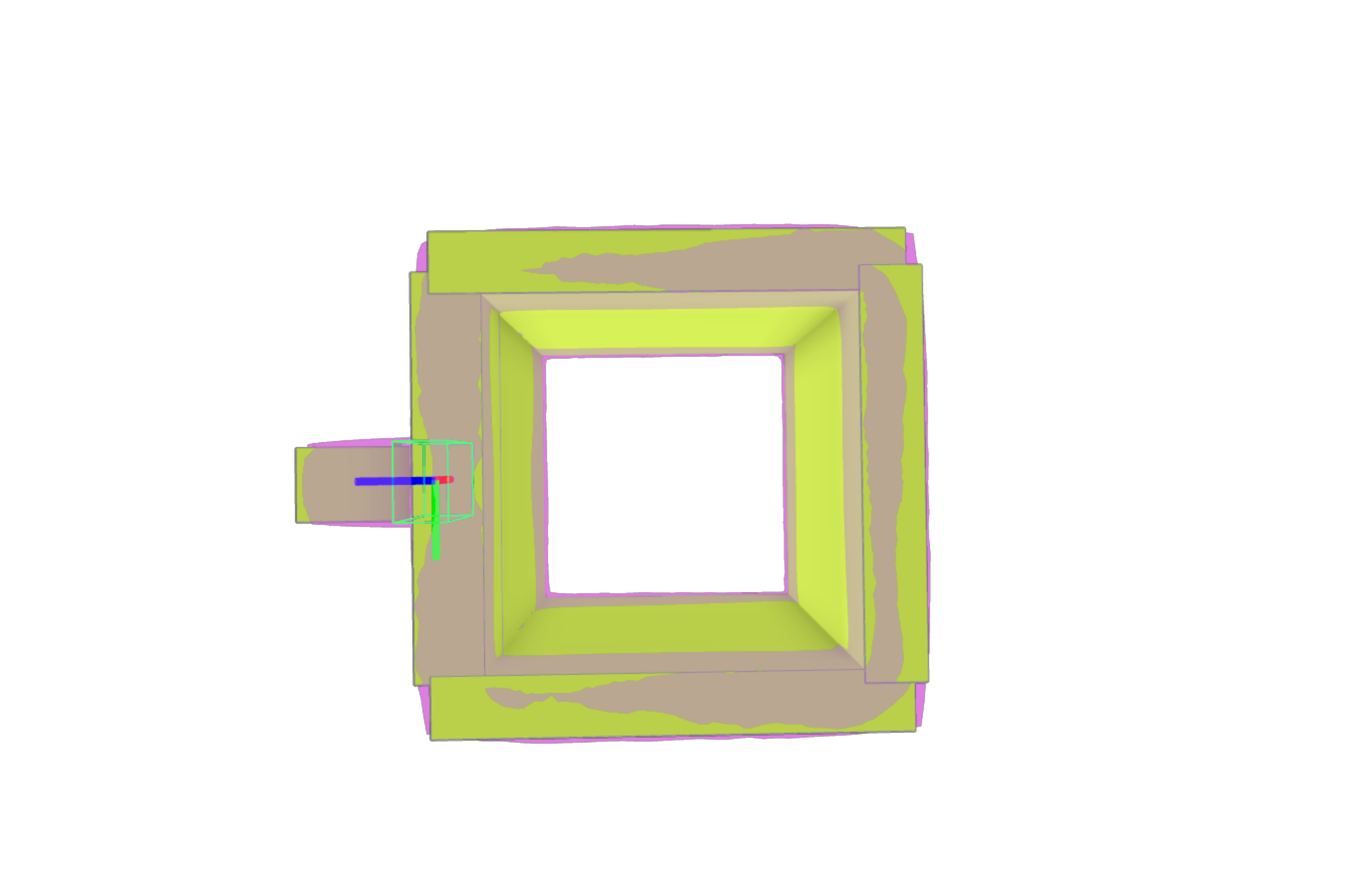}
    &
    \includegraphics[height=0.7in,trim=3.5in 1.8in 0 1.5in, clip]{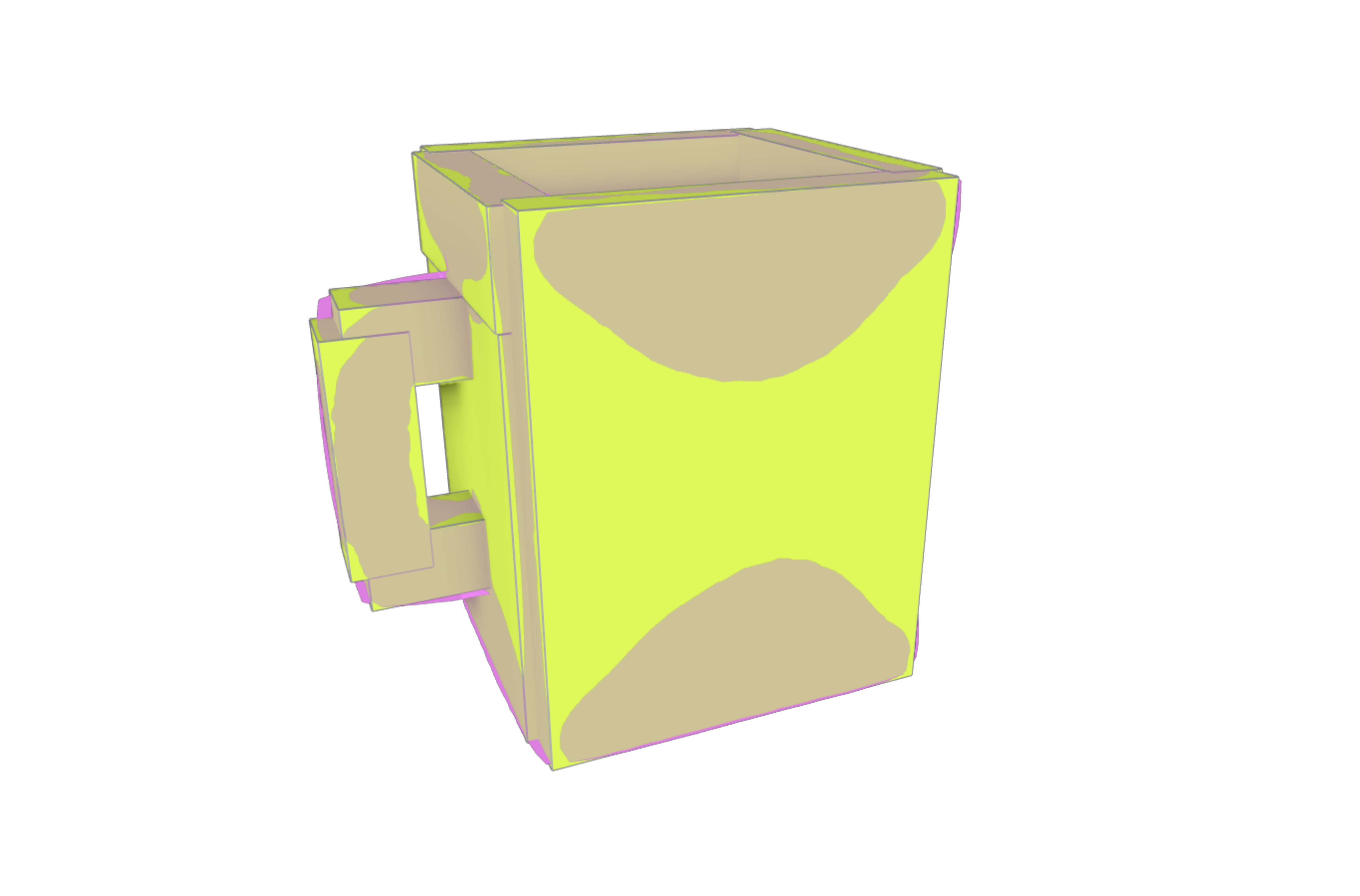} 
    \includegraphics[height=0.7in,trim=3.5in 1in 0 1.5in, clip]{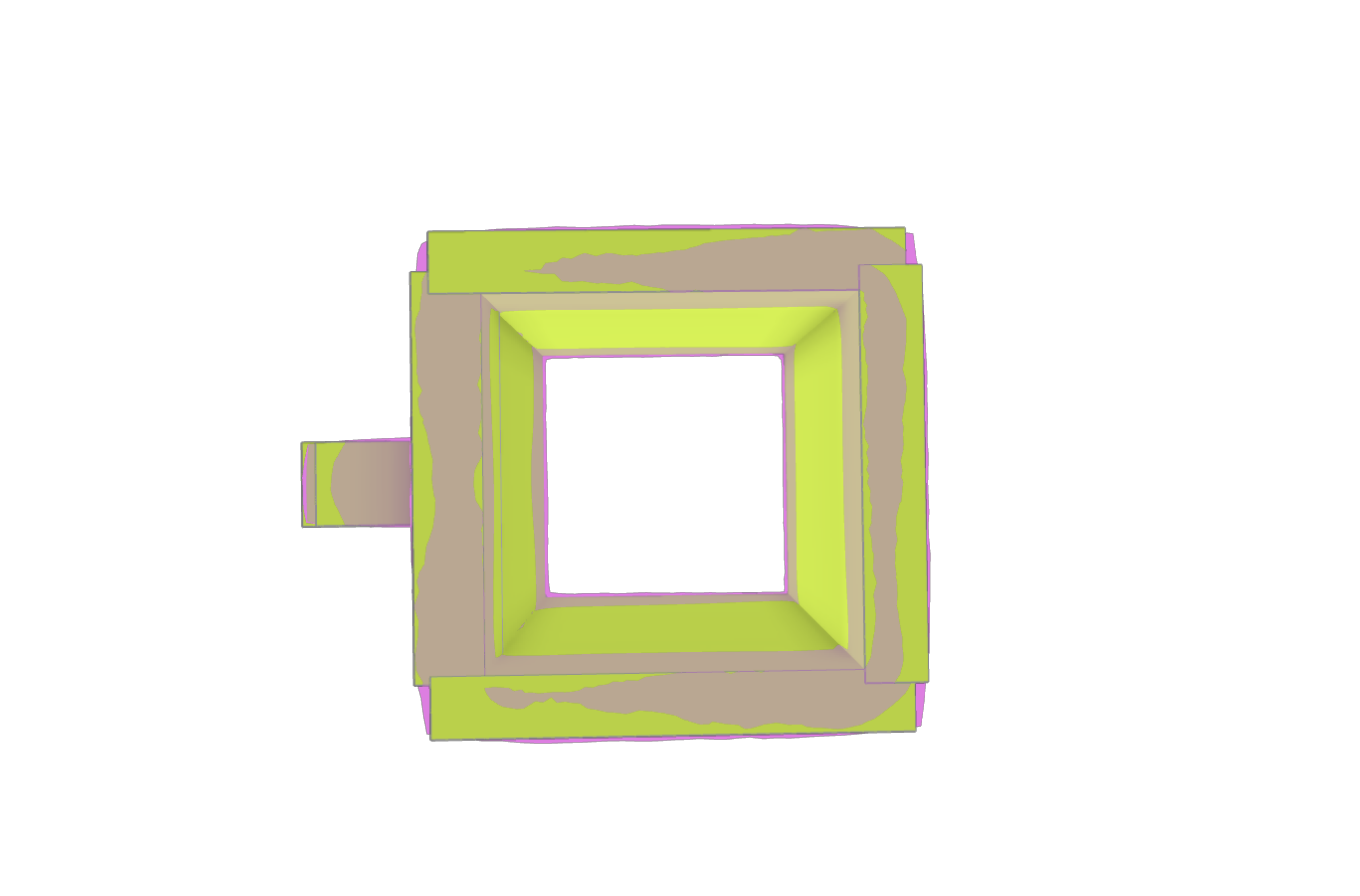}
      \\
    {\footnotesize (e) After \textit{Subtract}}& {\footnotesize (f) \textit{Add} and \textit{Reoptimize}} & {\footnotesize (g) \textit{Subtract} again } & {\footnotesize (h) Final PolyCube}
  \end{tabular}
  \caption{An example of constructing the PolyCube in the decomposition stage using the \textit{cup} model.
    A front view and a top-down view are shown for each entry. The surface of the deformed mesh is shown in translucent pink.
    (c) A single cuboid covers the hole of the cup after \textit{Reoptimize}.
    (d) The user sees a green editable region suggested by the system indicating that this region can be subtracted out.
    (e) The PolyCube after performing \textit{Subtract}.
    (f) The user can then proceed to add new cuboids and \textit{Reoptimize}, which may over-cover the cup handle.
    (g) Another \textit{Subtract} can be performed to recover the handle to get (h).
    Further \textit{Edit} operations can be done to reduce the number of corners.
    Alternatively, the user can avoid using \textit{Subtract} for this model by either directly editing the cuboids or using volume-based \textit{Add} to put a cuboid on each wall of the deformed cup.
  }
\label{fig:decomposition_editing}
\end{figure}

\begin{figure}
  \centering
    \hspace*{-0.1in}
  \begin{tabular}{m{1cm} m{1cm} m{1cm} m{1cm} m{1cm}}
    \includegraphics[height=1.3in,trim=7.0in 1in 2.9in 0, clip]{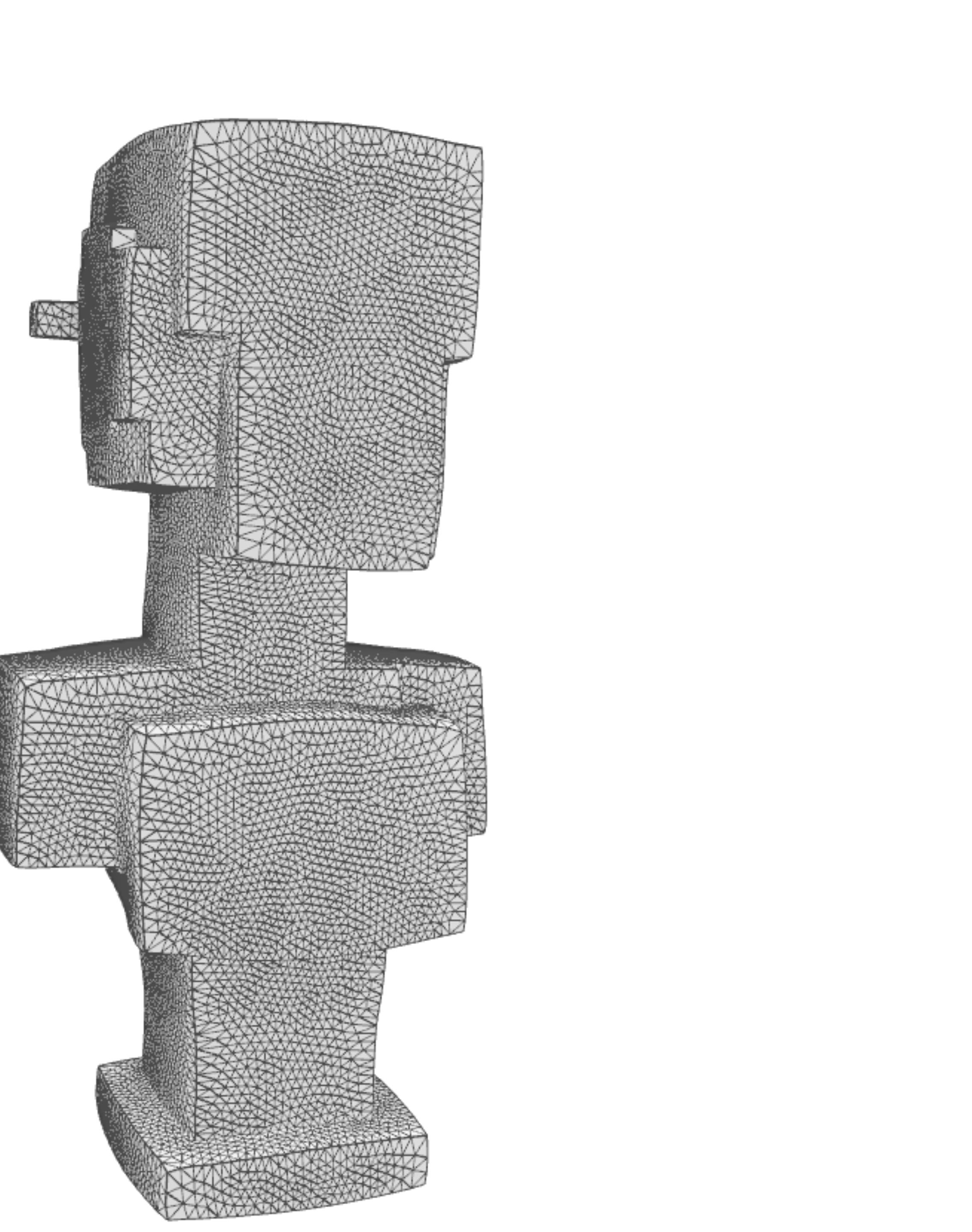} &
    \includegraphics[height=1.3in,trim=7.0in 1in 2.9in 0, clip]{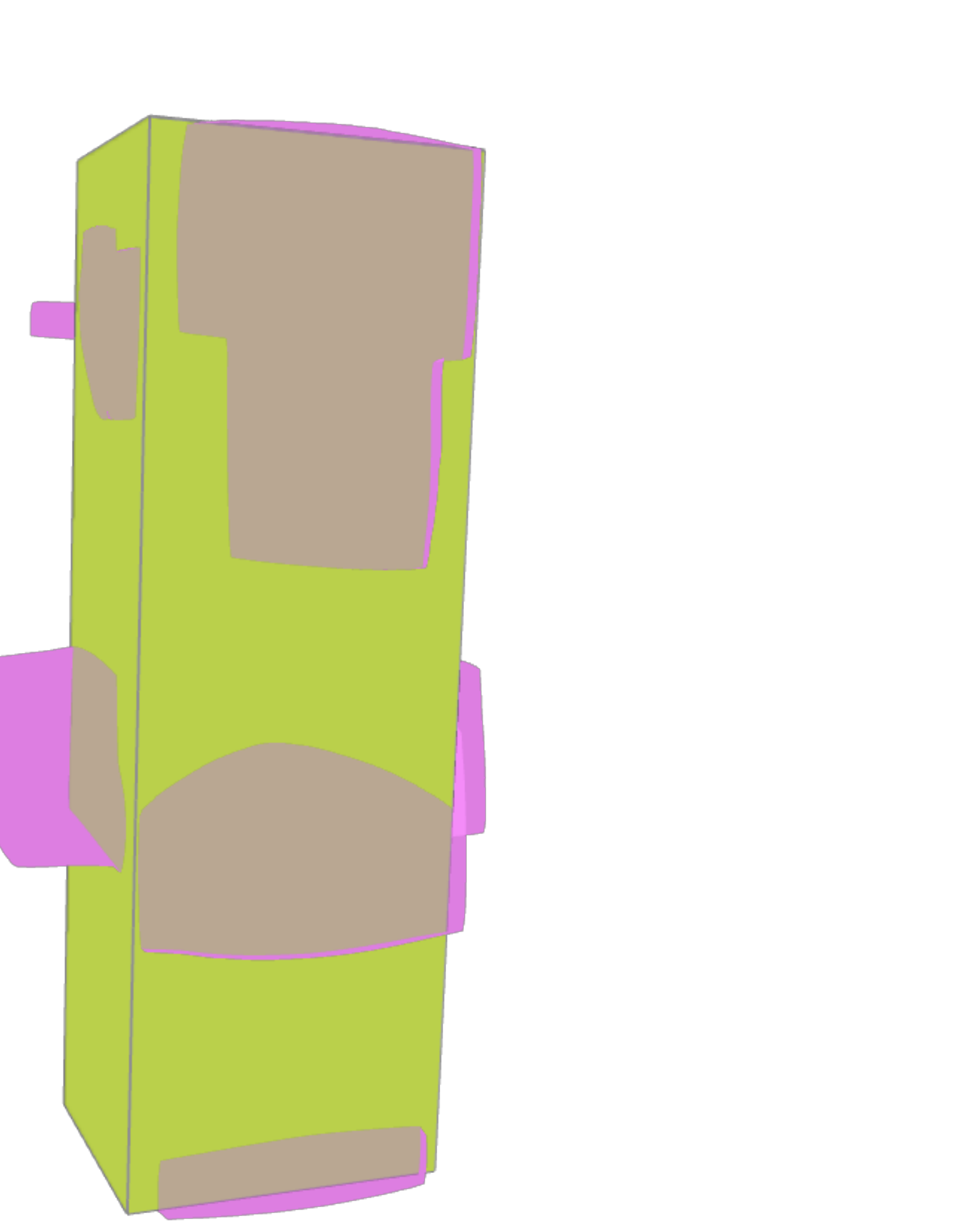} &
    \includegraphics[height=1.3in,trim=7.0in 1in 2.9in 0, clip]{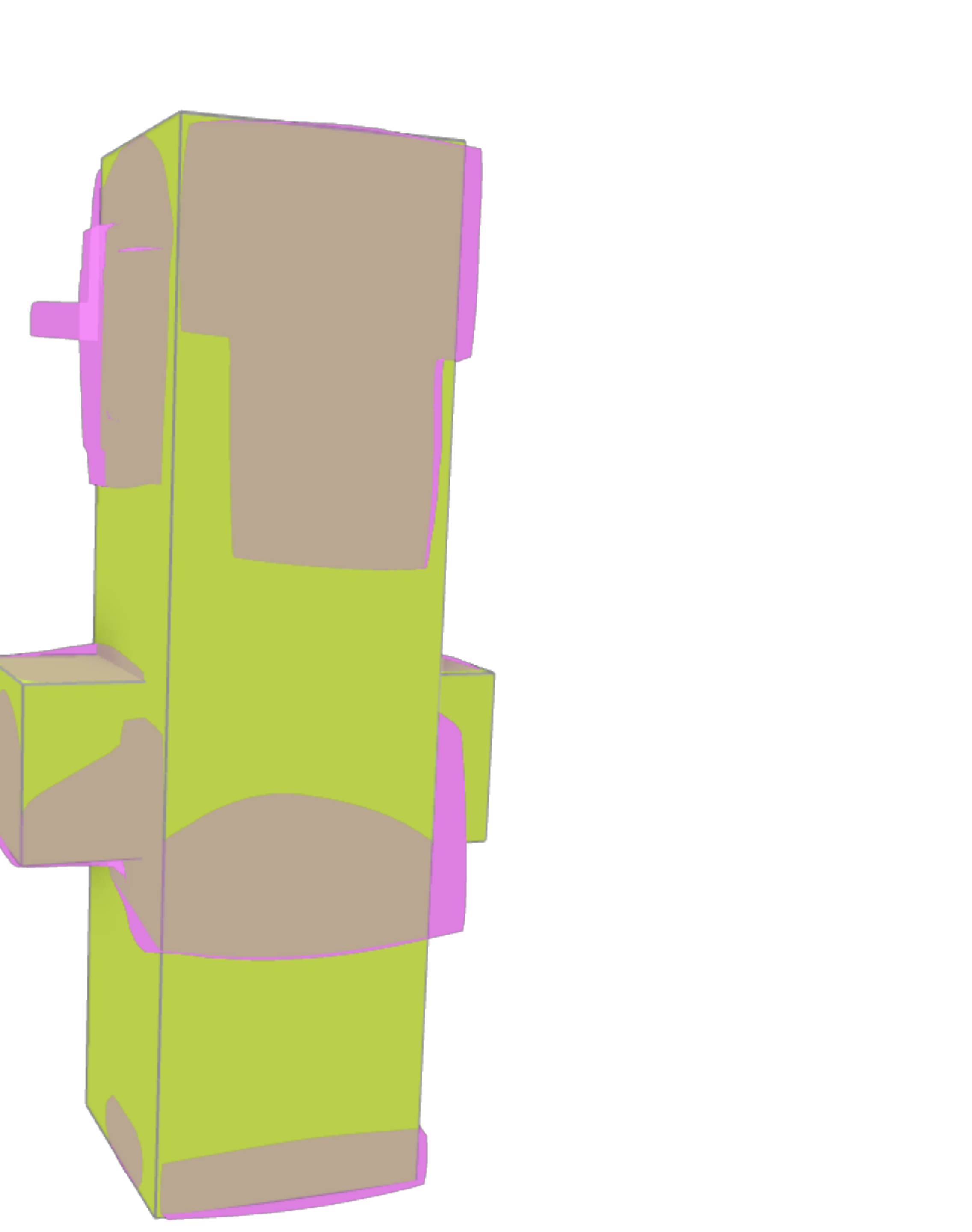} &
    \includegraphics[height=1.3in,trim=7.0in 1in 2.9in 0, clip]{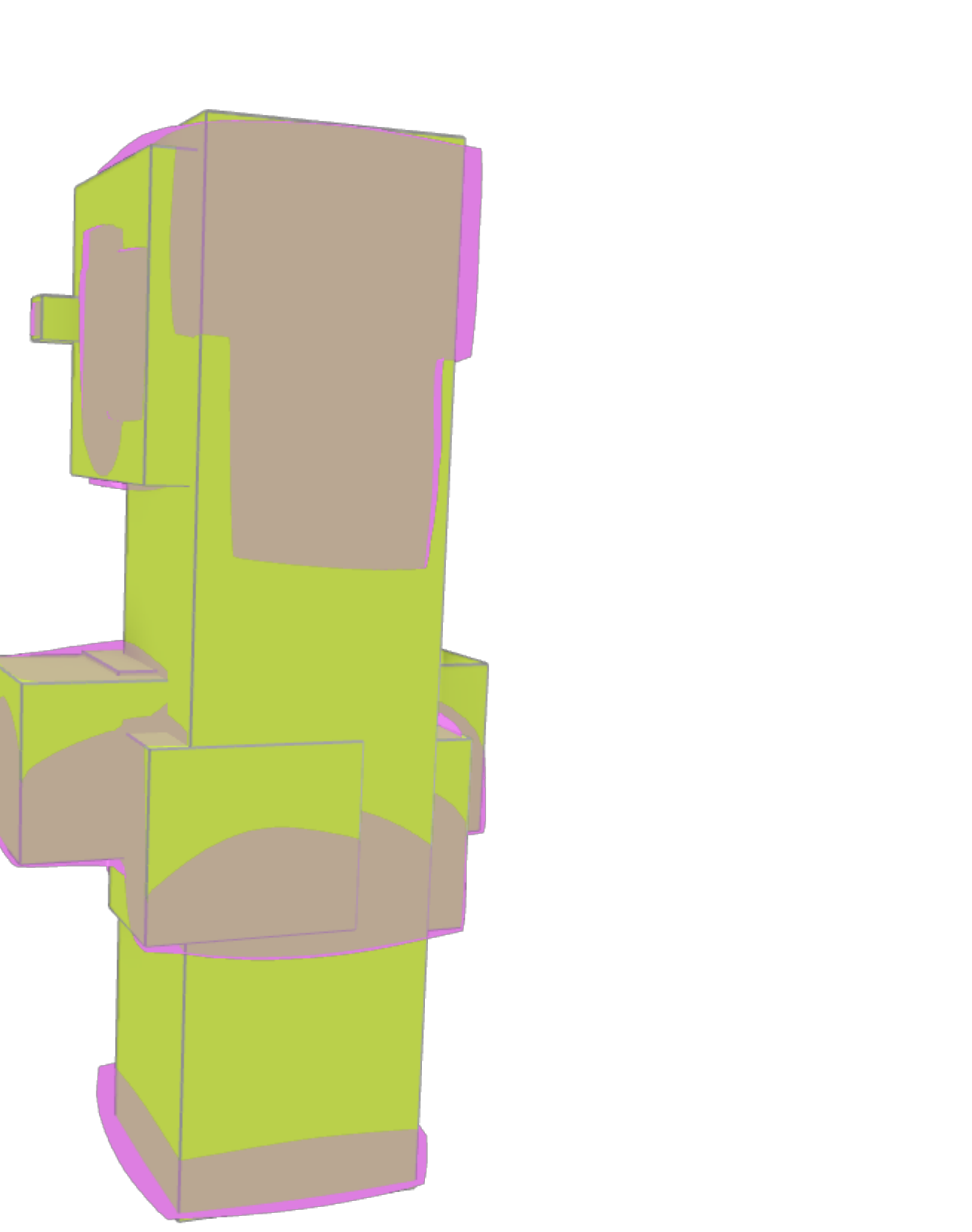} &
    \includegraphics[height=1.3in,trim=7.0in 1in 2.9in 0, clip]{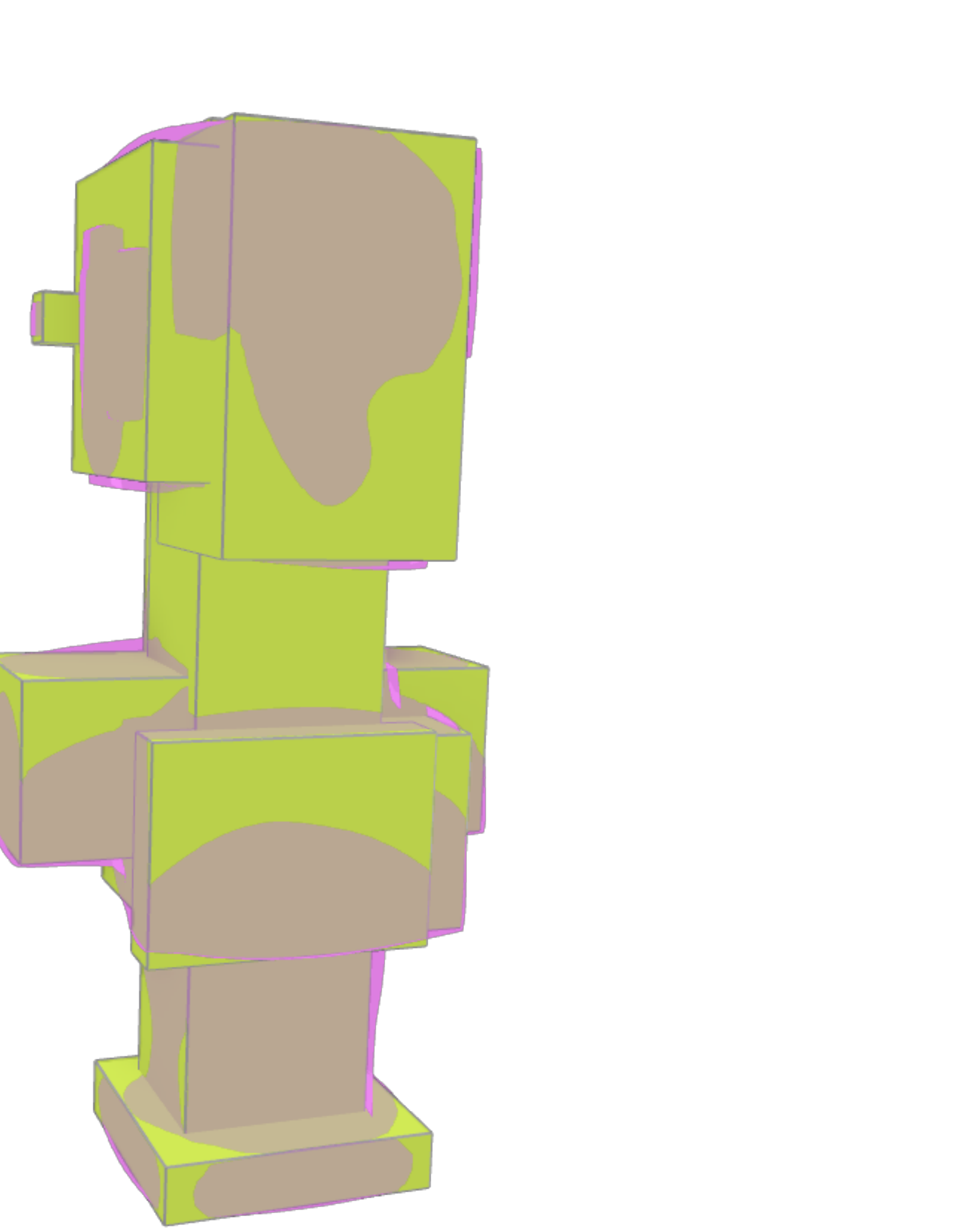}  \\
    \includegraphics[height=1.3in,trim=7.0in 1in 2.9in 0, clip]{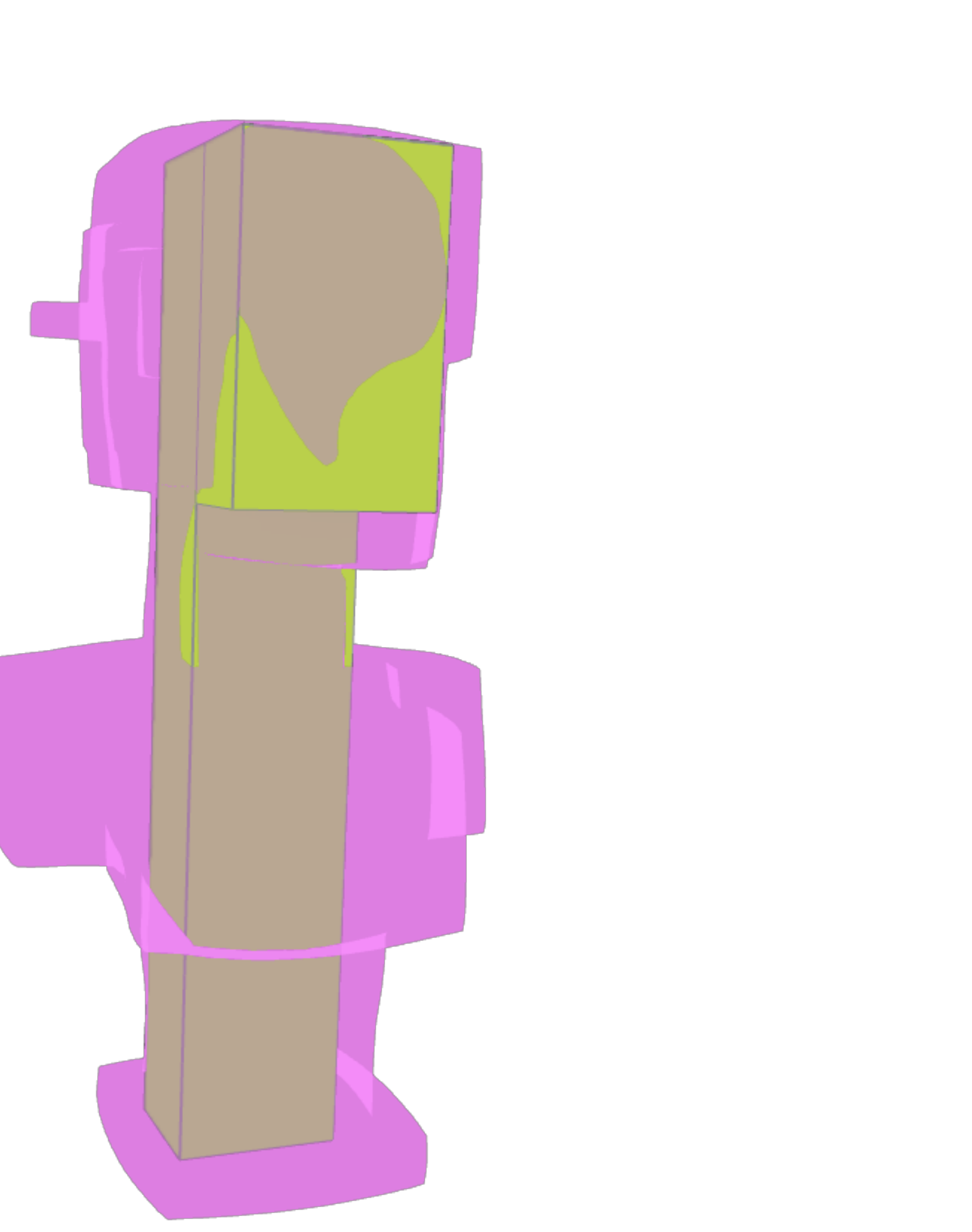} &
    \includegraphics[height=1.3in,trim=7.0in 1in 2.9in 0, clip]{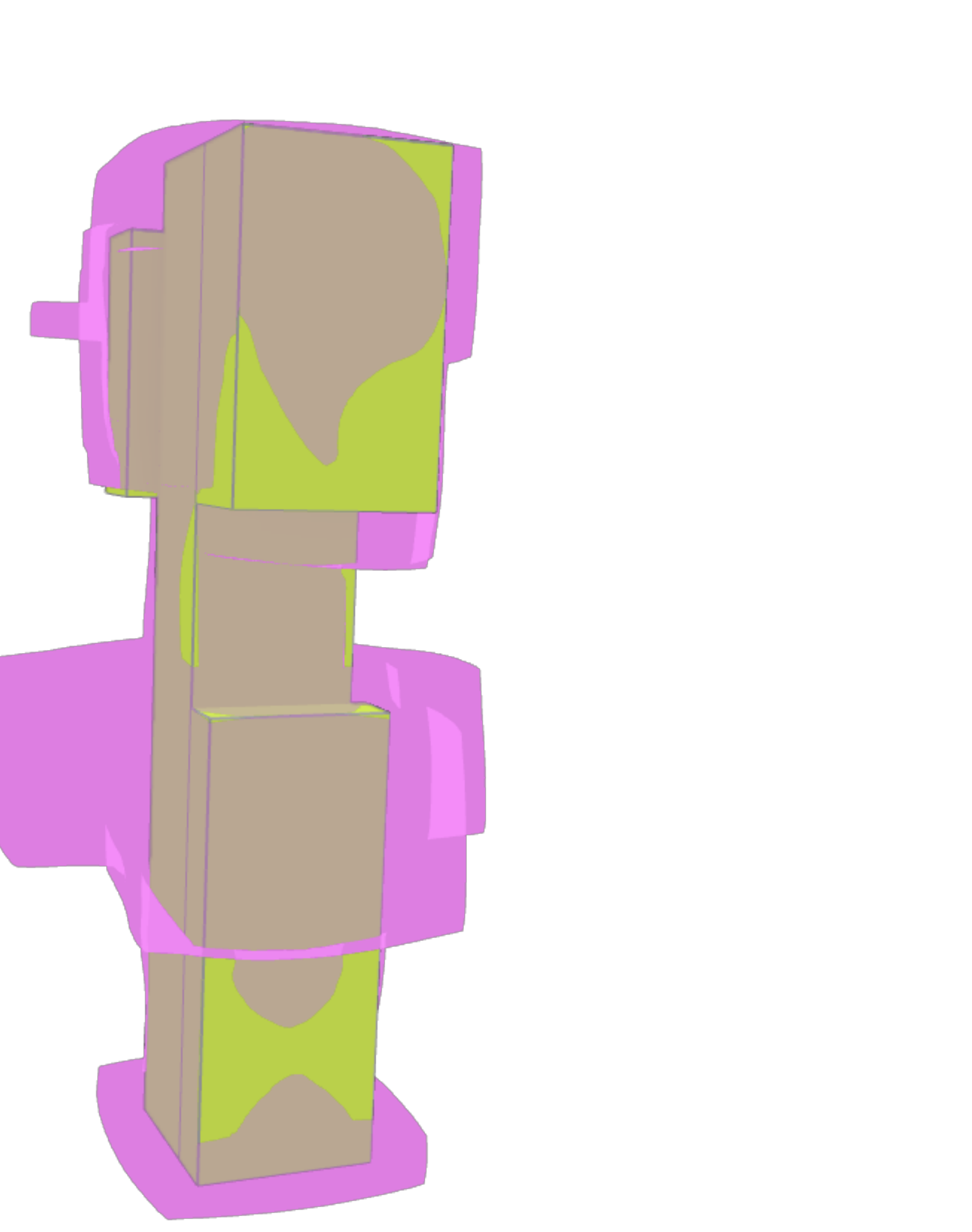} &
    \includegraphics[height=1.3in,trim=7.0in 1in 2.9in 0, clip]{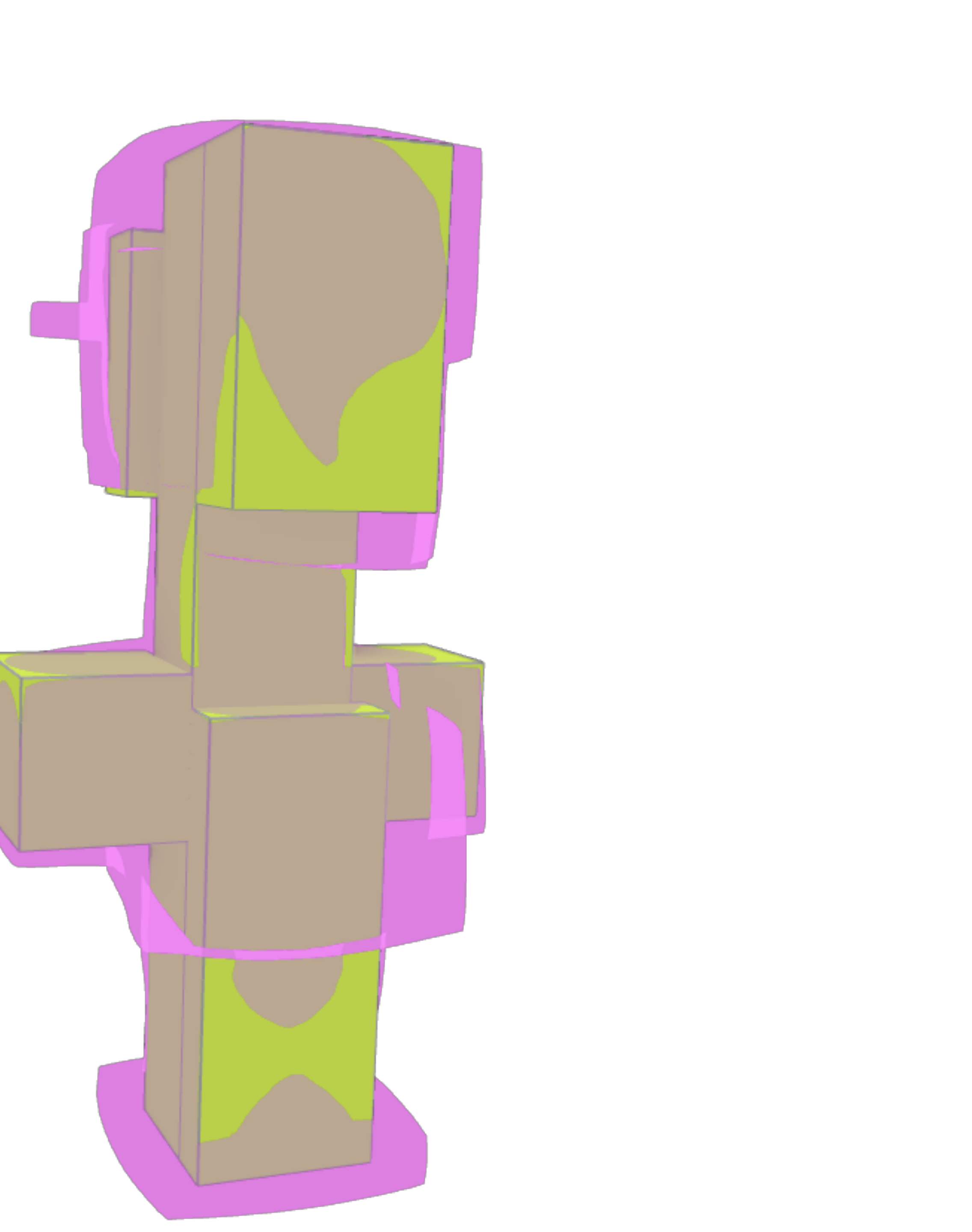} &
    \includegraphics[height=1.3in,trim=7.0in 1in 2.9in 0, clip]{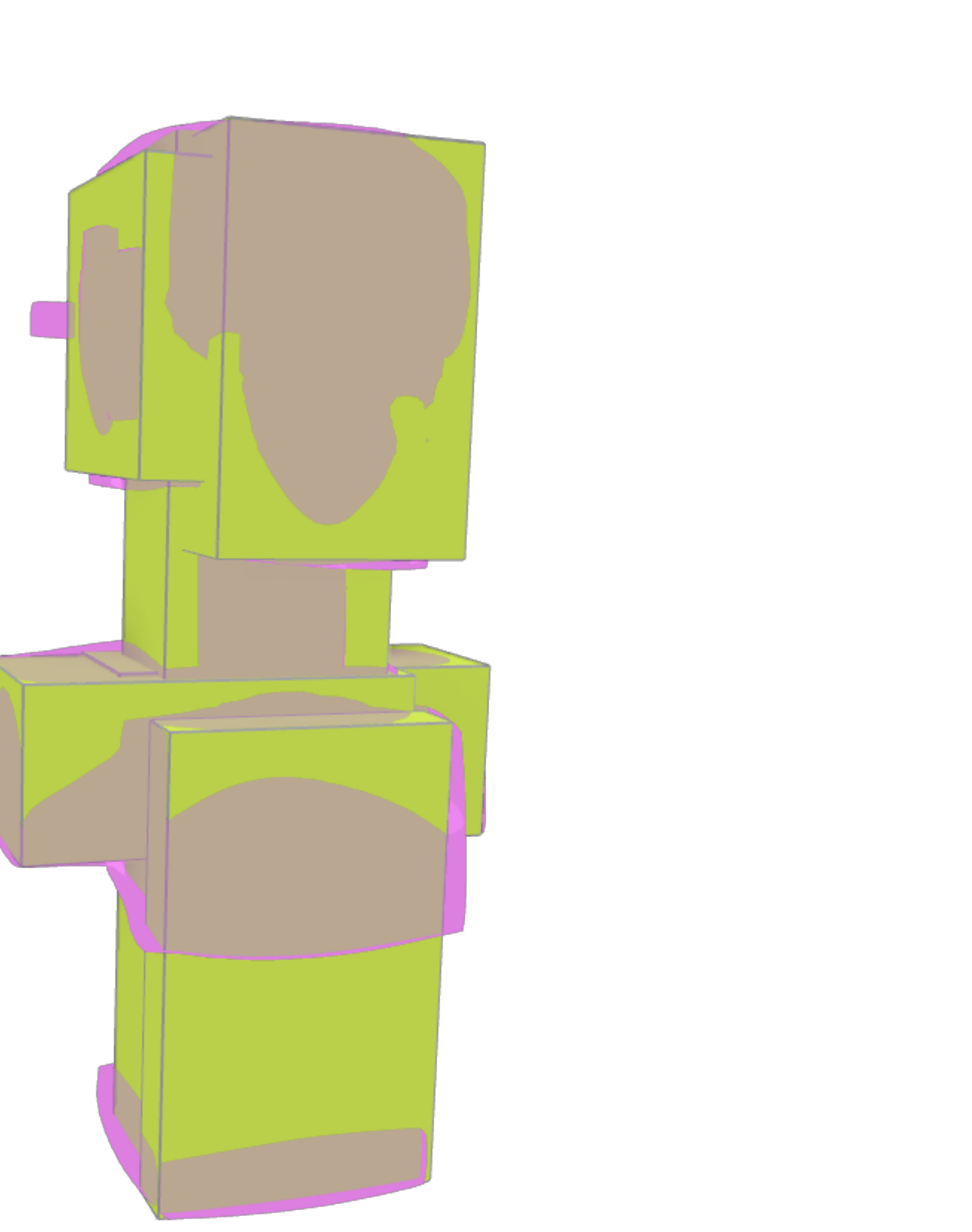} &
    \includegraphics[height=1.3in,trim=7.0in 1in 2.9in 0, clip]{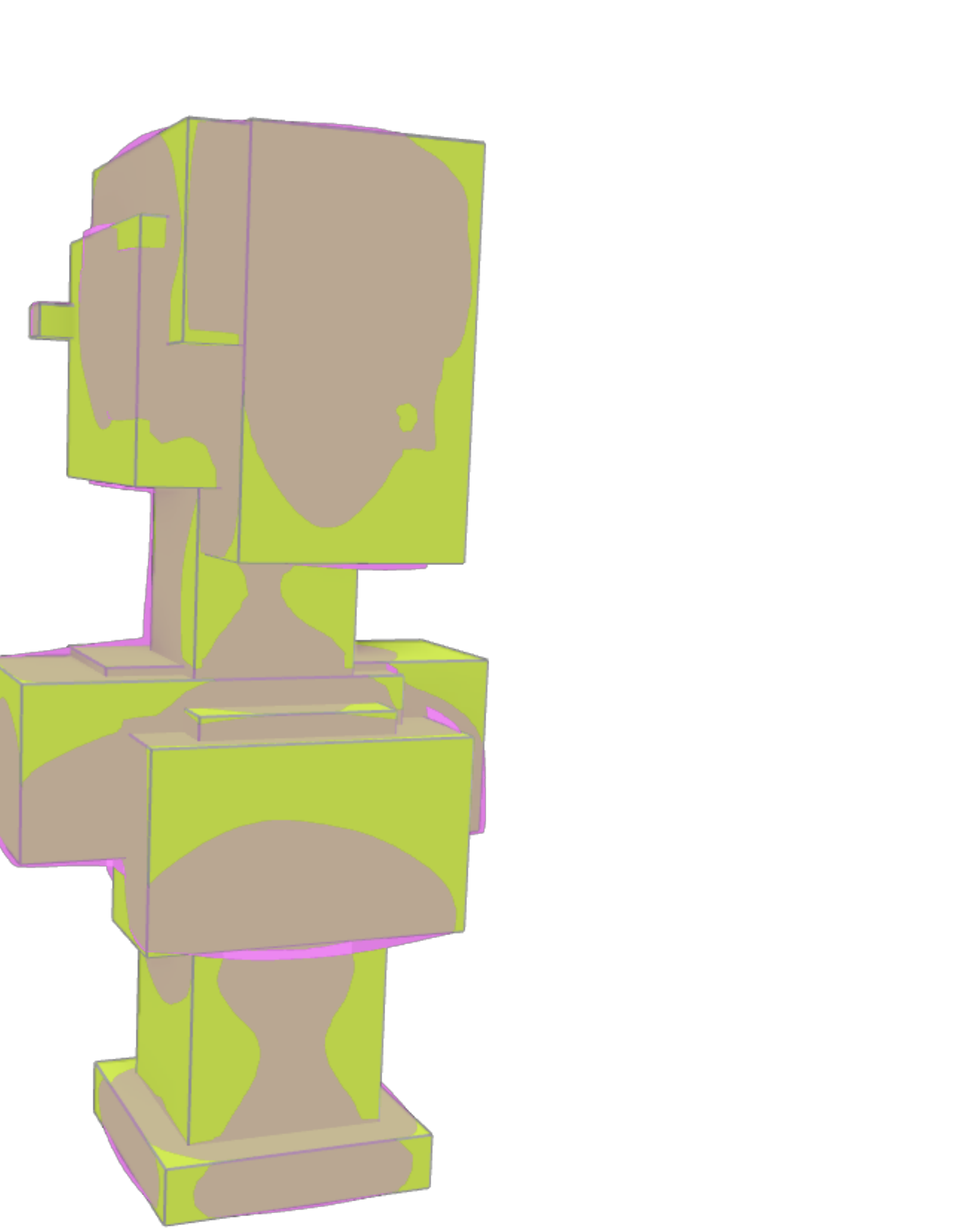} 
  \end{tabular}
  \caption{Comparison of two cuboid adding heuristics. The surface of the deformed mesh is shown in translucent pink. No user editing is involved in this example.
    In each row, the number of PolyCubes increases from left to right.
    In the top row we show the deformed \emph{buste} model followed by progressively generated PolyCubes after alternating distance-based \textit{Add} and \textit{Reoptimize} with 1, 3, 6, 11 cuboids, respectively.
The first three columns of the bottom row are PolyCubes after performing volume-based \textit{Add} 2, 4, 6 times. 
While avoiding over-coverage, volume-based \textit{Add} cannot capture small details such as braids, so we resort back to distance-based \textit{Add} and \textit{Reoptimize}, as shown in the last two columns.}
\label{fig:cuboid_add_comparison}
\end{figure}

\begin{figure}
  \centering
  \hspace*{-0.4in}
  \newcolumntype{C}[1]{>{\centering\arraybackslash}m{#1}}
  \begin{tabular}{C{2.2cm} C{2.4cm} C{2.4cm}}
    \includegraphics[height=1.0in,trim=5.0in 2in 0 2in, clip]{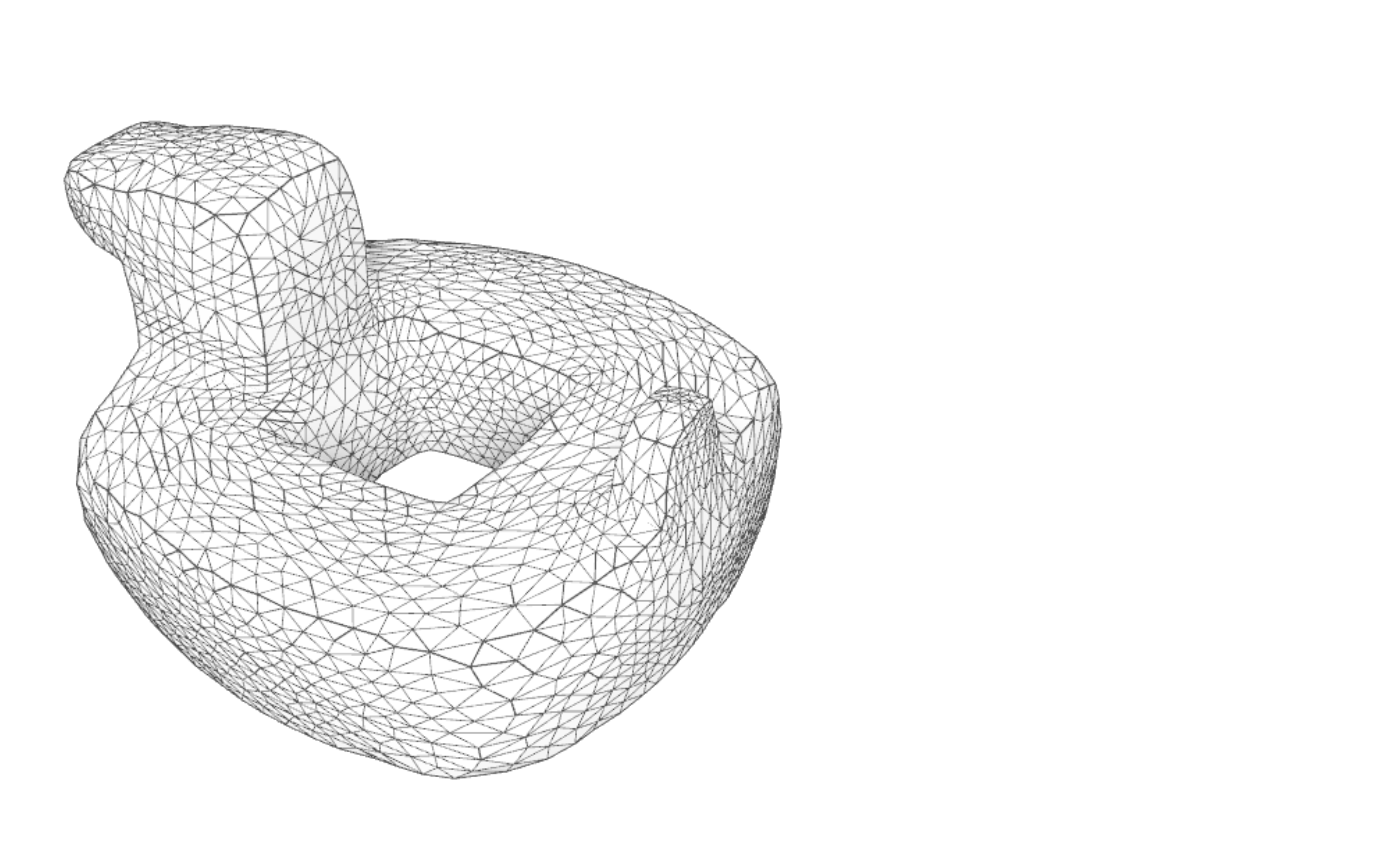} &
    \input{annotated/bob_bad_local_min.tex} &
    \input{annotated/bob_bad_local_min_fixed.tex} \\
    {\footnotesize (a) Deformed mesh}                             & {\footnotesize (b) \textit{Reoptimize} stuck} & {\footnotesize (c) Carving out the body} \\
    \input{annotated/bob_extra_stairs.tex} &
    \includegraphics[height=1.0in,trim=5.0in 1.8in 0 2in, clip]{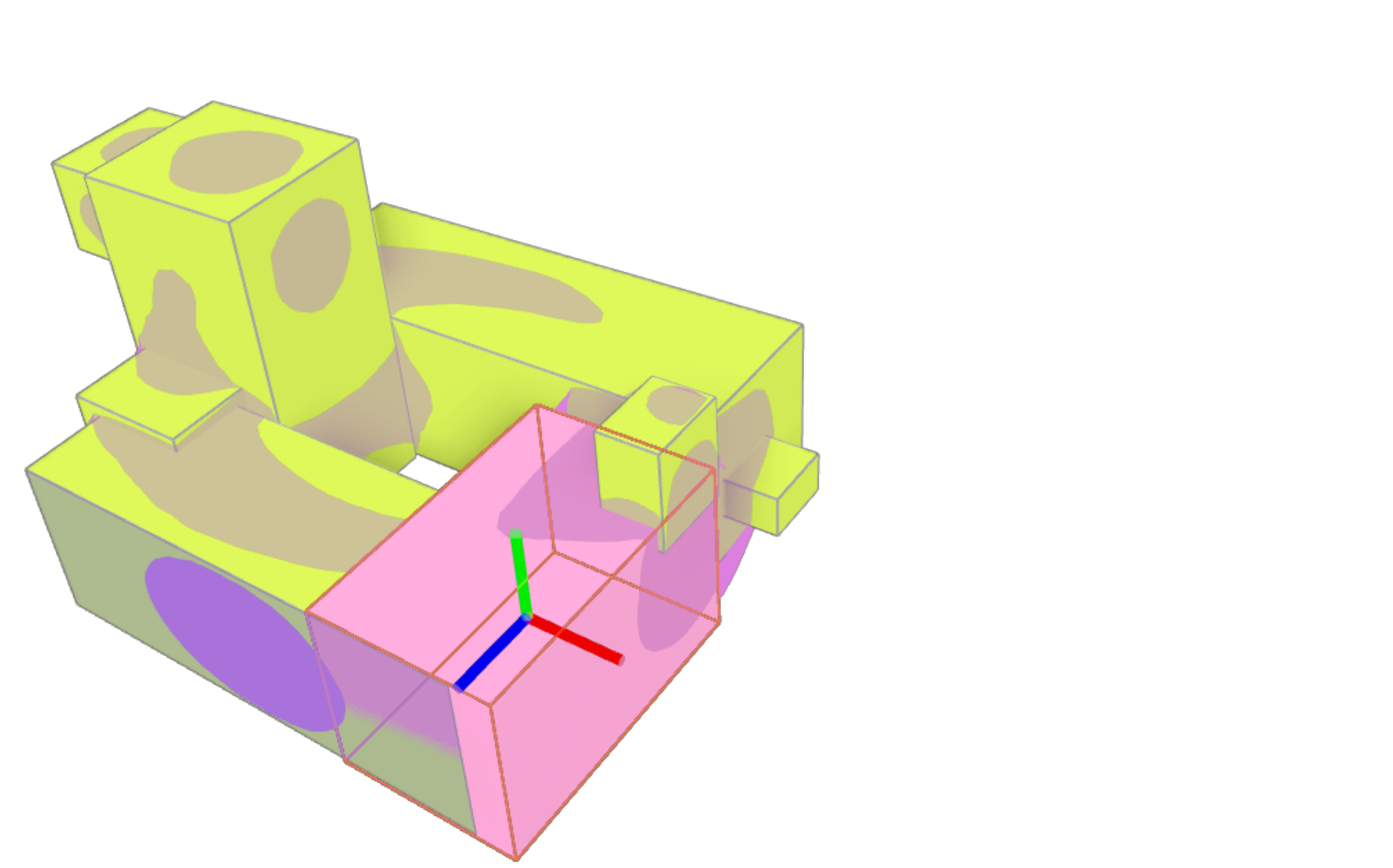} &
    \includegraphics[height=1.0in,trim=5.0in 1.8in 0 2in, clip]{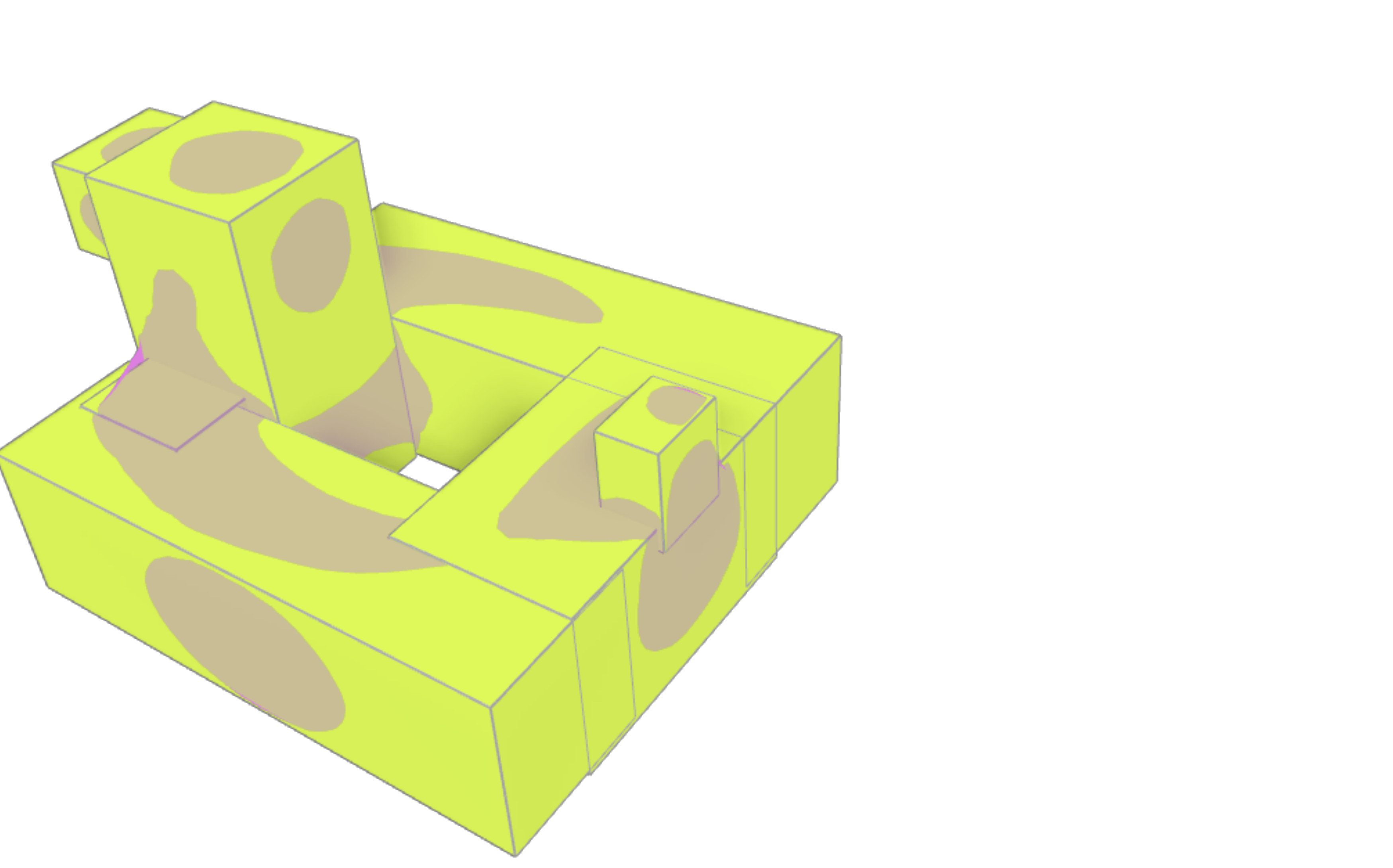}  
    \\
    {\footnotesize (d) Extra stairs} & {\footnotesize (e) \textit{Edit} to extend cuboid} & {\footnotesize (f) Final PolyCube}
  \end{tabular}
  \caption{An example where user interaction is helpful. 
    From a deformed \emph{bob} model (a), the optimization gets stuck in a local minimum (b).
    To resolve this, the user can remove the center cuboid (marked with red cross) and then add a new one near the tail (c). The optimization finds a configuration (d) that minimizes the approximation error but has unnecessary stairs. The user can remove the stairs by snapping the surrounding cuboids' faces together using \textit{sticky mode} (e) to get the final PolyCube (f).}
\label{fig:cuboid_user_fix}
\end{figure}

The user progressively builds the PolyCube by adding and modifying constituent cuboids using a combination of manual editing and automatic adjustment via our continuous PolyCube optimization (\cref{sec:opt_polycube}). This process continues until the PolyCube reaches a satisfactory level of complexity and fidelity to the near-PolyCube shape deformed from the input.
At any point, the user can perform one of the following operations (see also \cref{fig:decomposition_editing}): 
\begin{enumerate}[align=left]
  \item[\textit{Add}]
    A new cuboid is placed in the scene according to one of two automatic heuristics: a distance-based heuristic places a cuboid at an uncovered point of the deformed mesh that is furthest away from any existing cuboid, while a volume-based heuristic places a cuboid of largest volume that is inside the deformed mesh and outside any existing cuboid (see \cref{fig:cuboid_add_comparison} for comparison).
  \item[\textit{Edit}] 
    The user can resize or reposition any existing cuboid using mouse-based controls. The user can also toggle \textit{sticky mode} in which translational motion automatically snaps to align faces of the selected cuboid to that of nearby cuboids (\cref{fig:cuboid_user_fix}(e)).
    Additionally, a cuboid can be removed or duplicated, and cuboid parameters can be fine-tuned through input fields.
  \item[\textit{Subtract}]
    The system can suggest a large cubic region that is over-covered by the current set of cuboids, i.e., a region that is outside the deformed mesh but contained in the union of cuboids (in \cref{fig:decomposition_editing}(d)(g)).
    After optional user edits to the region, it is subtracted from any intersecting cuboid by splitting each cuboid into up to six non-disjoint cuboids.
    This is useful for recovering small topological features from over-covered regions like holes.
  \item[\textit{Reoptimize}]
    When cuboids are roughly in the right place, the user can choose to automatically optimize the parameters of all cuboids to best approximate the deformed mesh (\cref{sec:opt_polycube}).
    A cuboid can be \emph{locked} to prevent it from being optimized (e.g., grey cuboids in \cref{fig:pipeline_stages}(c)) if the user is satisfied with it.
\end{enumerate} 
\rev{The implementation of each heuristic is detailed in \cref{sec:app_heuristics}.}

We found two successful strategies requiring minimal user intervention:
(1) alternate between \textit{Add} with the distance-based heuristic and \textit{Reoptimize}, or (2) apply \textit{Add} repeatedly with the volume-based heuristic to cover significant regions, \textit{Reoptimize}, and then make fine adjustment to small regions.
See \cref{fig:cuboid_add_comparison} for a comparison of these strategies.

Allowing user interaction rather than fully automating this stage offers two distinct advantages.
First, for a fixed number of cuboids, the optimization from \cref{sec:opt_polycube} can get stuck in local minima; it is often easy for the user \rev{to} modify the configuration to suggest a better optimum (\cref{fig:cuboid_user_fix} (b)(c)).
Second, there can be different configurations of cuboids that yield similar approximation error, yet one may be preferred over the others depending on the user's application.
For example, it is often desirable to sacrifice a small amount of approximation error for a simpler PolyCube structure, e.g., one with fewer stairs, even for a fixed number of cuboids (\cref{fig:cuboid_user_fix} (d)).
Thanks to our choice of representation, editing the PolyCube structure is intuitive and transparent by editing individual cuboids.

\subsection{Discretization stage}
\label{stage:discretization}
In this stage, the PolyCube made in the previous stage is voxelized into a hex mesh.
To accomplish this, we first snap all cuboid corners to a regular grid of a user-specified edge length.
Then, the user can edit the voxelized PolyCube by adding or removing voxels, either one at at time or by editing an entire layer.
Once the user is satisfied, a layer of global padding is added in the same way as in \citep{gregson2011all}. %

\begin{figure}
  \centering
  \hspace*{-0.4in}
  \newcolumntype{C}[1]{>{\centering\arraybackslash}m{#1}}
  \begin{tabular}{C{3.0cm}  C{3.0cm} }
    \includegraphics[height=1in,trim=2.0in 1.1in 0 1in, clip]{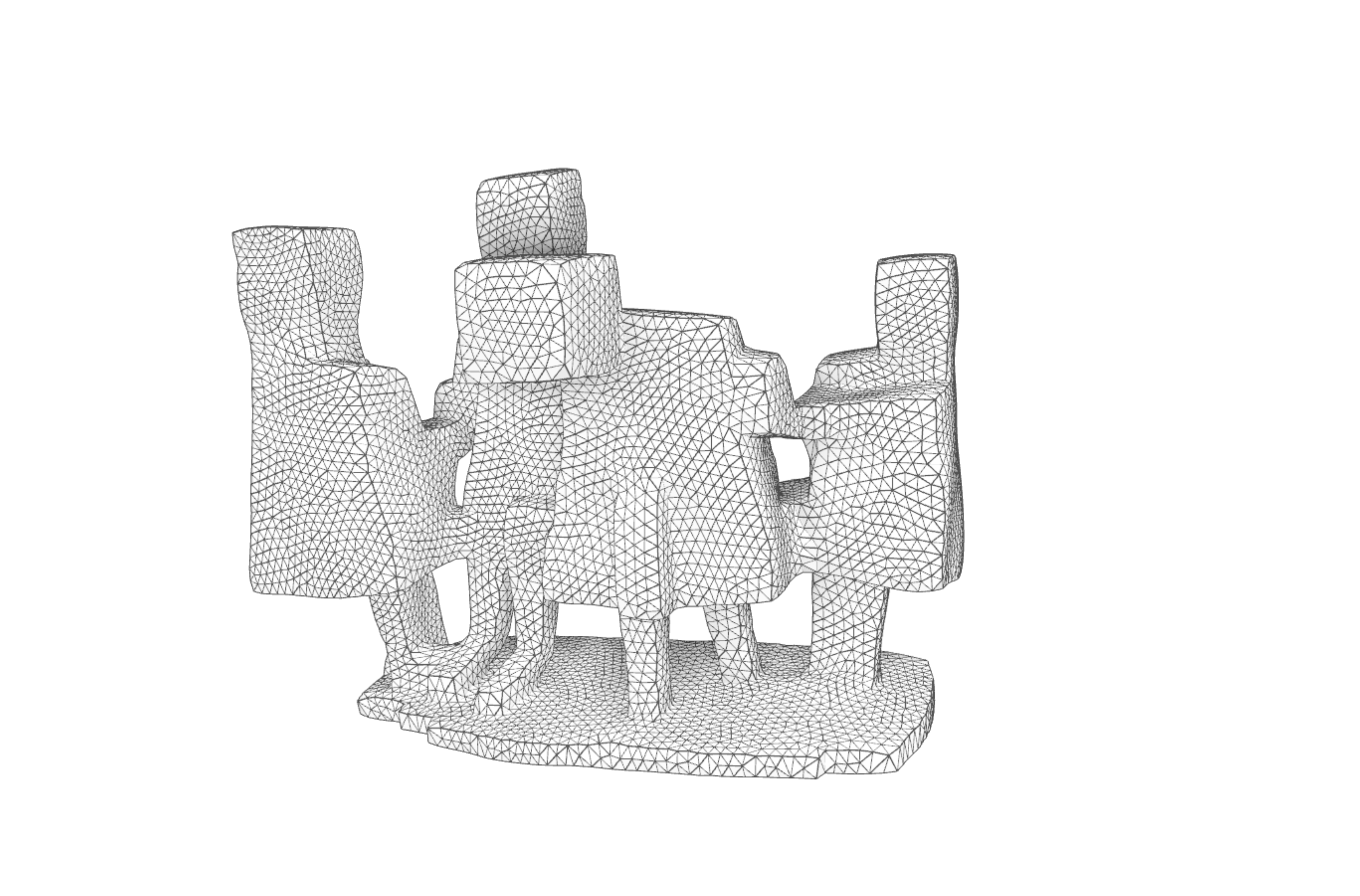}  &
    \includegraphics[height=1in,trim=2.0in 1.1in 0 1in, clip]{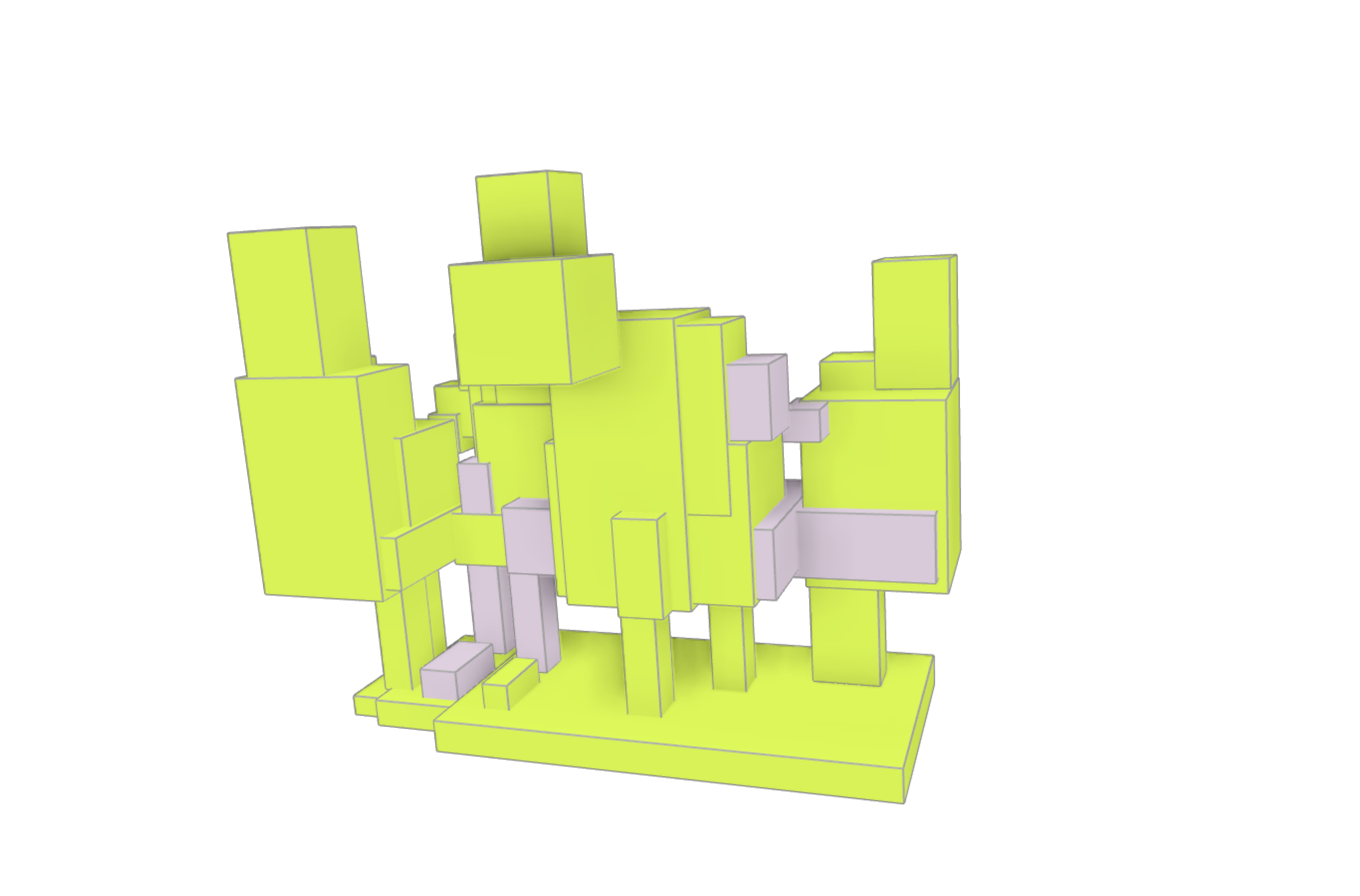}  
    \\
    (a) & (b) \\
\input{annotated/dancing_children_discretized.tex}
       & \input{annotated/dancing_children_discretized_fixed.tex}  \\
    (c) & (d)
  \end{tabular}
  \caption{Fixing topological problems and removing unnecessary stairs. 
    (a) Deformed mesh. (b) PolyCube after the decomposition stage (grey indicates fixed cuboids). (c) Direct discretization by snapping PolyCube corners to an integer grid may result in disconnected regions (shown in orange boxes) even if the cuboids are not disconnected. There can also be unnecessary stairs (shown in blue boxes).
    (d) By adding and removing individual voxels or layers of voxels, the user can fix the topological problems and remove unnecessary stairs.
  }

\label{fig:dig_extrude_example}
\end{figure}

User interaction is helpful in two ways. 
First, snapping PolyCube corners to integers may erase small topological features \citep{protais2020robust} that the user can easily identify and fix.
Second, the user has one more opportunity to simplify the PolyCube structure, e.g., by removing unnecessary stairs. See \cref{fig:dig_extrude_example} for an example.

\subsection{Hexahedralization stage}
\label{stage:hexahedralization}
In the final stage, the PolyCube hex mesh from the previous stage is deformed to obtain the output hex mesh. 
The deformation matches the surface of the PolyCube hex mesh with the input surface while retaining \rev{hex} element quality. 

\begin{figure}
  \centering
  \newcolumntype{C}[1]{>{\centering\arraybackslash}m{#1}}
  \begin{tabular}{C{2.5cm} C{0.1cm} C{2.0cm} C{0.1cm} C{2.0cm}}
    \includegraphics[height=1in,trim=4.0in 1.1in 0 1in, clip]{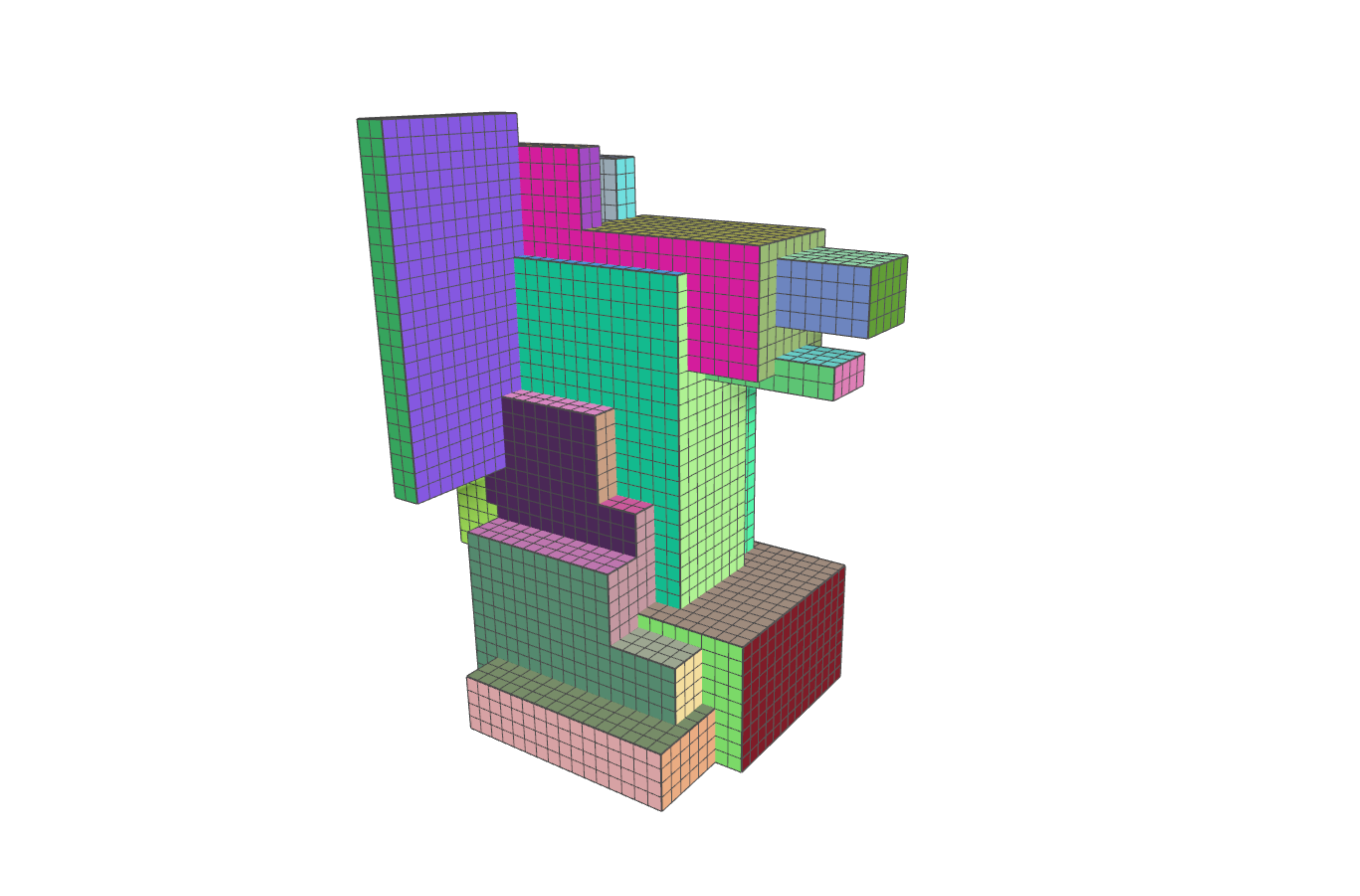}  &
    \includegraphics[height=1in,trim=6.0in 1.1in 0 1in, clip]{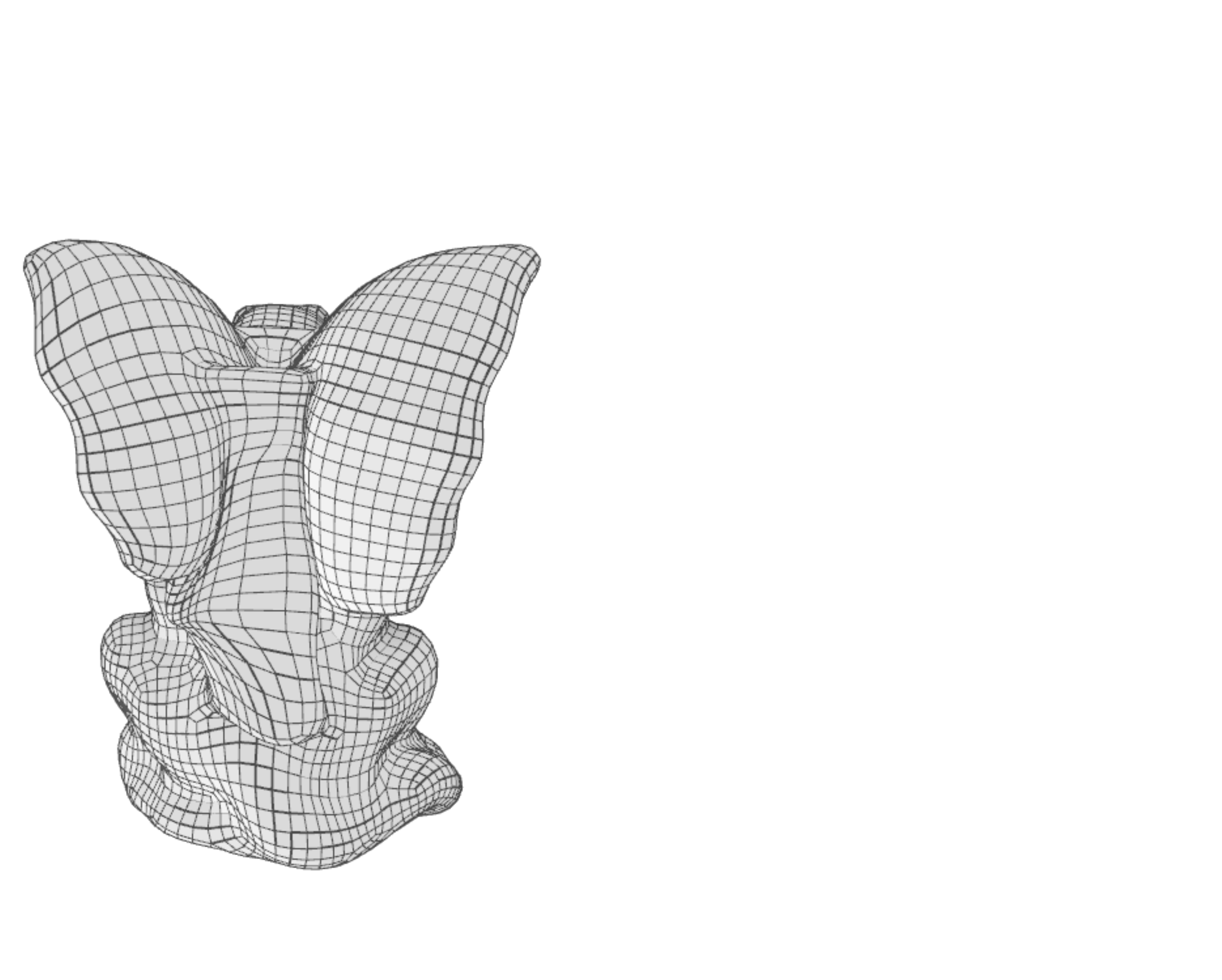}   &
    \includegraphics[height=1in,trim=4.0in 1.1in 0 1in, clip]{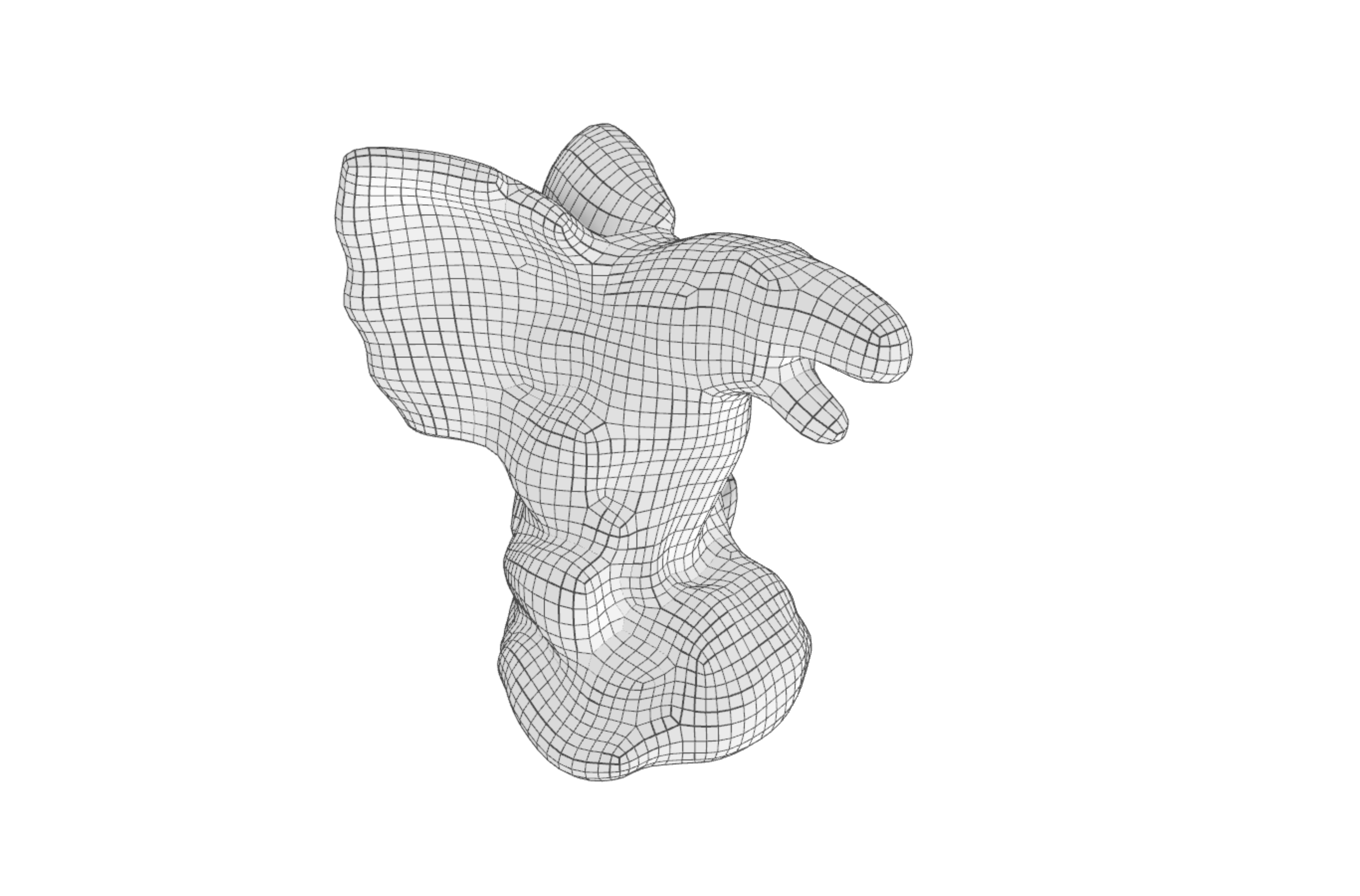}   &
    \includegraphics[height=1in,trim=6.0in 1.1in 0 1in, clip]{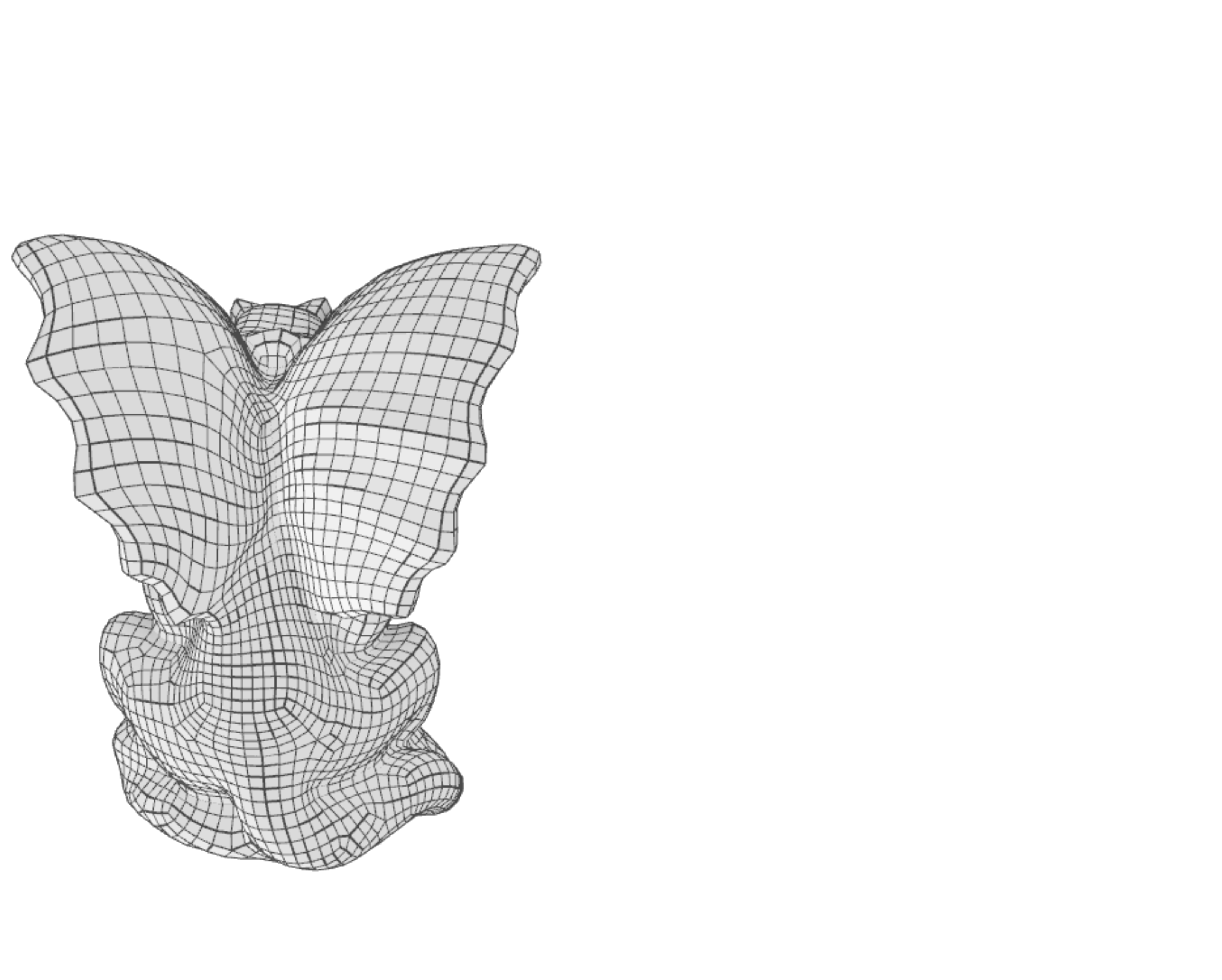} &
    \includegraphics[height=1in,trim=4.0in 1.1in 0 1in, clip]{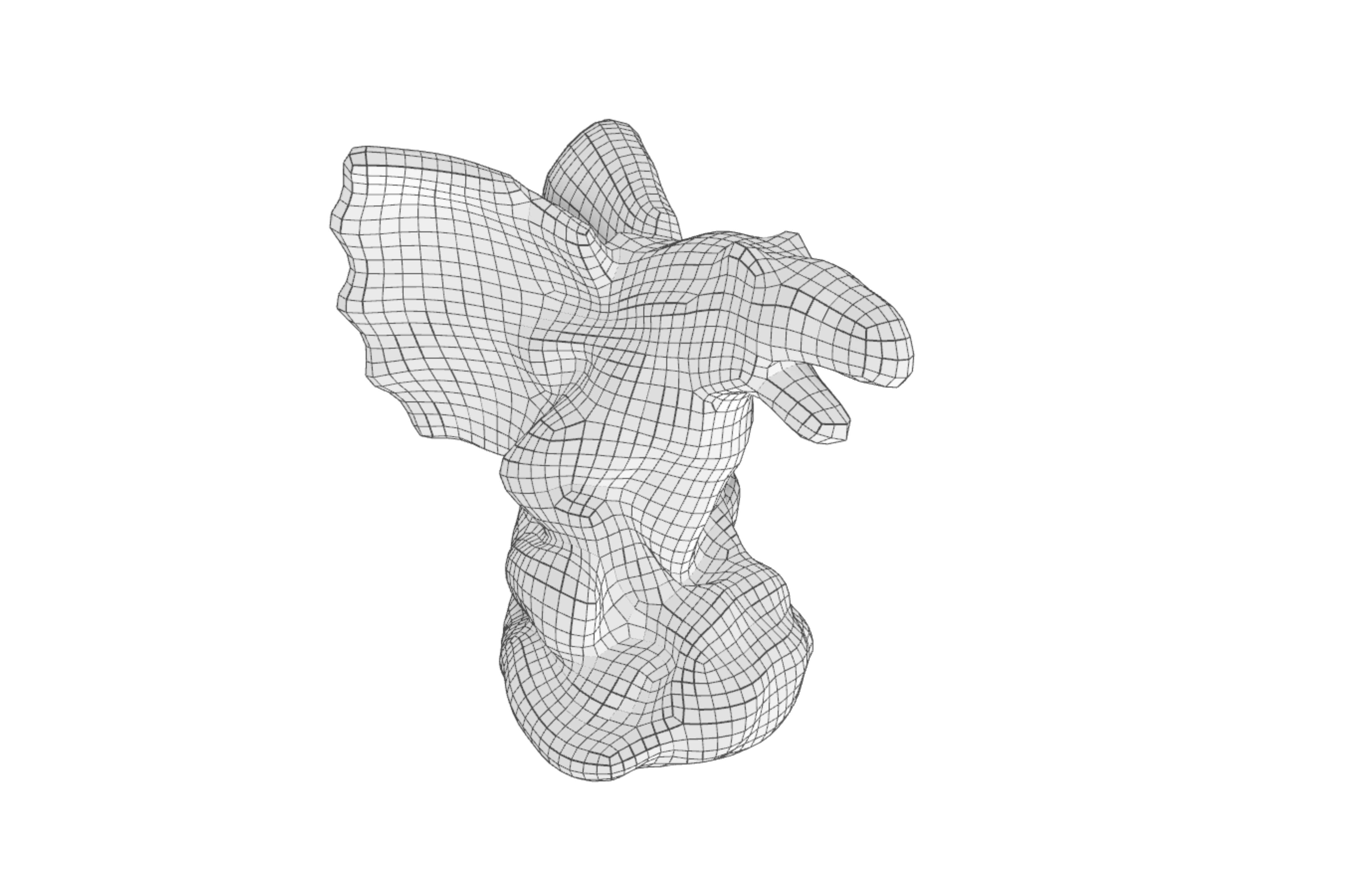} \\
    {\footnotesize (a) PolyCube hex mesh} & \multicolumn{2}{c}{{\footnotesize (b) Pullback}} & \multicolumn{2}{c}{{\footnotesize (c) Optimized}}
  \end{tabular}
  \caption{Demonstration of the hexahedralization stage using the \textit{gargoyle} model.
    (a) PolyCube hex mesh from the discretization stage.
    (b) Visualization of the front and the back of the initialized hex mesh obtained from our inversion-free pullback (scaled Jacobians: $J_{\min}= 0.014$, $J_\avg = 0.781$; Hausdorff distance: $d_{\max} = 46.292$).
    (c) Optimized version of (b) with a large \textit{details} parameter while improving worst hex element quality ($J_{\min}= 0.253$, $J_\avg = 0.771$, $d_{\max} = 19.095$).
    Notice how the details around the wings and the ears are recovered during optimization.
  }

  \label{fig:hexahedralization_flow}
\end{figure}

\begin{figure}
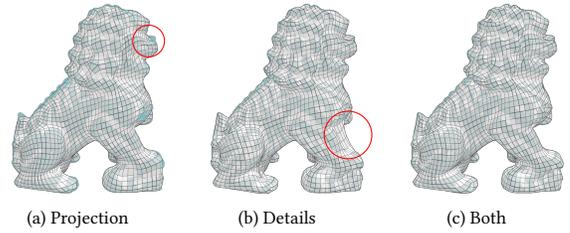

  \centering
  \newcolumntype{C}[1]{>{\centering\arraybackslash}m{#1}}
  \begin{tabular}{C{2.3cm} C{2.3cm} C{2.3cm}}
    \input{annotated/chinese_lion_projection_only} &
    \input{annotated/chinese_lion_details_only} &
    \input{annotated/chinese_lion_projection_details} \\
    {\footnotesize (a) Projection} & {\footnotesize (b) Details} & {\footnotesize (c) Both}
  \end{tabular}
  \caption{
    \rev{Comparison of surface metrics in the hexahedralization stage on the \textit{Chinese lion} model.
    (a) Setting a large projection parameter and a small details parameter makes the surface vertices of the resultant hex mesh stay on the input surface, but fine details are not recovered ($d_{\max} = 22.263, d_{\avg} = 1.297$).
    (b) Finer details, such as the lion's nose, are better recovered using a large details parameter, but there is over-coverage at the front leg due to a small projection parameter ($d_{\max} = 44.952, d_{\avg} = 0.861$).
    (c) When using large parameters for both projection and details, our resulting surface approximates the input surface the best ($d_{\max} = 9.342, d_{\avg} = 1.297$).
  }
  }

  \label{fig:hexahedralization_surface_metrics}
\end{figure}

\begin{figure}
  \centering
  \newcolumntype{C}[1]{>{\centering\arraybackslash}m{#1}}
  \begin{tabular}{C{2.5cm} C{2.5cm} C{2.5cm}}
  \input{annotated/camel_distortion.tex} &
  \input{annotated/camel_sj.tex} &
    \includegraphics[height=1in,trim=3.5in 0.8in 0 1in, clip]{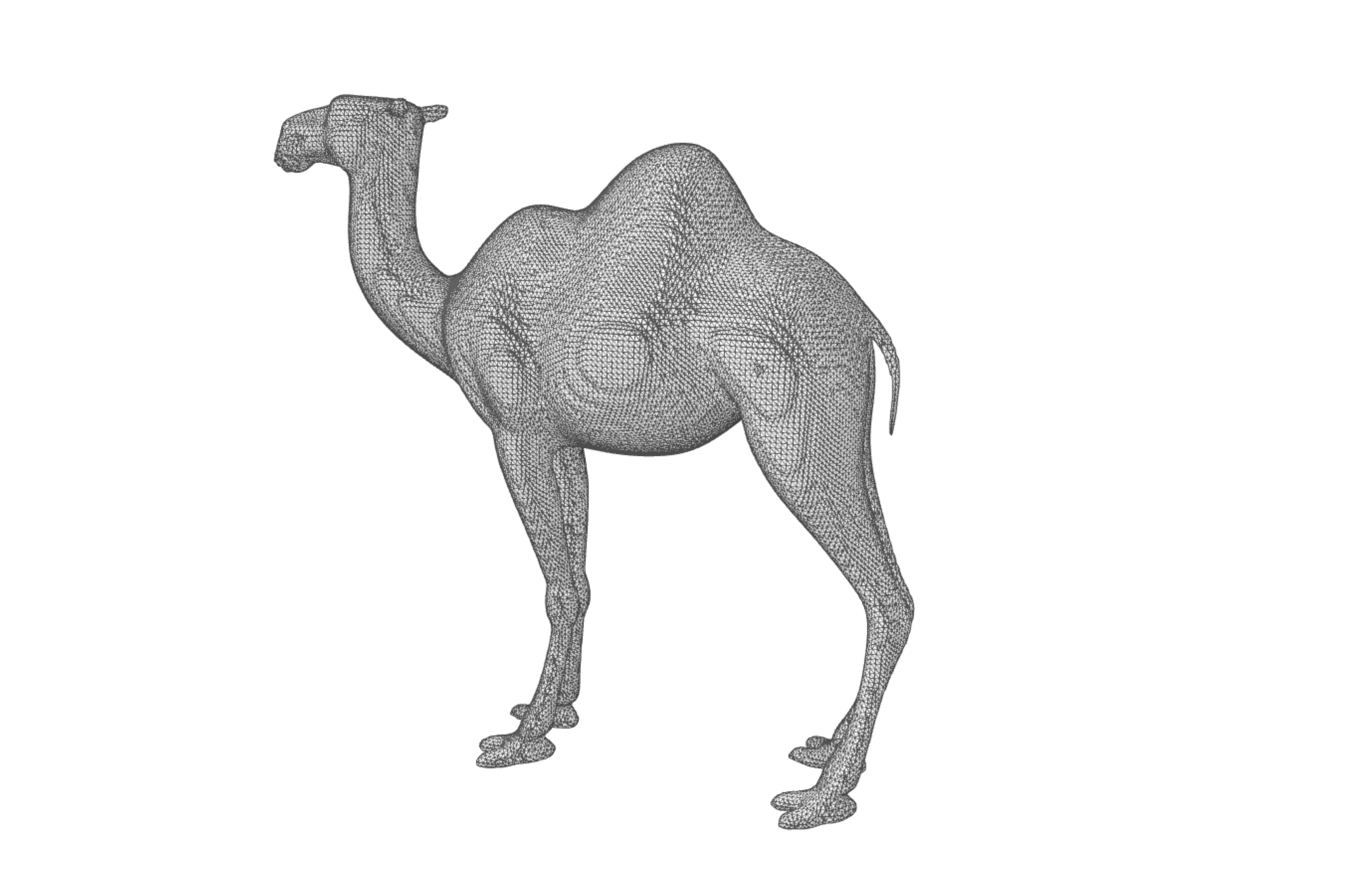}  \\
    {\footnotesize (a) Volume preservation} & {\footnotesize (b) Scaled Jacobian} & {\footnotesize (c) Input mesh}
  \end{tabular}
  \caption{Trade-off between volume preservation (\textit{authalic} parameter) and scaled Jacobian (\textit{custom} parameter).
    (a) Result when using a high \textit{authalic} parameter for hex-meshing the \textit{camel} model.
    The tail and toes are bloated compared to the input to preserve the volume of each hex element.
    (b) Result when using a low \textit{authalic} parameter but with a high \textit{custom} parameter to improve the scaled Jacobian.
    The shin part of the camel has more cubic hexes with varying volume, which are preferred by the scaled Jacobian.
    (c) Input tet mesh for comparison.
  }

  \label{fig:hexahedralization_tradeoff_j_sj}
\end{figure}

The system initializes the output hex mesh by pulling back the PolyCube hex mesh in an inversion-free manner (\cref{sec:pullback}, \cref{fig:hexahedralization_flow}(a)).
Next, the user guides the system to optimize the hex mesh while aligning its surface to that of the input mesh.
The user has control over a range of quality parameters, divided into surface and hex element quality metrics.
\rev{For surface metrics, the user can control the smoothness of the surface, the closeness of the surface to the input, and the level of detail of the surface (\cref{fig:hexahedralization_surface_metrics}).
  For hex element quality metrics, the user can choose how much they want the elements to be angle-preserving or volume-preserving, or they can supply an application-dependent custom metric (\cref{fig:hexahedralization_tradeoff_j_sj}). 
We choose the scaled Jacobian as the custom metric, but other types can be easily incorporated %
(\cref{sec:constrained_smoothing}).
The user may also choose to optimize over the worst elements or over the average.
\rev{These controllable parameters are summarized in \cref{tbl:hexahedralization_parameters}.}
Similar to \cref{stage:deformation}, the user can change the parameters before continuing the optimization of the mesh (\cref{fig:hexahedralization_flow}(c)).
}

\rev{
As these metrics compete with each other, we leave it to the user to explore the trade-off landscape (e.g., \cref{fig:trade_off_aut_sj,fig:trade_off_detail}).
To guide user exploration, a hex quality metric (scaled Jacobian by default) of the current hex mesh and the Hausdorff distance to the input mesh surface are displayed in a window to help the user choose the desired trade-off.
Additionally, the input mesh surface is shown as a translucent shell as a visual aid (\cref{fig:hexahedralization_surface_metrics}).
}

\begin{figure}
   \hspace*{-0.3in}
  \newcolumntype{C}[1]{>{\centering\arraybackslash}m{#1}}
  \begin{tabular}{C{1.5cm} C{1.5cm} C{1.5cm} C{1.5cm}}
    \input{annotated/armadillo_crossed.tex} & 
    \includegraphics[height=0.8in,trim=6.1in 2.8in 8in 3in, clip]{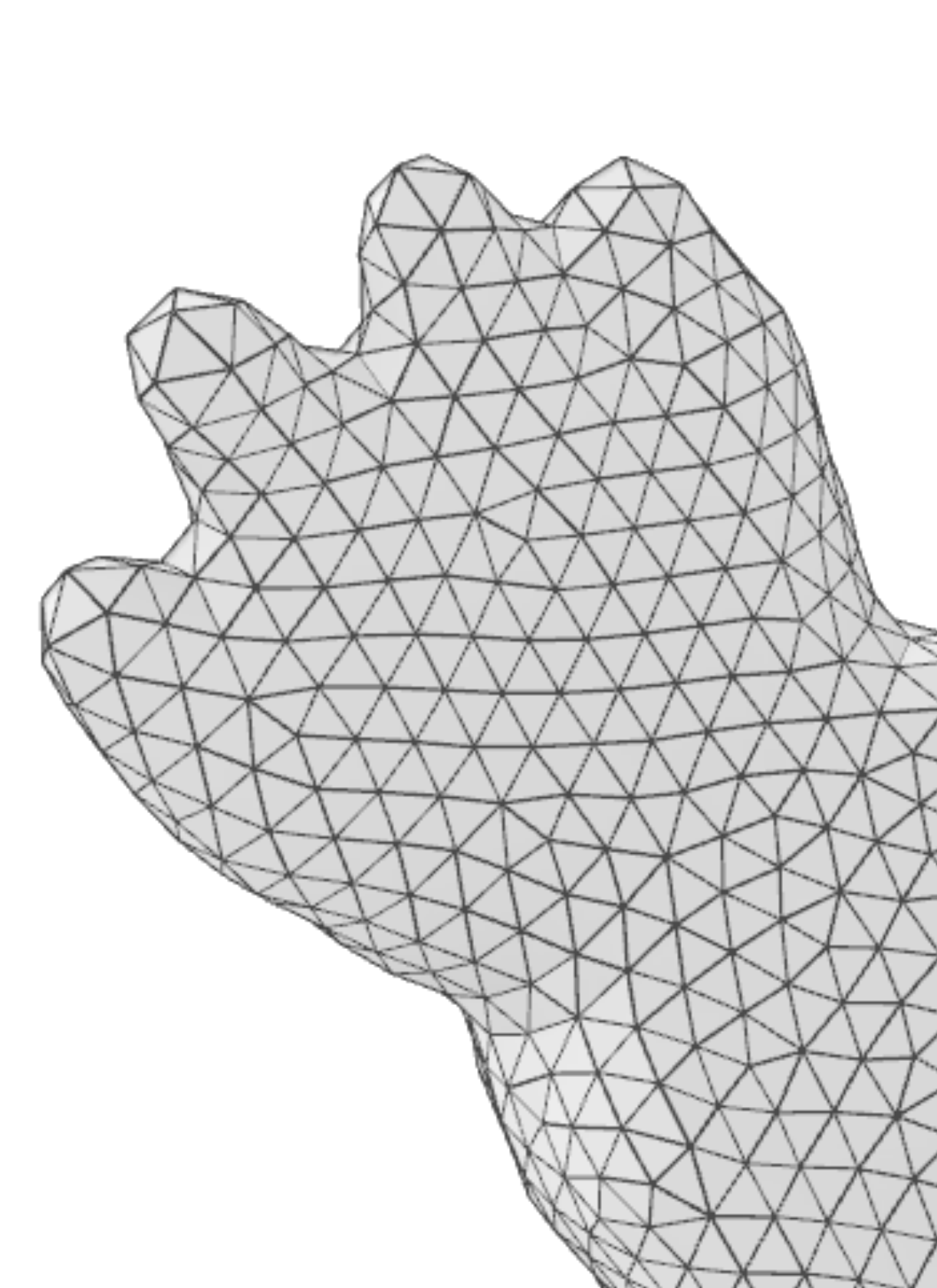}  & 
    \input{annotated/armadillo_fix_1.tex} & 
    \input{annotated/armadillo_fix_2.tex}  
                                                                                                  \\
    {\footnotesize (a)} & {\footnotesize (b)} & {\footnotesize (c)} & {\footnotesize (d)} \\
    \input{annotated/armadillo_fix_3.tex}  &
    \input{annotated/armadillo_fix_4.tex}  &
    \includegraphics[height=0.8in,trim=6.1in 2.8in 8in 3in, clip]{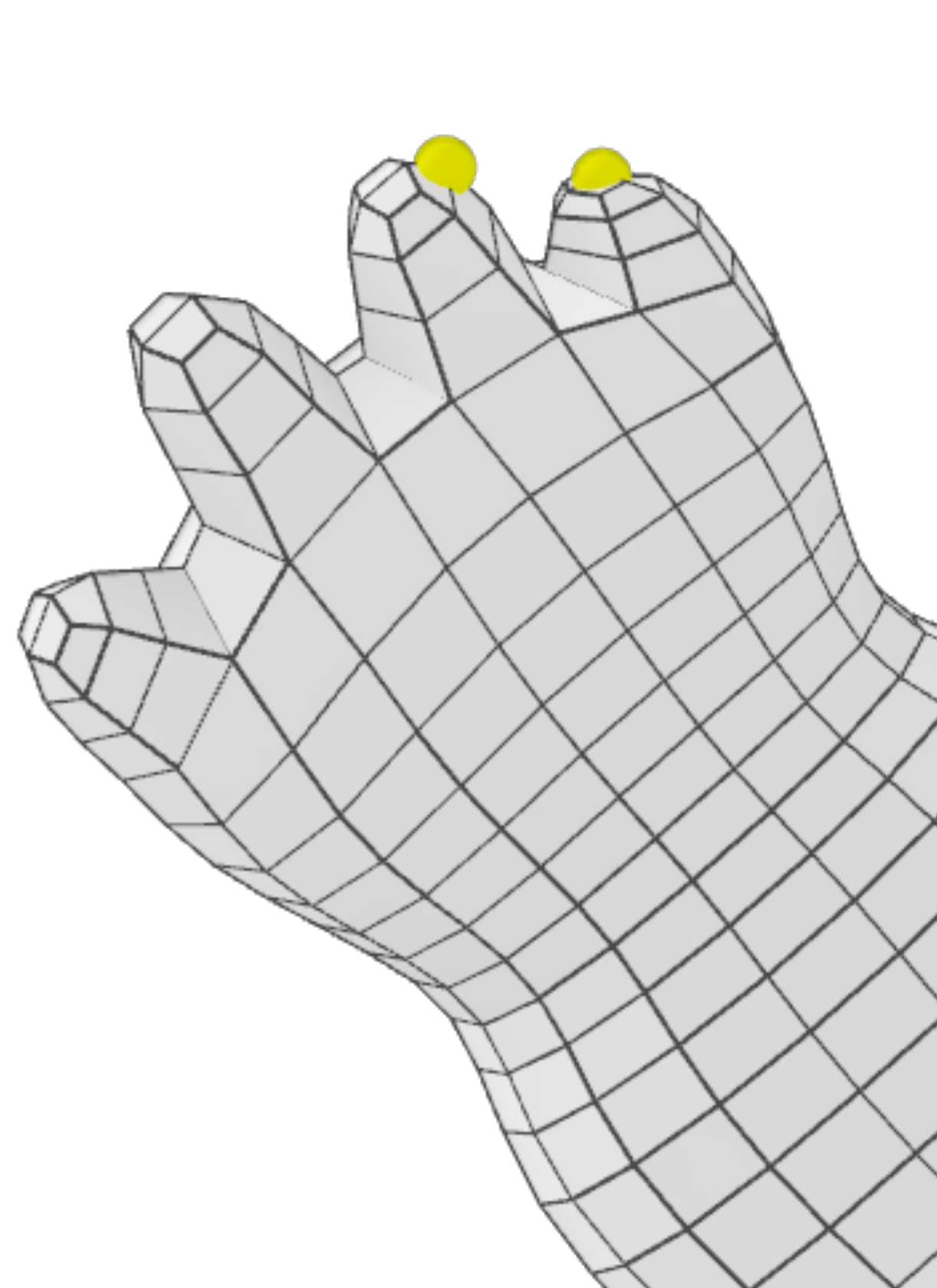} &
    \includegraphics[height=0.8in,trim=6.1in 2.8in 8in 3in, clip]{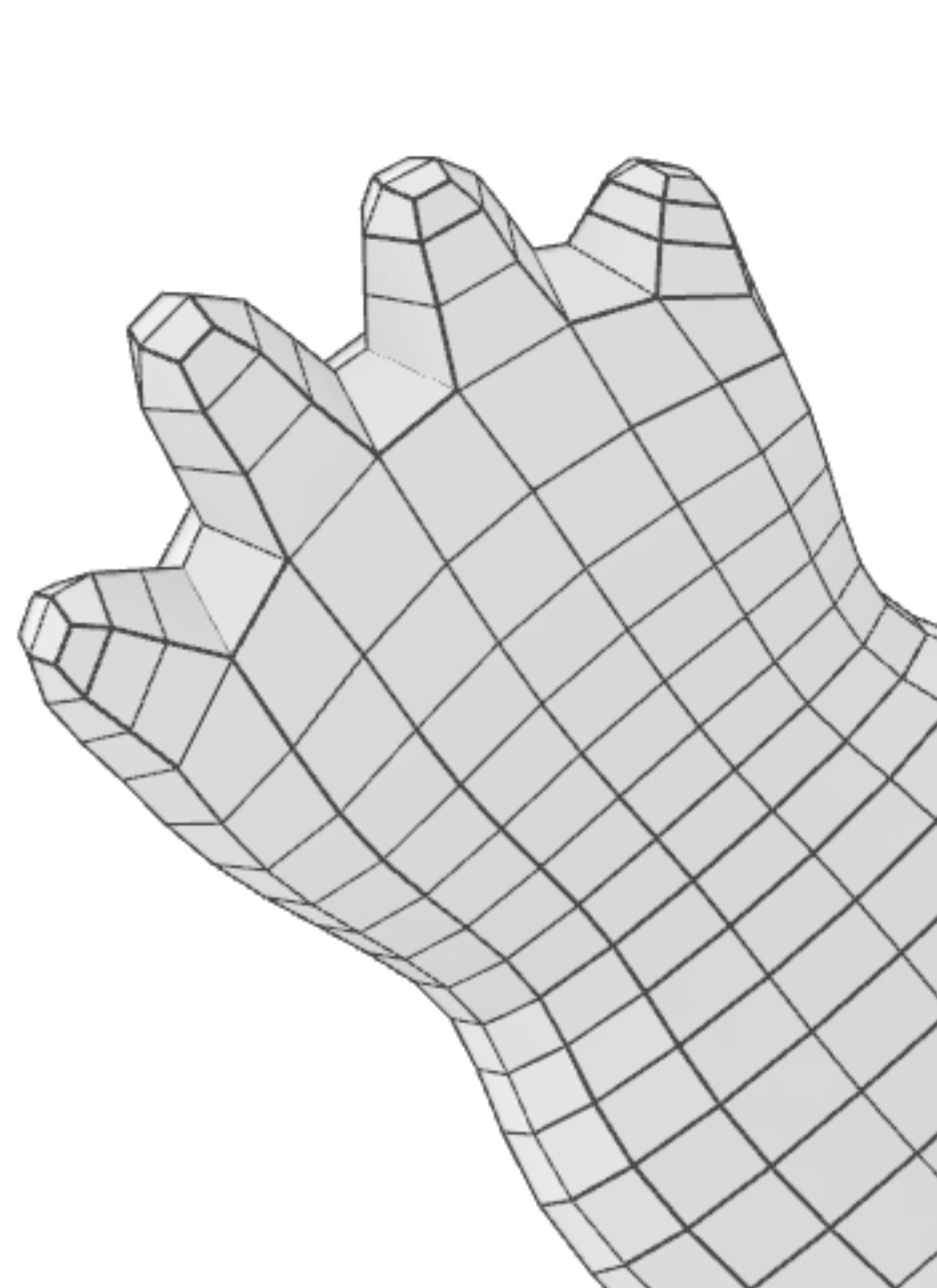} \\
    {\footnotesize (e)} & {\footnotesize (f)} & {\footnotesize (g)} & {\footnotesize (h)} \\
  \end{tabular}
  \caption{Fine-grained surface control using landmarks.
    (a) Optimized \textit{armadillo} model has fingers intersecting.
    (b) Input tet mesh for reference. It suggests we should shift the fingers in the red box the right to align them with the input.
    (c)-(d) We put a landmark on the rightmost finger and then move it to the right.
    (e) Reoptimization then moves the rightmost finger into roughly the correct position in order to reduce distortion.
    (f)-(g) We repeat the process with the other finger. 
    (h) Finally we clear the landmarks and reoptimize again.
  }

  \label{fig:hexahedralization_untangle_fingers}
\end{figure}

To enable fine-grained control over the surface, we provide a tool for the user to mark and reposition surface vertices as \emph{landmarks}, which are then fixed during the optimization.
This is helpful for guiding the optimization, e.g., by preventing points from getting projected onto the wrong side of the surface (\cref{fig:hexahedralization_untangle_fingers}).

The user has the additional option of choosing how the surface vertices are parameterized during optimization using one of three modes: \textit{free}---surface vertices are free to move off the input surface, \textit{constrained}---surface vertices are constrained to move only along the input surface, and \textit{fixed}---surface vertices are fixed during the optimization.
See the corresponding mathematical formulation in \cref{sec:constrained_smoothing}.
The \textit{free} mode is useful for most situations since it can effectively alleviate or prevent foldovers, while \textit{constrained} and \textit{fixed} modes are good for final refinement.

For visualization, we provide the user with tools to filter away hex elements, including a slicing plane and a quality threshold, similar to Hexalab \cite{bracci2019hexalab}.
For instance, the user can easily identify where the bad hex elements are using the quality threshold filter to fix them using landmark tools or by going back to the previous stage to make topological changes to the voxelized PolyCube.

\begin{table}
  \caption{User-controlled parameters during the hexahedralization stage. \rev{The symbol column corresponds to the mathematical notation used in the energy terms from \cref{sec:constrained_smoothing}.}}
\begin{tabular}{|c|c|c|}
  \hline
  parameter & description & symbol \\
  \hline
  \textit{smoothness} & smoothness of result surface & $\lambda_{\text{lap}}$\\ 
  \textit{projection} & closeness of result surface to input & $\lambda_{m\sto 0}$\\
  \textit{details} & level of recovered surface details & $\lambda_{0\sto m}$ \\
  \hline 
  \textit{conformal} & angle-preserving level & $\lambda_{\text{angle}}$ \\
  \textit{authalic} & volume-preserving level & $\lambda_{\text{vol}}$ \\
  \textit{custom} & custom mesh quality metric & $\lambda_{\text{custom}}$\\
  \hline
\end{tabular}
\label{tbl:hexahedralization_parameters}
\end{table}

\section{Optimization}
In this section, we present our mathematical formulation of the optimization problems faced in the pipeline stages above.
We use Adam \citep{kingma2014adam} as the optimizer for all problems described in this section to achieve interactive speed.

\paragraph{Notation.}
We denote a tetrahedral mesh as $\mathcal{M} = (V, T)$ where $V$ and $T$ are the sets of vertices (represented as elements in $\RR^3$) and tetrahedra (represented as 4-tuples of indices), respectively.
We use $\partial \mathcal{M}$ to denote the boundary triangular mesh induced by $\mathcal{M}$ as $\partial \mathcal{M} \defeq (\partial V, F)$ where $\partial V$ and $F$ are the surface vertices restricted from $V$ and surface faces, respectively.
We use similar notations for hexahedral meshes. For instance, $\mathcal{M} = (V, H)$ represents a hex mesh with a vertex set $V$ and a hex set $H$, and $\partial \mathcal{M} \defeq (\partial V, Q)$ denotes its surface quadrilateral mesh with vertices $\partial V$ and quads $Q$.

We use $\vol(V, t)$ to denote the volume of tetrahedron $t \in T$ computed using the corresponding vertices in $V$, and we write $\vol(\mathcal{M}) \defeq \sum_{t \in T} \vol(V, t)$.
Similarly, we define $\area(V, f)$ to be the area of the face $f \in F$ and $\area(\partial \mathcal{M}) \defeq \sum_{f\in F} \area(V, f)$.
We call vectors $(\pm 1, 0, 0)$, $(0, \pm 1, 0)$, $(0, 0, \pm 1)$ the \emph{major axis directions}.

An $\RR^3$-valued piecewise-linear map $f$ on a tetrahedral mesh $\mathcal{M}=(V,T)$ is determined uniquely by $f(V)$, its values on the vertices $V$.
Hence, we parameterize such maps using the mapped vertex positions, and, with slight abuse of notation, we use $f$ to denote both the map on the mesh domain and the discrete map on vertices.
We denote the (constant) Jacobian of $f$ in a tetrahedron $t \in T$ as $J_t(f)$, which is a linear function of $f(V)|_t$, the 4 mapped vertex positions of $t$.

\subsection{Deformation to a near-PolyCube}
\label{sec:opt_deformation}
In the deformation stage (\cref{stage:deformation}), the input shape is deformed into a near-PolyCube shape to guide the placement of cuboids in the decomposition stage (\cref{stage:decomposition}).
The deformation map is later used in the hexahedralization stage (\cref{stage:hexahedralization}) to 
initialize
the PolyCube hex mesh.
Hence, we want the deformation map to be of low distortion and inversion-free while encouraging axis-alignment.

Let $\mathcal{M}_0 = (V_0, T_0)$ denote the input tetrahedral mesh.
We look for a low-distortion piecewise-linear deformation $f_d: V_0 \to \RR^3$ that maps the input mesh $\mathcal{M}_0$ to the PolyCube-like deformed shape $\mathcal{M}_d \defeq (V_d, T_d)$, with $V_d \defeq f_d(V_0)$ and $T_d \defeq T_0$. %
For $t \in T_0$, let $J_t \defeq J_t(f_d)$ denote the Jacobian of $f_d$ restricted to the tetrahedron $t$. 
To obtain $f_d$, we minimize a deformation energy defined as $E \defeq E_{\text{iso}} + E_{\text{align}}$, where $E_{\text{iso}}$ is the distortion energy defined in \cref{eqn:iso_energy} and $E_{\text{align}}$ is the axis-alignment energy defined in \cref{eqn:alignment_energy}. %
We take the identity map as the initial $f_d$.
\paragraph{Distortion energy}
We want a smooth distortion energy that encourages angle and volume preservation of the deformation map.
As our initial $f_d$ is inversion-free, we also want our distortion energy to blow up when inversion happens.
We choose the regularized distortion energy introduced by \citet{garanzha2021foldover}:
\begin{equation}
  E_{\text{iso}} \defeq \sum_{t \in T_0}  \frac{\vol(V_0, t)}{\vol(\mathcal{M})}\left( \lambda_{\text{angle}} \frac{\tr J_t^\top J_t}{\left(R_\epsilon(\det J_t)\right)^{2/3}} + \lambda_{\text{vol}} \frac{\det^2 J_t + 1}{R_\epsilon(\det J_t)} \right), \label{eqn:iso_energy}
\end{equation}
where $\lambda_{\text{angle}}, \lambda_{\text{vol}}$ are constants and $R_\epsilon: \RR \to \RR_{> 0}$ is a regularizer that forces the energy to blow up when $\det J_t$ is close to zero or negative:%
\begin{equation}
  R_\epsilon(x) \defeq \frac{x+\sqrt{x^2 + \epsilon^2}}{2}.
\end{equation}
The first term in the summand of \cref{eqn:iso_energy} favors angle-preserving maps, while the second term favors volume-preserving maps.
We find $\lambda_{\text{angle}} = \lambda_{\text{vol}} = 1.0$ and $\epsilon=10^{-3}$ sufficient in most cases, but we let the user change them if needed.

\paragraph{Alignment energy}\label{par:cubeness_energy}
We add an alignment energy for the surface vertices $\partial V_d$ to encourage deformation into an axis-aligned near-PolyCube. 
This energy has two terms, one favoring cubeness and the other favoring a smooth transition of normals on nearby faces, similar to the one used by \citet{fang2016all}:
\begin{equation}
  E_{\text{align}} \defeq \lambda_{\text{cube}}E_{\text{cube}} + \lambda_{\text{smooth}} E_{\text{smooth}},
  \label{eqn:alignment_energy}
\end{equation}
where
\begin{align}
  E_{\text{cube}} &\defeq \sum_{f \in \partial F_0} \frac{\area(V_0, f)}{\area(\partial \mathcal{M}_0)}\Phi(\hat{n}_f), \\ 
  E_{\text{smooth}} &\defeq \sum_{\substack{f_i, f_j \in \partial F_0\\ f_i, f_j \text{ adjacent}}}  \frac{\area(V_0, f_i) + \area(V_0, f_j)}{3\area(\partial \mathcal{M}_0)}\norm{\hat{n}_{f_i} - \hat{n}_{f_j}}^2_2. 
  \label{eqn:smooth_normals}
\end{align}
Here we use $\hat{n}_f$ to denote the unit normal obtained by normalizing $(v_1 - v_0) \times (v_2 - v_0)$, where $v_0,v_1,v_2$ are deformed vertex positions of the face $f$. 
We choose a smooth cubeness function $$\Phi(n) = n_x^2n_y^2 + n_y^2n_z^2 + n_z^2n_x^2$$ from \citet{fu2016efficient} to penalize deviation of normals from the major axis directions.
We found this energy better at globally orienting the shape (see \cref{fig:cubeness_adjustment} (b)) and more stable during optimization compared to the $\ell_1$-norm from \citet{huang2014l1}. %
To prevent collapsing to a point, we use the original vertex positions to calculate the area weights.

\subsection{Continuous PolyCube optimization}
\label{sec:opt_polycube}
In the decomposition stage (\cref{stage:decomposition}), the user guides the construction of a PolyCube using cuboids whose union approximates the deformed near-PolyCube shape $\mathcal{M}_d$.
Such construction is facilitated by the a continuous PolyCube optimization scheme described below that automatically adjusts existing cuboids' parameters.

Our formulation is inspired by the distance-field-based approach by \citet{smirnov2020deep}, where the discrepancy between two shapes is measured by comparing their distance fields. 
For an arbitrary set $S \subset \RR^3$, the \emph{signed distance field} of $S$ is defined to be 
\[
d_S(x) \defeq (-1)^{\mathbbm{1}_{x \in S}}\inf_{y \in S} \norm{x-y}_2.\]
In particular, $d_S(x) \ge 0$ if $x \notin S$, and $d_S(x) \le 0$ if $x \in S$.
For an axis-aligned cuboid $\mathcal{C}$ with center $c \in \mathbb{R}^3$ and side lengths $h \in \mathbb{R}^3$, its signed distance field is given by 
\[
  d_{\mathcal{C}}(p) = \norm{\max(d, 0)}_2 + \min(\max(d_x, d_y, d_z), 0)
,\]
where $d = (|p_x - c_x|, |p_y-c_y|, |p_z-c_z|) - h$ \citep{smirnov2020deep}. 

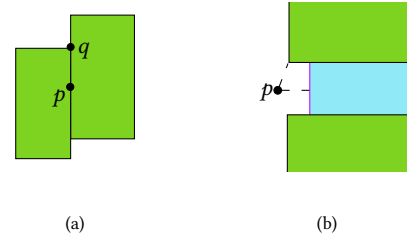
\begin{figure}
  \newcolumntype{C}[1]{>{\centering\arraybackslash}m{#1}}
  \begin{tabular}{C{3cm} C{3cm}}
    \begin{tikzpicture}[x=0.75pt,y=0.75pt,yscale=-0.5,xscale=0.5]
\draw  [fill={rgb, 255:red, 126; green, 211; blue, 33 }  ,fill opacity=1 ] (286,84.5) -- (350.36,84.5) -- (350.36,209.4) -- (286,209.4) -- cycle ;
\draw  [fill={rgb, 255:red, 126; green, 211; blue, 33 }  ,fill opacity=1 ] (230.6,118) -- (286,118) -- (286,229.4) -- (230.6,229.4) -- cycle ;
\draw  [fill={rgb, 255:red, 0; green, 0; blue, 0 }  ,fill opacity=1 ] (282,157) .. controls (282,155.01) and (283.61,153.4) .. (285.6,153.4) .. controls (287.59,153.4) and (289.2,155.01) .. (289.2,157) .. controls (289.2,158.99) and (287.59,160.6) .. (285.6,160.6) .. controls (283.61,160.6) and (282,158.99) .. (282,157) -- cycle ;
\draw  [fill={rgb, 255:red, 0; green, 0; blue, 0 }  ,fill opacity=1 ] (282.5,116.7) .. controls (282.5,114.77) and (284.07,113.2) .. (286,113.2) .. controls (287.93,113.2) and (289.5,114.77) .. (289.5,116.7) .. controls (289.5,118.63) and (287.93,120.2) .. (286,120.2) .. controls (284.07,120.2) and (282.5,118.63) .. (282.5,116.7) -- cycle ;

\draw (266,157) node [anchor=north west][inner sep=0.75pt]    {$p$};
\draw (291,109.6) node [anchor=north west][inner sep=0.75pt]    {$q$};

\end{tikzpicture} & 
    \begin{tikzpicture}[x=0.75pt,y=0.75pt,yscale=-0.5,xscale=0.5]
  \clip (270, 60) rectangle + (160, 170);
\draw  [color={rgb, 255:red, 212; green, 0; blue, 255 }  ,draw opacity=1 ][fill={rgb, 255:red, 143; green, 233; blue, 247 }  ,fill opacity=1 ] (332.44,-31.92) -- (538.71,-31.92) -- (538.71,322.17) -- (332.44,322.17) -- cycle ;
\draw  [fill={rgb, 255:red, 126; green, 211; blue, 33 }  ,fill opacity=1 ] (311.52,-10.72) -- (547,-10.72) -- (547,120.34) -- (311.52,120.34) -- cycle ;
\draw  [fill={rgb, 255:red, 126; green, 211; blue, 33 }  ,fill opacity=1 ] (309.8,173.2) -- (545.28,173.2) -- (545.28,304.26) -- (309.8,304.26) -- cycle ;
\draw  [color={rgb, 255:red, 0; green, 0; blue, 0 }  ,draw opacity=1 ][fill={rgb, 255:red, 0; green, 0; blue, 0 }  ,fill opacity=1 ] (296.29,148.46) .. controls (296.29,146.33) and (298.02,144.6) .. (300.16,144.6) .. controls (302.29,144.6) and (304.02,146.33) .. (304.02,148.46) .. controls (304.02,150.6) and (302.29,152.33) .. (300.16,152.33) .. controls (298.02,152.33) and (296.29,150.6) .. (296.29,148.46) -- cycle ;
\draw  [dash pattern={on 4.5pt off 4.5pt}]  (300.16,148.46) -- (310.52,121.34) ;
\draw  [dash pattern={on 4.5pt off 4.5pt}]  (300.16,148.46) -- (332,148.46) ;

\draw (280.64,139.19) node [anchor=north west][inner sep=0.75pt]    {$p$};

\end{tikzpicture} \\
  {\footnotesize (a)} & {\footnotesize (b)}
  \end{tabular}
  \caption{
    (a) 
    An illustration showing that the gap between $\widetilde{d_{\mathcal{P}}}$ and $d_{\mathcal{P}}$ can be arbitrarily large. 
    Consider the PolyCube comprised of the two green cuboids. 
    For the point $p$, $\widetilde{d_{\mathcal{P}}}(p) = 0$, but $d_{\mathcal{P}}(p) = - \norm{q - p} < 0$.
(b) An example showing that the discrepancy energy $E_+$ alone may create gaps.
In this scenario, the light blue indicates $\mathcal{M}_d$, and the PolyCube consists of two green cuboids.
The distance from $p$ to $\mathcal{M}_d$ and to $\mathcal{P}$ is the same, so $(d_{\mathcal{M}_d}(p) - \widetilde{d}_{\mathcal{P}}(p))^2 = 0$, and $p$ could be a local optimum of $E_+$ (\cref{eqn:df_positive}). 
  This explains why $E_-$ is needed to close the gaps.
}
\label{fig:sdf_illustration}
\end{figure}

With slight abuse of notation, we use $\mathcal{M}_d$ to denote the deformed shape as a subset of $\RR^3$, so that $d_{\mathcal{M}_d}$ denotes its signed distance field.
Let $\mathcal{P} \defeq \bigcup_{i=1}^{k} \mathcal{C}_i$ denote a PolyCube consisting of $k$ cuboids $\mathcal{C}_1,\ldots,\mathcal{C}_k$.
Defining the signed distance field of $\mathcal{P}$ in terms of the signed distance fields of $\mathcal{C}_i$'s is difficult.
In the case $x \notin \mathcal{P}$, however, $d_{\mathcal{P}}$ is given by a straightforward expression: %
\[
  d_{\mathcal{P}}(x) = \min_{1\le i \le k} d_{\mathcal{C}_i}(x)\ \forall x \notin \mathcal{P}. %
\]
Define $\widetilde{d}_{\mathcal{P}}(x) \defeq \min_{i=1}^k d_{\mathcal{C}_i}(x)$.
We have $\widetilde{d}_{\mathcal{P}}(x) \le d_{\mathcal{P}}(x)$, but the gap can be arbitrary large (\cref{fig:sdf_illustration}(a)).
Even though $\widetilde{d}_{\mathcal{P}}\neq d_{\mathcal{P}}$, we have $\{x: \widetilde{d}_{\mathcal{P}}(x) \le 0\} = \{x: d_{\mathcal{P}}(x) \le 0\}$, so they represent the same interior shape.

Let $A \subset \RR^3$ be a finite set of points that we call \textit{anchors} on which we compare $d_{\mathcal{M}_d}$ and $d_{\mathcal{P}}$.
We design our energy to be a combination of two terms: $E \defeq \lambda_+ E_+ + \lambda_- E_-$, where $E_+$ and $E_-$ are defined by \cref{eqn:df_positive} and \cref{eqn:df_negative} respectively.

\paragraph{Discrepancy energy $E_+$}
For $p \in \RR^3 \setminus \mathcal{M}_d$, we want cuboids to avoid covering $p$, i.e., $d_{\mathcal{C}_i}(p) \ge 0$ for every $i$. 
But if this is the case, then $\widetilde{d}_{\mathcal{P}}(p) = d_{\mathcal{P}}(p)$, so we can compare $d_{\mathcal{P}}(p)$ with $d_{\mathcal{M}_d}(p)$ by using $\widetilde{d}_{\mathcal{P}}(p)$ in place of $d_{\mathcal{P}}(p)$.
Thus we define
\begin{equation}  
  E_{+} \defeq \sum_{p \in A \setminus \mathcal{M}_d} \left(d_{\mathcal{M}_d}(p) - \widetilde{d}_{\mathcal{P}}(p)\right)^2.
  \label{eqn:df_positive}
\end{equation}
\paragraph{Gap-closing energy $E_-$}
For $p \in \mathcal{M}_d$, we want $p$ to be contained in at least one cuboid, i.e., $\widetilde{d}_{\mathcal{P}}(p) \le 0$.
Therefore, we let
\begin{equation}  
  E_{-} \defeq \sum_{p \in A \cap \mathcal{M}_d} \mathbbm{1}_{\widetilde{d}_{\mathcal{P}}(p) \ge 0} \left(\widetilde{d}_{\mathcal{P}}(p)\right)^2
  \label{eqn:df_negative}
\end{equation}
to help close the gaps caused by using only $E_+$ (\cref{fig:sdf_illustration}(b)).

\begin{figure}
  \hspace*{-0.1in}
  \newcolumntype{C}[1]{>{\centering\arraybackslash}m{#1}}
    \begin{tabular}{C{3cm} C{3cm}}
      \includegraphics[height=0.7in,trim=5.0in 3.in 0 4in, clip]{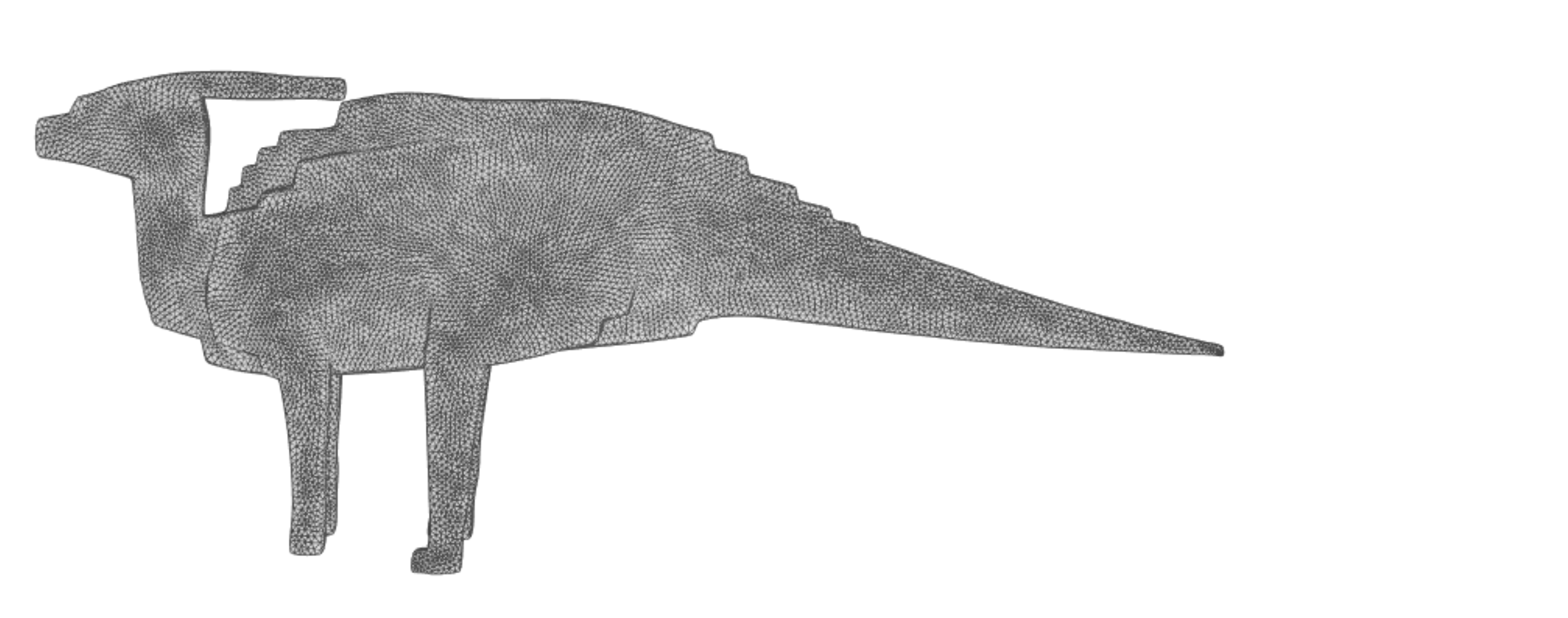} &
      \input{annotated/dilo_p_only.tex}
      \\
      {\footnotesize (a) Deformed mesh} & {\footnotesize (b) $\lambda_+ =1, \lambda_- = 0$} \\
      \input{annotated/dilo_m_only.tex} &
      \input{annotated/dilo_balanced.tex}  \\
      {\footnotesize (c) $\lambda_+ = 0, \lambda_- = 1$ } & {\footnotesize (d) $\lambda_+=0.1, \lambda_-=1$}
    \end{tabular}
  \caption{
    Comparison of PolyCube optimization parameters $\lambda_+, \lambda_-$.
    (a) Deformed mesh of \textit{dilo}.
    (b) If we use only discrepancy energy (\cref{eqn:df_positive}), then small gaps are created in the circled regions.
    (c) If we use only gap-closing energy (\cref{eqn:df_positive}), then gaps from (b) are not be created, but the PolyCube is too coarse and has undesired intersections (red circle).
    (d) With the right parameters, we can close the gaps without making the PolyCube too coarse.
  }
  \label{fig:pm_adjustment}
\end{figure}

In our experiments, we choose anchors to be a combination of a uniform grid and perturbed points from the surface of $\mathcal{M}_d$, though the user can modify the anchors if desired.
In practice, we found that it is good to start with $\lambda_+ = \lambda_- = 1.0$ to prevent over-coverage in the beginning and reduce $\lambda_+$ while increasing $\lambda_-$ before reoptimizing to close the gaps.
Hence we let the user adjust the parameters $\lambda_+, \lambda_-$ (see \cref{fig:pm_adjustment}).
An alternative to fix the gaps is through manually editing and locking the cuboids (\cref{stage:decomposition}).

\subsection{Inversion-free pullback of PolyCube}
\label{sec:pullback}
In the beginning of the hexahedralization stage (\cref{stage:hexahedralization}), we want to find a volumetric map that deforms the PolyCube hex mesh from the discretization stage (\cref{stage:discretization}) back into the input mesh geometry. %
We start with a two-step initialization to an inversion-free map from the PolyCube hex mesh to the input volume. 
The idea is to use the deformation map from the deformation stage as a guide to pull back the PolyCube hex mesh gradually, while incorporating a barrier function  to prevent foldovers.

In the first step, we deform the PolyCube mesh into the near-PolyCube mesh from the deformation stage to account for the discrepancy introduced in discretization (\cref{fig:pullback}(d)).
In the second step, we deform back to the input mesh using barycentric pullback as a guide (\cref{fig:pullback}(f)).
We do not change the mesh connectivity throughout.

Let $\mathcal{M}_p \defeq (V_p, H_p)$ denote the hexahedral mesh of the voxelized PolyCube from the discretization stage, whose hex elements are regular cubes with the same length on all sides.
\subsubsection{Deforming to $\mathcal{M}_d$}\label{sec:pullback_step_1}
In the first step, we look for a map $f_{d'}: V_p \to \RR^3$ that sends $\mathcal{M}_p$ to $\mathcal{M}_{d'} \defeq (V_{d'}, H_p)$ so that $\partial \mathcal{M}_{d'}$ is close to $\partial \mathcal{M}_d$.
Let $T_p$ denote the collection of tetrahedra formed by putting one tetrahedron on each of the 8 corners of every hex in $H_p$.
For $t \in T_p$, like in \cref{sec:opt_deformation}, we use $J_t \defeq J_t(f_{d'})$ to denote the Jacobian of $f_{d'}$ on tetrahedron $t$.
Note that there are cases where $\det J_t > 0$ for all 8 tetrahedra in a hex but the induced trilinear mapping is not locally injective \citep{zhang2005subtetrahedral}.
We have not observed such cases in our experiments, but, if necessary, we can augment $T_p$ to include the 32 tetrahedra defined in \citep{zhang2005subtetrahedral} for each hex to make the induced trilinear map inversion-free everywhere. 

We initialize $f_{d'}$ to be the identity map and optimize $f_{d'}$ by minimizing $E \defeq E_{\text{hex-iso}} + E_{\text{prox}} + E_{\text{lap}}$, where each energy term is described below.
\paragraph{Distortion energy}
Since we want to prevent foldovers, we use a barrier-like distortion energy similar to \cref{eqn:iso_energy} but with uniform weights, due to the fact that volumes of all hexes in $H_p$ (and by extension all tets in $T_p$) are the same:
\begin{equation}
  E_{\text{hex-iso}} \defeq \sum_{t \in T_p} \left( \lambda_{\text{angle}} \frac{\tr J_t^\top J_t}{\left(R_\epsilon(\det J_t)\right)^{2/3}} + \lambda_{\text{vol}} \frac{\det^2 J_t + 1}{R_\epsilon(\det J_t)} \right).
  \label{eqn:hex_iso_energy}
\end{equation}
\paragraph{Proximity energy}
We introduce a term that measures the bi-directional distance between $\partial \mathcal{M}_{d'}$ and $\partial \mathcal{M}_d$:
\begin{equation}
  E_{\text{prox}} \defeq \lambda_{d'\sto d} E_{d' \sto d} + \lambda_{d\sto d'} E_{d \sto d'},
  \label{eqn:prox_energy}
\end{equation}
with
\begin{align}
  E_{d' \sto d} &\defeq \sum_{v \in \partial V_{d'}} \norm{v - \proj(v, \partial \mathcal{M}_d)}_2^2, \label{eqn:prox_energy_d}\\
  E_{d \sto d'} &\defeq \int_{\partial \mathcal{M}_d} \norm{v - \proj(v, \partial \mathcal{M}_{d'})}_2^2 \dd v, \label{eqn:prox_energy_dd}
\end{align}
where $\proj(v, \partial \mathcal{M})$ denotes the projected point of $v$ onto the surface $\partial \mathcal{M}$.
We choose uniform weights in \cref{eqn:prox_energy_d} because the quad surface before and after applying $f_{d'}$ should consist of quads with similar areas.
To evaluate \cref{eqn:prox_energy_dd} during each gradient step, we use a uniformly sampled batch of $|\partial V_{d'}|$ points on $\partial \mathcal{M}_d$ before computing the gradient.
\paragraph{Smoothness energy}
We finally add a Laplacian-smoothing energy that helps maintain a smooth surface during the deformation:
\begin{equation}
  E_{\text{lap}} \defeq \lambda_{\text{lap}} \sum_{v \in \partial V_{d'}} \norm{v - \frac{1}{\abs{N(v)}}\sum_{u \in N(v)} u}_2^2,
  \label{eqn:laplacian_smooth_energy}
\end{equation}
where $N(v)$ denotes the 1-ring neighborhood of $v$ on $\partial V_{d'}$.

In this step, we set all weights $\lambda_*$ to be $1$ and $\epsilon = 10^{-4}$.
\cref{fig:pullback}(d) shows an example of this step.

\subsubsection{Deforming back to $\mathcal{M}_0$}\label{sec:pullback_step_2}
\begin{figure}
  \centering
  \newcolumntype{C}[1]{>{\centering\arraybackslash}m{#1}}
    \begin{tabular}{C{2.2cm} C{2.2cm} C{2.2cm} }
      \includegraphics[height=1.7in,trim=6.4in 1.in 0 1in, clip]{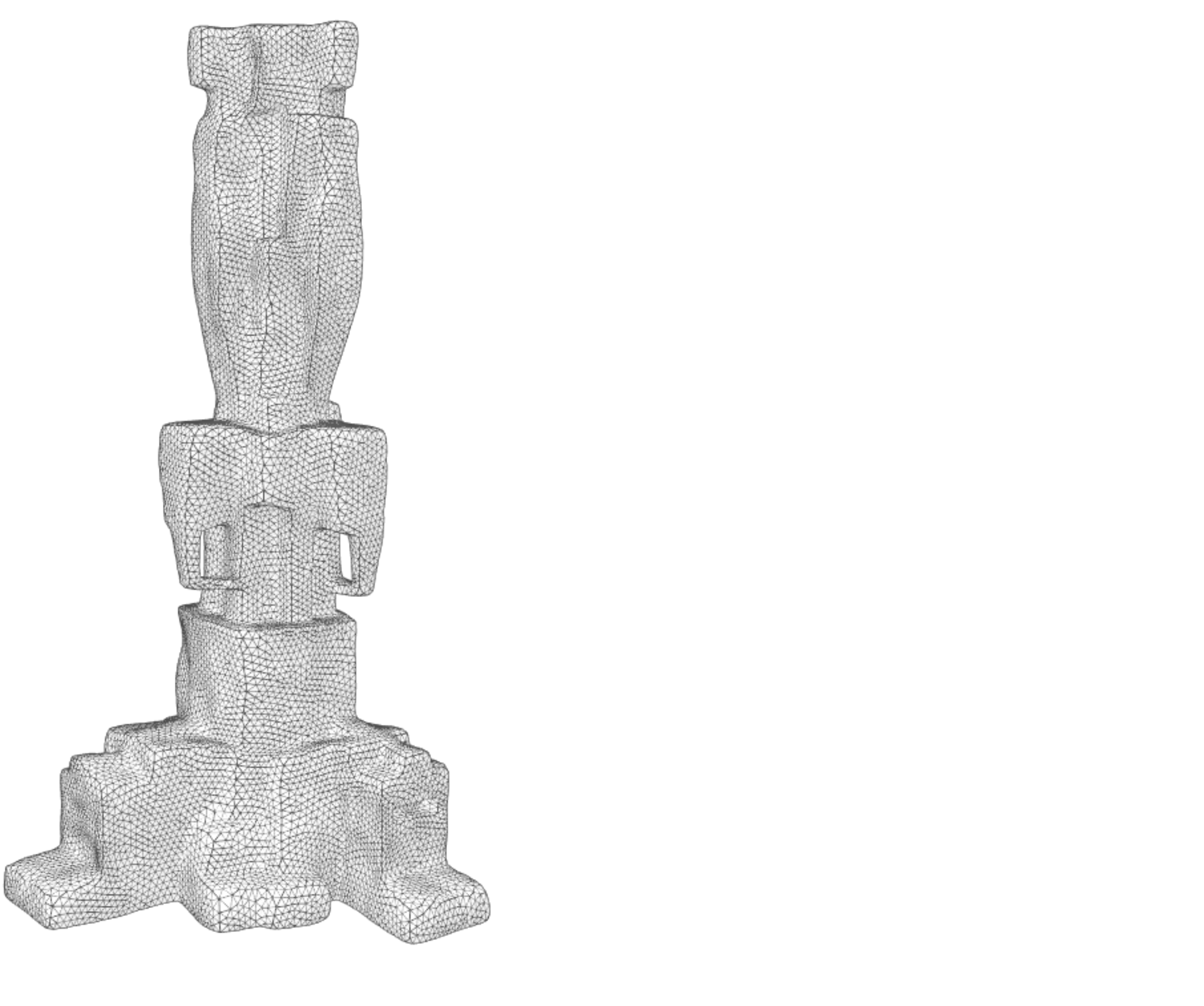} 
      &
      \includegraphics[height=1.7in,trim=6.4in 1.in 0 1in, clip]{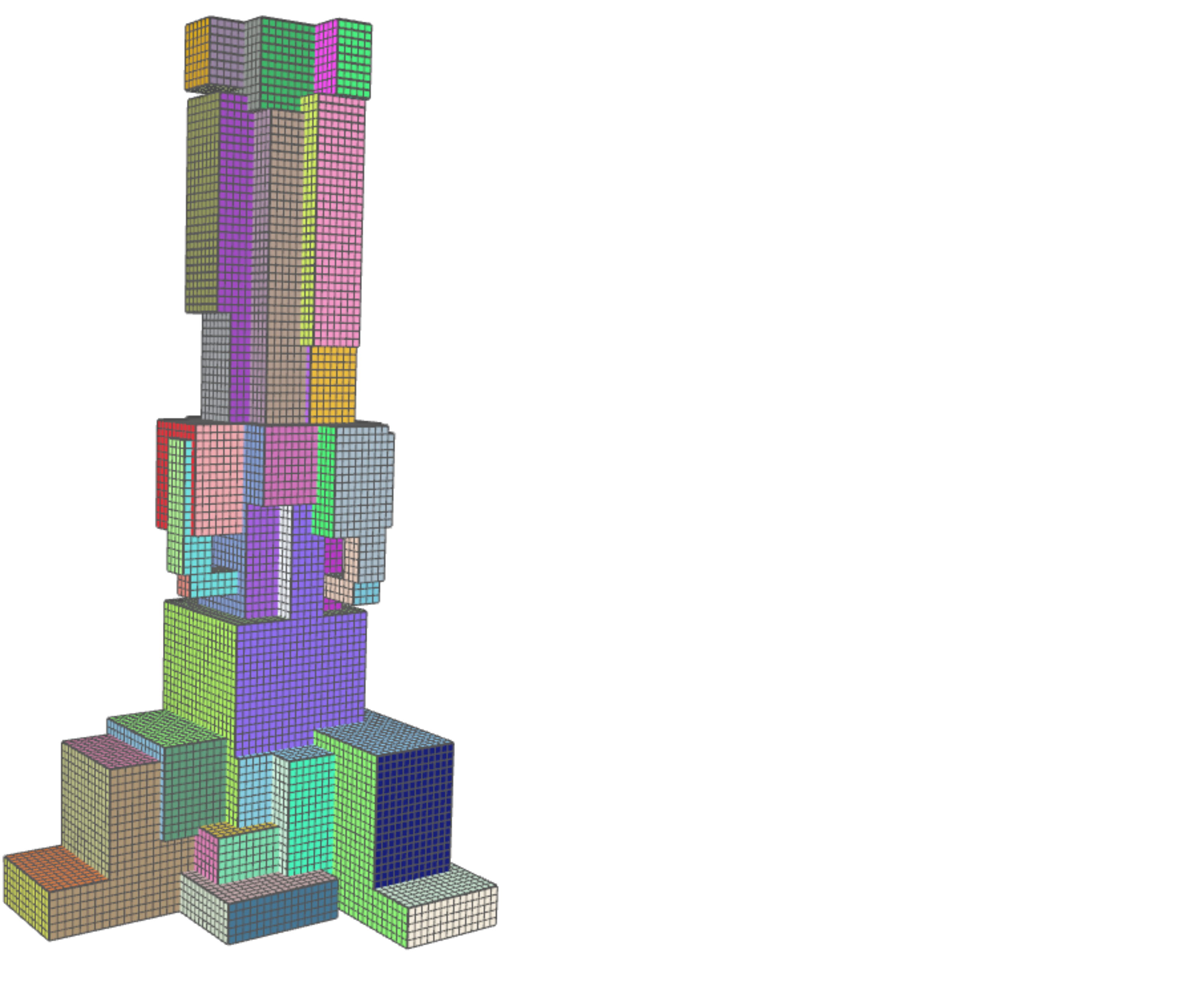} 
      &
      \input{annotated/thai_statue_direct_pullback.tex}
      \\
      {\footnotesize (a) $\mathcal{M}_d$}
      &
      {\footnotesize (b) $\mathcal{M}_p=(V_p, H_p)$}
      &
      {\footnotesize (c) $(\pull(V_p), H_p)$}
      \\
      \input{annotated/thai_statue_gradual_1.tex}
      &
      \input{annotated/thai_statue_indirect_pullback.tex}
      &
      \includegraphics[height=1.7in,trim=6.4in 1.in 0 1in, clip]{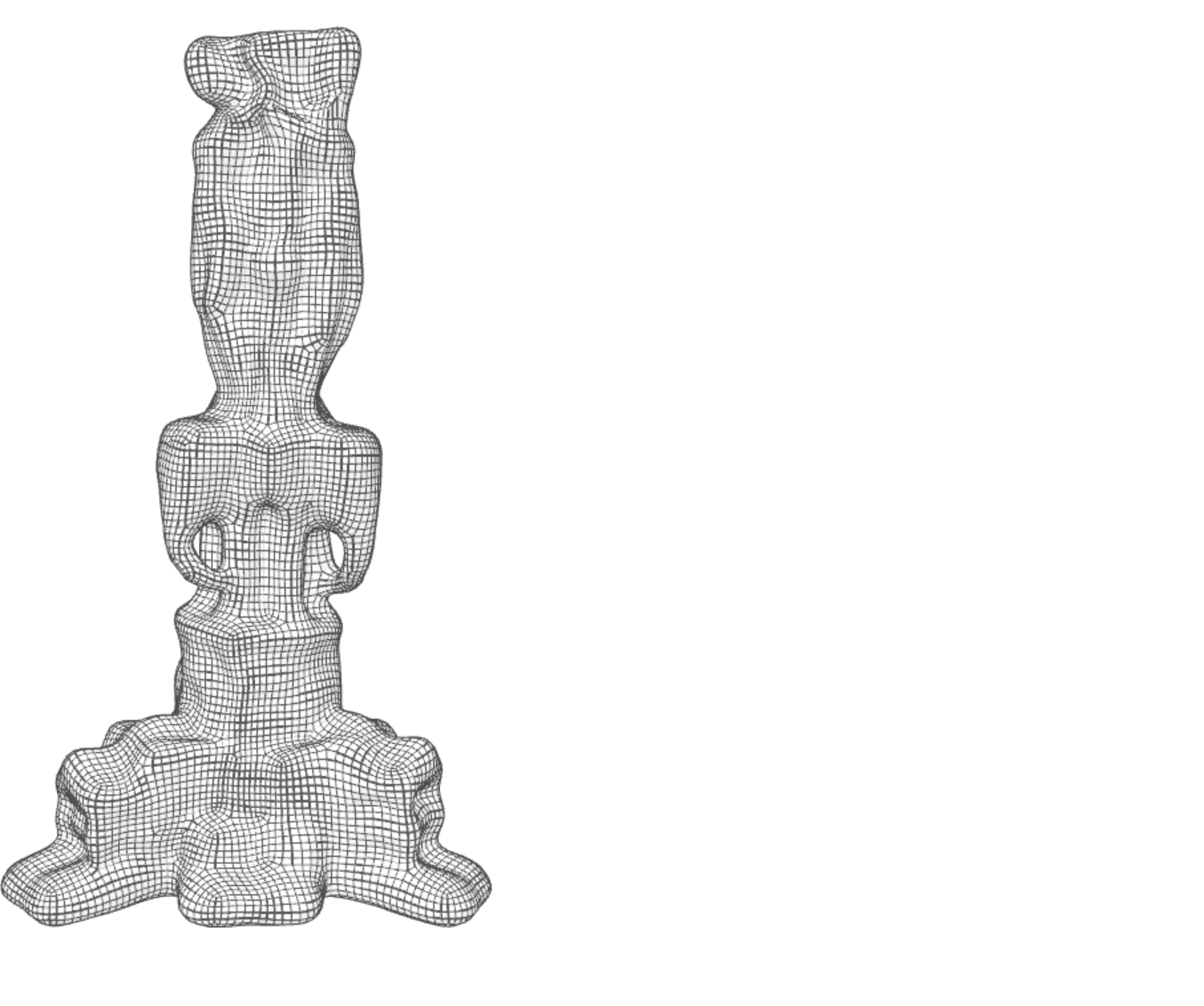} 
      \\
      {\footnotesize (d) $\mathcal{M}_{d'}=(V_{d'}, H_p)$}
      &
      {\footnotesize (e) $(\pull(V_{d'}), H_p)$}
      &
      {\footnotesize (f) $\mathcal{M}_m=(V_m, H_p)$}
    \end{tabular}
  \caption{
    Comparison of pullback strategies.
    (a) Deformed mesh of \textit{Thai statue}.
    (b) Generated PolyCube with distinctly colored charts.
    (c) Directly projecting and pulling back results in many inverted hexes as well as large distortion.
    (d) Result of the first step of our inversion-free pullback (\cref{sec:pullback_step_1}). 
    (e) If we project and pull back directly after first step, we can still obtain inverted hexes ($J_{\min} = -0.883$).
    (f) Result of the second step of our inversion-free pullback (\cref{sec:pullback_step_2}) with no inverted hexes ($J_{\min} = 0.048$ before optimize mesh quality using \cref{sec:constrained_smoothing}). 
  }
  \label{fig:pullback}
\end{figure}

Now we have a hex mesh $\mathcal{M}_{d'} = (V_{d'}, H_p)$ that is close to the near-PolyCube mesh $\mathcal{M}_d$.
In this step, we look for a map $f_m: V_{p} \to \RR^3$ that extends $f_{d'}$ from \cref{sec:pullback_step_1} to obtain $\mathcal{M}_m = (V_m, H_p)$ with $V_m \defeq f_m(V_p)$, so that its surface $\partial \mathcal{M}_m$ approximates the input surface $\partial \mathcal{M}_0$.
For $v \in \RR^3$, let $\pull(v) \in \mathcal{M}_0$ denote the result of projecting $v$ to the closest tetrahedron in $\mathcal{M}_d$ and then pull back to $\mathcal{M}_0$ via $f_d: \mathcal{M}_0 \to \mathcal{M}_d$ from \cref{sec:opt_deformation} using barycentric coordinates.
If $\mathcal{M}_{d'}$ is exactly $\mathcal{M}_d$, then $(\pull(V_{d'}), H_p)$ gives an inversion-free hex mesh of the input mesh.
However, this is often not the case, and na\"ively using projection and then pulling back can result in foldovers or points being projected onto the wrong side of the surface (\cref{fig:pullback} (e)). 

Instead, we use $\pull(V_{d'})$ only as a guide and deform $\mathcal{M}_{d'}$ to gradually reduce the distance between $V_m$ and $\pull(V_{d'})$ while avoiding inversion.
We find $f_m$ by initializing it to be $f_{d'}$ and then minimizing $E \defeq E_{\text{hex-iso}} + E_{\text{pullback}} + E_{\text{lap}}$, where $E_{\text{hex-iso}}$ and $E_{\text{lap}}$ are the same as \cref{eqn:hex_iso_energy} and \cref{eqn:laplacian_smooth_energy} respectively, except the variables are now $V_m$.
The new energy term $E_{\text{pullback}}$ is defined as
\begin{equation}
  E_{\text{pullback}} \defeq \lambda_{\text{pullback}} \sum_{v \in V_{m}} \norm{v - \pull(v_{d'})}_2^2,
  \label{eqn:pullback_energy}
\end{equation}
where $v_{d'}$ is position of the vertex in $V_{d'}$ with the same index as $v$.
We find $\lambda_{\text{pullback}} = 1$ sufficient.
The result of this step is shown in \cref{fig:pullback}(f).

\subsection{Constrained mesh quality optimization}\label{sec:constrained_smoothing}
After initializing an inversion-free hex mesh that approximates the input domain (\cref{sec:pullback}), the user can further improve the mesh quality in various ways.

To achieve this, we take $f_m$ from \cref{sec:pullback_step_2} and further optimize it.
In contrast to past hex-mesh quality improvement work \citep{livesu2015practical, fu2015computing}, where it is typically assumed that the surface vertices are either fixed or do not move around substantially, in our case the surface vertices $\partial V_m$ can move significantly during optimization to support more aggressive improvement strategies (e.g., \cref{fig:hexahedralization_untangle_fingers}). 
To constrain the surface vertices $\partial V_m$ to move along the input surface $\partial \mathcal{M}_0$, we use the bi-directional proximity energy $E_{\text{prox}}$ from \cref{eqn:prox_energy} but with $\partial \mathcal{M}_0$ in place of $\partial \mathcal{M}_d$ and with variables $V_m$ instead of $V_{d'}$:
\begin{equation}
  E_{\text{prox}} \defeq \lambda_{m \sto 0} E_{m \sto 0} + \lambda_{0\sto m} E_{0 \sto m},
  \label{eqn:prox_energy_final}
\end{equation}
where
\begin{align}
  E_{m \sto 0} &\defeq \sum_{v \in \partial V_{m}} \norm{v - \proj(v, \partial \mathcal{M}_0)}_2^2, \label{eqn:prox_energy_final_d} \\
  E_{0 \sto m} &\defeq \int_{\partial \mathcal{M}_0} \norm{v - \proj(v, \partial \mathcal{M}_{m})}_2^2 \dd v.\label{eqn:prox_energy_final_dd}
\end{align}
We further include the distortion energy $E_{\text{hex-iso}}$ from \cref{eqn:hex_iso_energy} and the smoothness energy $E_{\text{lap}}$ from \cref{eqn:laplacian_smooth_energy}, similar to \cref{sec:pullback_step_2}.

We also allow the user to add in a custom energy $E_{\text{custom}}$ that favors user-specific mesh quality.
For instance, if the user wants the resulting mesh to have a high scaled Jacobian (\cref{fig:hexahedralization_tradeoff_j_sj}(b)), then we set
\begin{equation}
  E_{\text{custom}} \defeq -\lambda_{\text{custom}} \sum_{t \in T_p} \det \widehat{J_t},
  \label{eqn:custom_energy_sj}
\end{equation}
where $\widehat{J_t}$ is obtained from $J_t$ by normalizing each column.
Any smooth metric that can be computed using vertex positions can be accommodated this way.

We let the user control the weights $\lambda_{\text{lap}}$, $\lambda_{m\sto 0}$, $\lambda_{0\sto m}$, $\lambda_{\text{angle}}$, $\lambda_{\text{vol}}$, $\lambda_{\text{custom}}$, which correspond to the interpretable parameters \textit{smoothness}, \textit{projection}, \textit{details}, \textit{conformal}, \textit{authalic}, \textit{custom}, respectively from \cref{tbl:hexahedralization_parameters}.

Our final energy is $E \defeq E_{\text{hex-iso}} + E_{\text{prox}} + E_{\text{lap}} + E_{\text{custom}}$.
One option we provide is to allow the user to optimize for the worst hex element quality instead of the average one.
This is inspired by the AMIPS energy from \citet{fu2015computing}, but instead of directly exponentiating the summands, we use the following log-sum-exp modification of \cref{eqn:hex_iso_energy} to improve numerical stability:
\begin{equation}
  E_{\text{hex-lse}} \defeq \log \left( \sum_{t \in T_p} \exp\left( \frac{\lambda_{\text{angle}} \tr J_t^\top J_t}{(R^{\det}_\epsilon(\det J_t))^{2/3}} + \frac{\lambda_{\text{vol}} (\det^2 J_t + 1)}{R^{\det}_\epsilon(\det J_t)} \right) \right).\label{eqn:iso_lse}
\end{equation}
We find that when $\mathcal{M}_m$ has no inverted hex, using $E_{\text{hex-lse}}$ in place of $E_{\text{hex-iso}}$ can effectively improve the worst hex element.
We provide similar options for the custom loss, for instance, if the user wants to improve the minimum scaled Jacobian (\cref{eqn:custom_energy_sj}).

The three modes of parameterizing surface vertices $\partial V_m$ described in \cref{stage:hexahedralization} correspond to the following:
\paragraph{Free} 
This is the default option where surface vertices are free to move in $\RR^3$ while relying on $E_\text{prox}$ (\cref{eqn:prox_energy_final}) to make them stay close to $\partial \mathcal{M}_0$.
\paragraph{Constrained} 
In this mode, we force $\partial V_m$ to be on $\partial \mathcal{M}_0$ during the optimization.
For each $v \in \partial V_m$, we create a latent variable $z \in \RR^3$, \rev{so that $v \defeq \proj(z, \partial \mathcal{M}_0)$}.
Let $Z$ denote the set of all latent variables.
\rev{Then, we use $\proj(z, \partial \mathcal{M}_0) \in \mathcal{M}_0$ in place of vertex $v$ when calculating the total energy}, and we update $z$ by differentiating through the projection operator during each gradient step.
A caveat is that there are regions where $\partial \proj(z,\partial \mathcal{M}_0) / \partial z$ vanishes, such as when $z$ is above a ridge formed by two neighboring faces.
To prevent gradient-based optimization from getting stuck in these situations, we use instead the constant non-zero gradient of $\partial \proj(z,\partial \mathcal{M}_0) / \partial z$ as if the closest triangle extends to a plane.
We also modify \cref{eqn:prox_energy_final_d} to 
\begin{equation}
  \sum_{z \in Z} \norm{z - \proj(z, \partial \mathcal{M}_0)}_2^2,
\end{equation}
so that the wandering latent variables will stay close to the input surface.
Comparing to alternatives like projected gradient descent, this parameterization makes it seamless to use momentum-based optimizers (\cref{sec:implementation}).
\paragraph{Fixed} 
In this mode the positions of $\partial V_m$ are fixed during the optimization.
This is useful when the user is happy with the surface but wants to further improve the interior mesh quality.

\section{Implementation Details}
\label{sec:implementation}
We implement our interactive system in C++ with Vulkan and GUI library ImGui.\footnote{https://github.com/ocornut/imgui}
Most of the optimization is powered by the C++ frontend of PyTorch \cite{pytorch} and runs on the GPU for interactive speed.
We utilize CPU-level parallelism to accelerate intensive computations like ray tracing for in-scene mouse control and hex element filtering.
We use Adam \citep{kingma2014adam} as our optimizer with a default learning rate of $10^{-3}$ in the deformation stage and $10^{-4}$ in the hexahedralization stage, and $\beta_1=\beta_2 = 0.9$, although the user can change these parameters as well as the number of gradient steps if needed. %
We implement custom CUDA functions to enable fast signed distance fields computation and point-to-mesh projection with back-propagation support to make optimization of terms like \cref{eqn:prox_energy_final} possible on meshes with about $10^5$ hexes.
\rev{See \cref{sec:app_cuda} for more details.}

For rendering, we use multi-sampled anti-aliasing and screen-space ambient occlusion in a deferred pipeline to strike a balance between visual clarity and efficiency, as we are also heavily utilizing the GPU for optimization.
All figures in the paper are generated directly by our rendering pipeline. 

Our system can run comfortably on mid-range NVIDIA graphics cards, such as the GTX 1080 and RTX 2060. %
The implementation of our system is available at \url{https://github.com/lingxiaoli94/interactive-hex-meshing}.

\section{Results}
\label{sec:results}
\begin{figure}
  \newcolumntype{C}[1]{>{\centering\arraybackslash}m{#1}}
  \hspace*{-0.1in}
  \begin{tabular}{C{1.3cm}  C{1.3cm}  C{1.3cm}  C{1.3cm} C{1.3cm}}
    \includegraphics[height=1.2in,trim=6.6in 1.in 0 1in, clip]{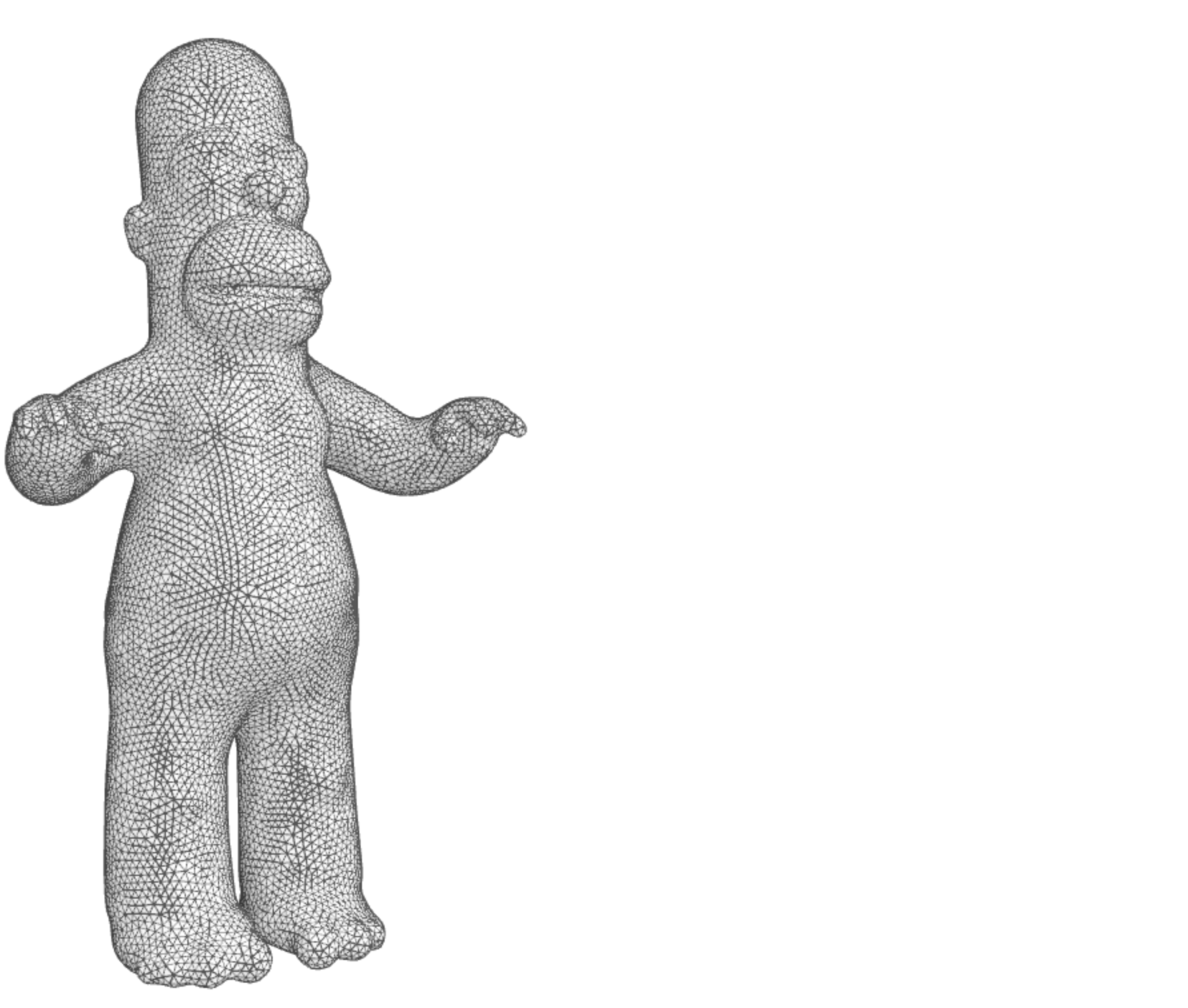} &
    \includegraphics[height=1.2in,trim=6.6in 1.in 0 1in, clip]{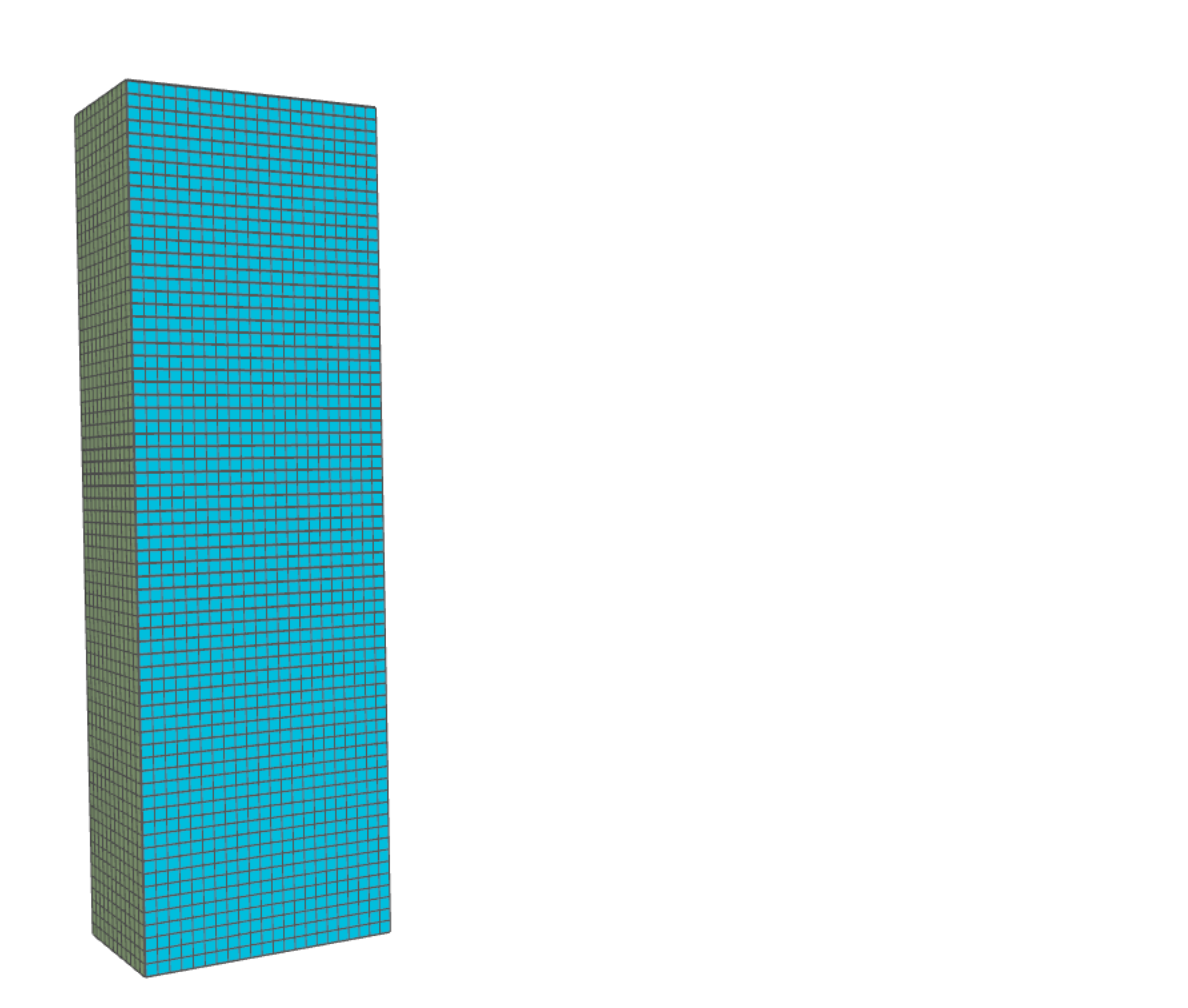} &
    \includegraphics[height=1.2in,trim=6.6in 1.in 0 1in, clip]{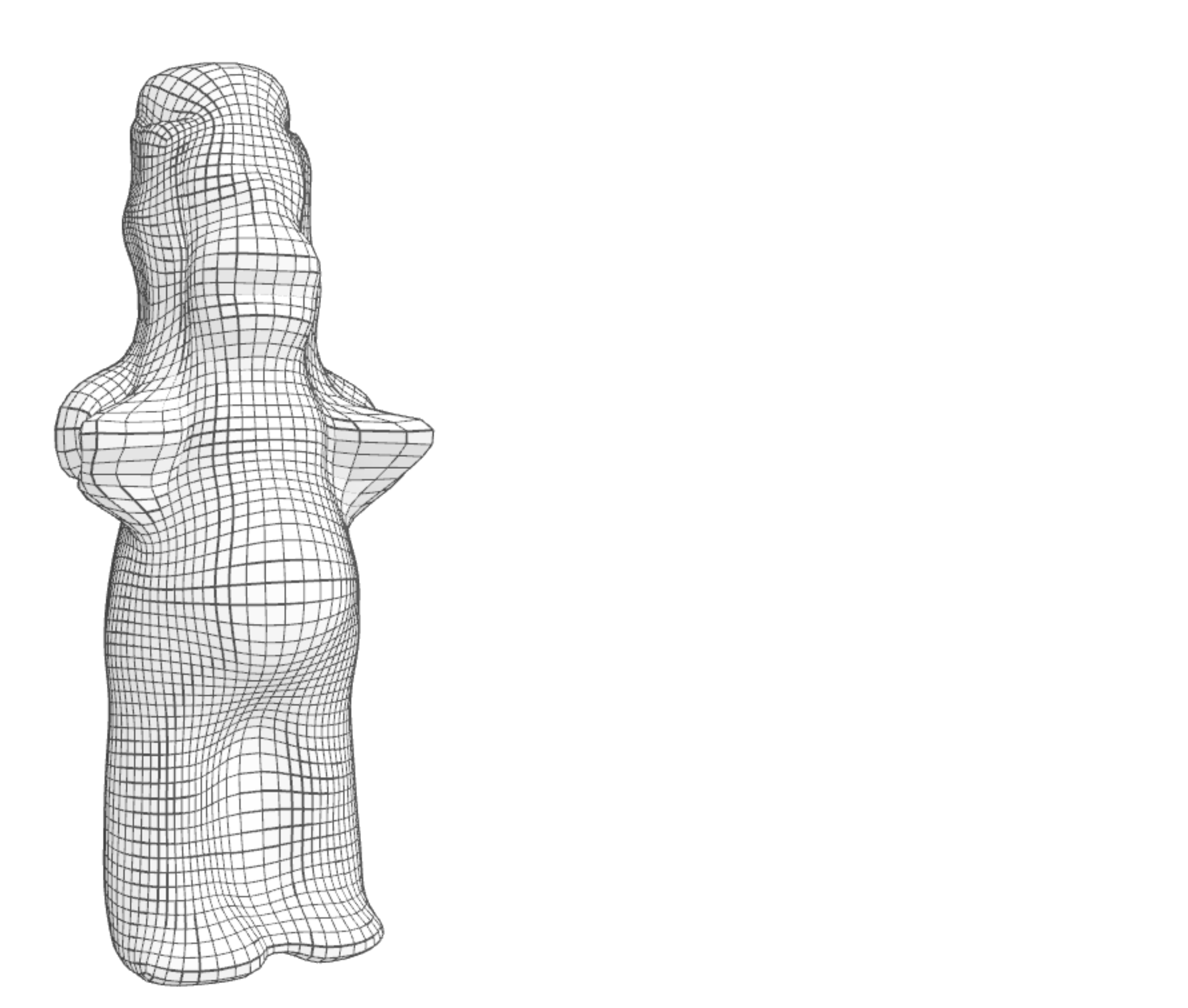} &
    \includegraphics[height=1.2in,trim=6.6in 1.in 0 1in, clip]{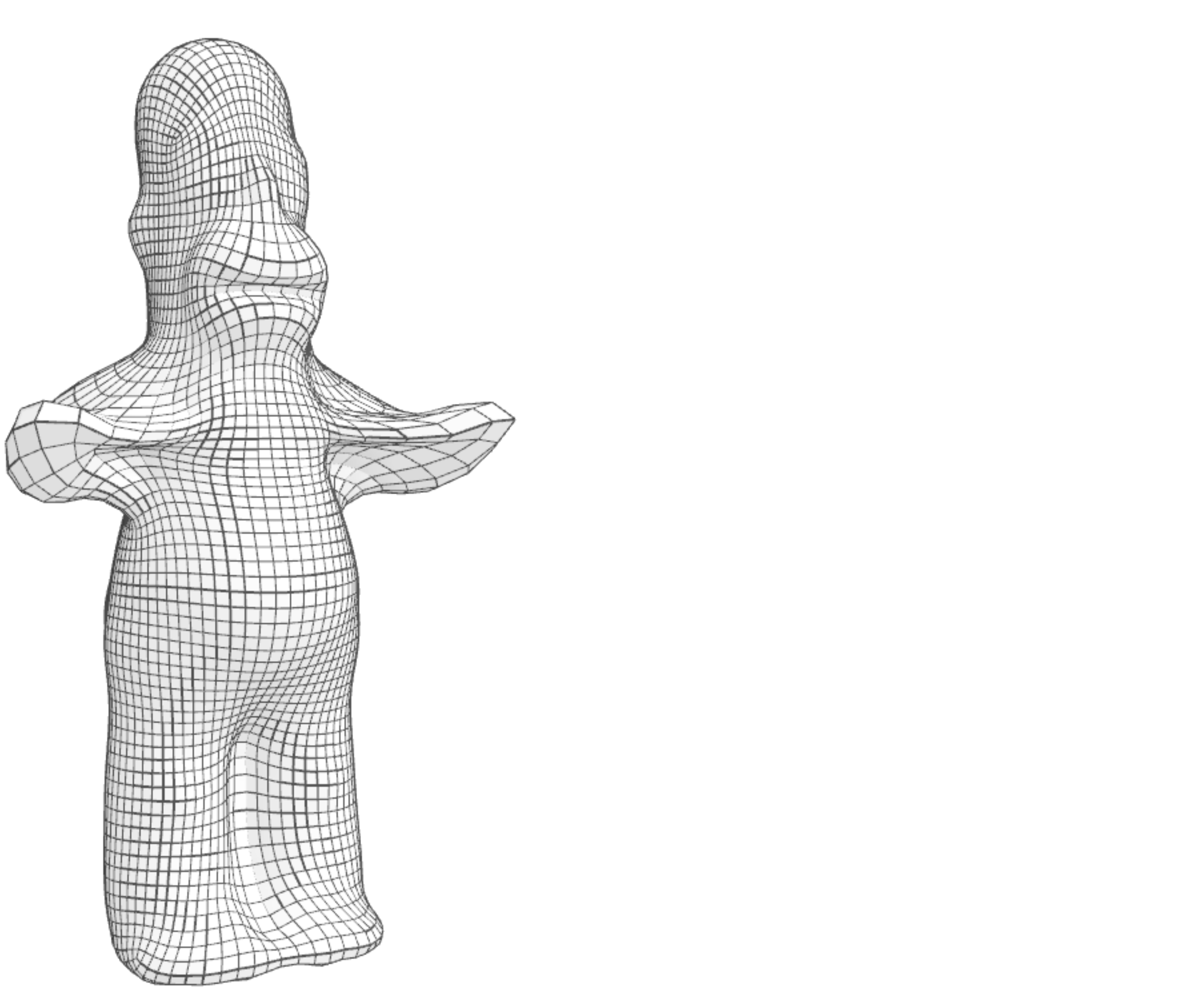} &
    \includegraphics[height=1.2in,trim=6.6in 1.in 0 1in, clip]{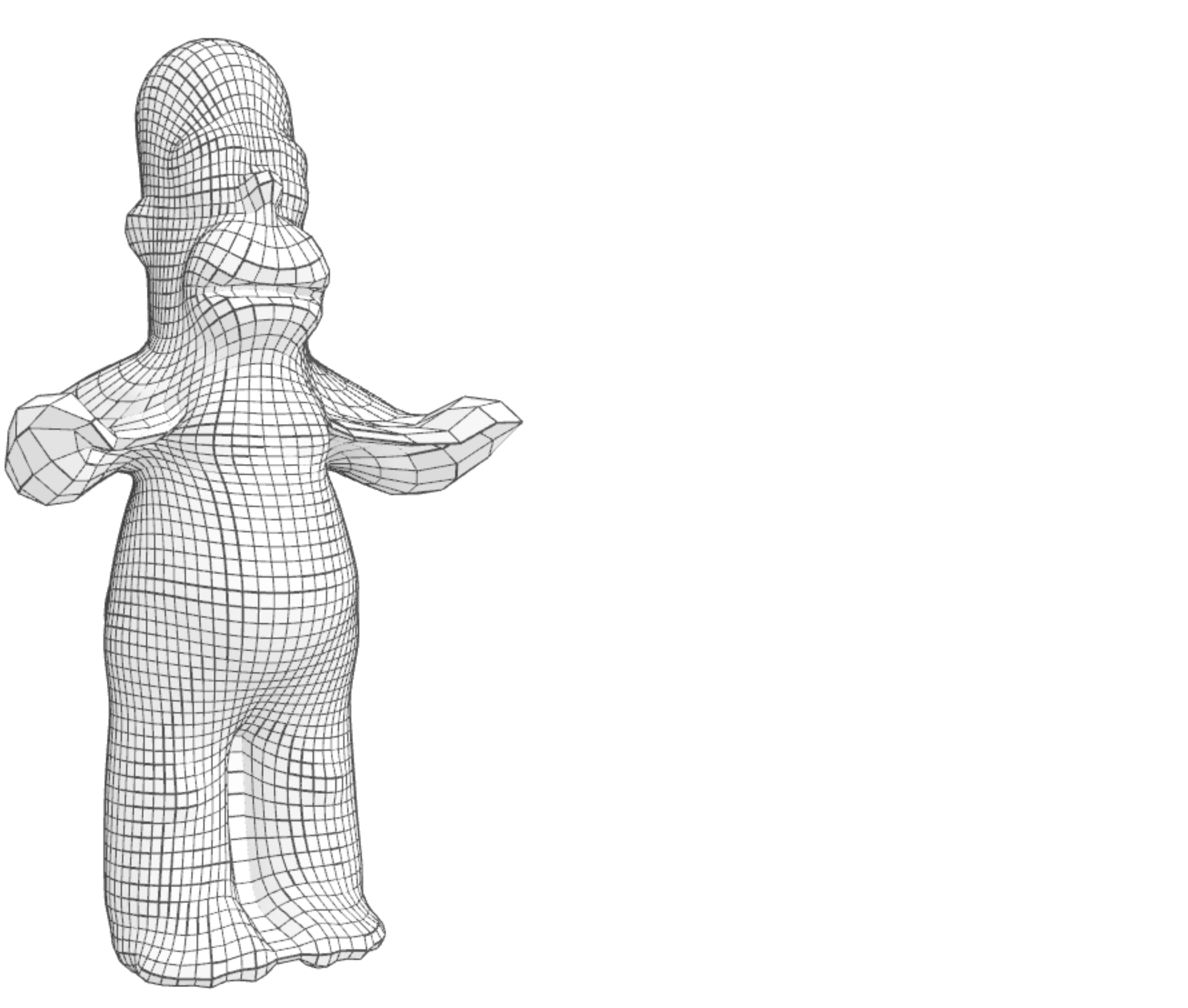}  \\
    \hspace*{-0.1in}
    \includegraphics[height=1.0in,trim=5.2in 1.in 0 1in, clip]{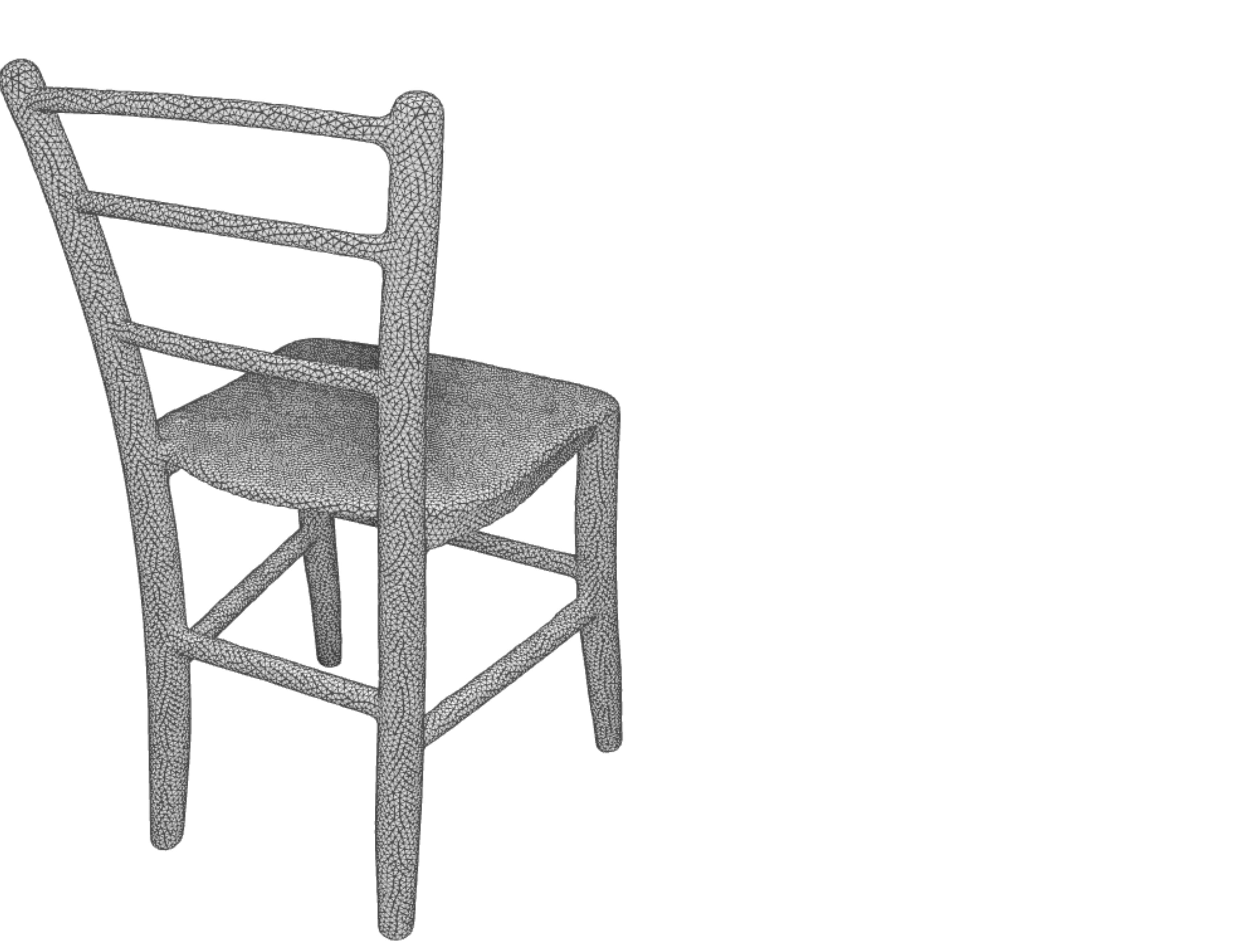} &
    \hspace*{-0.1in}
    \input{annotated/chair_discretized.tex} &
    \hspace*{-0.1in}
    \includegraphics[height=1.0in,trim=5.2in 1.in 0 1in, clip]{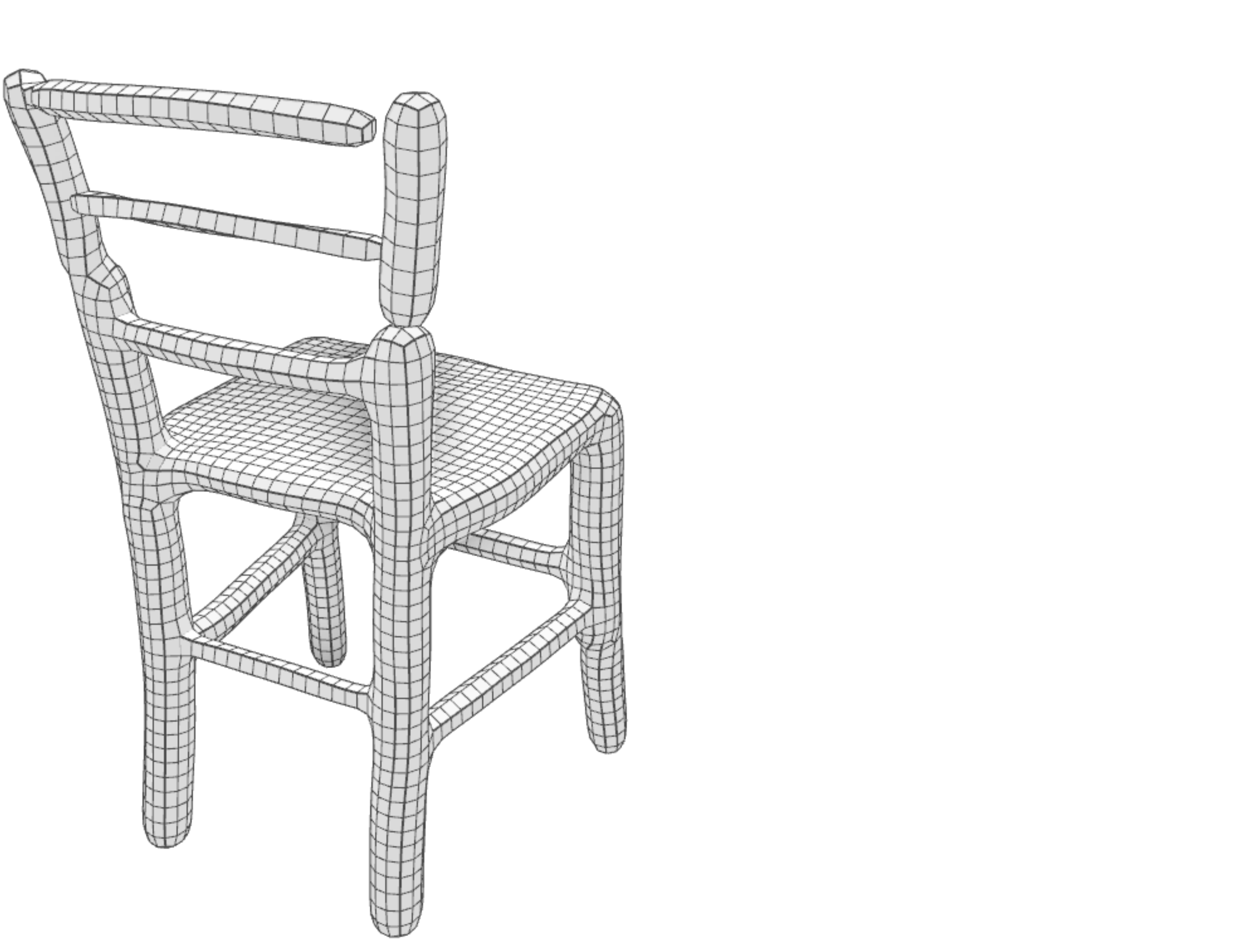} &
    \hspace*{-0.1in}
    \includegraphics[height=1.0in,trim=5.2in 1.in 0 1in, clip]{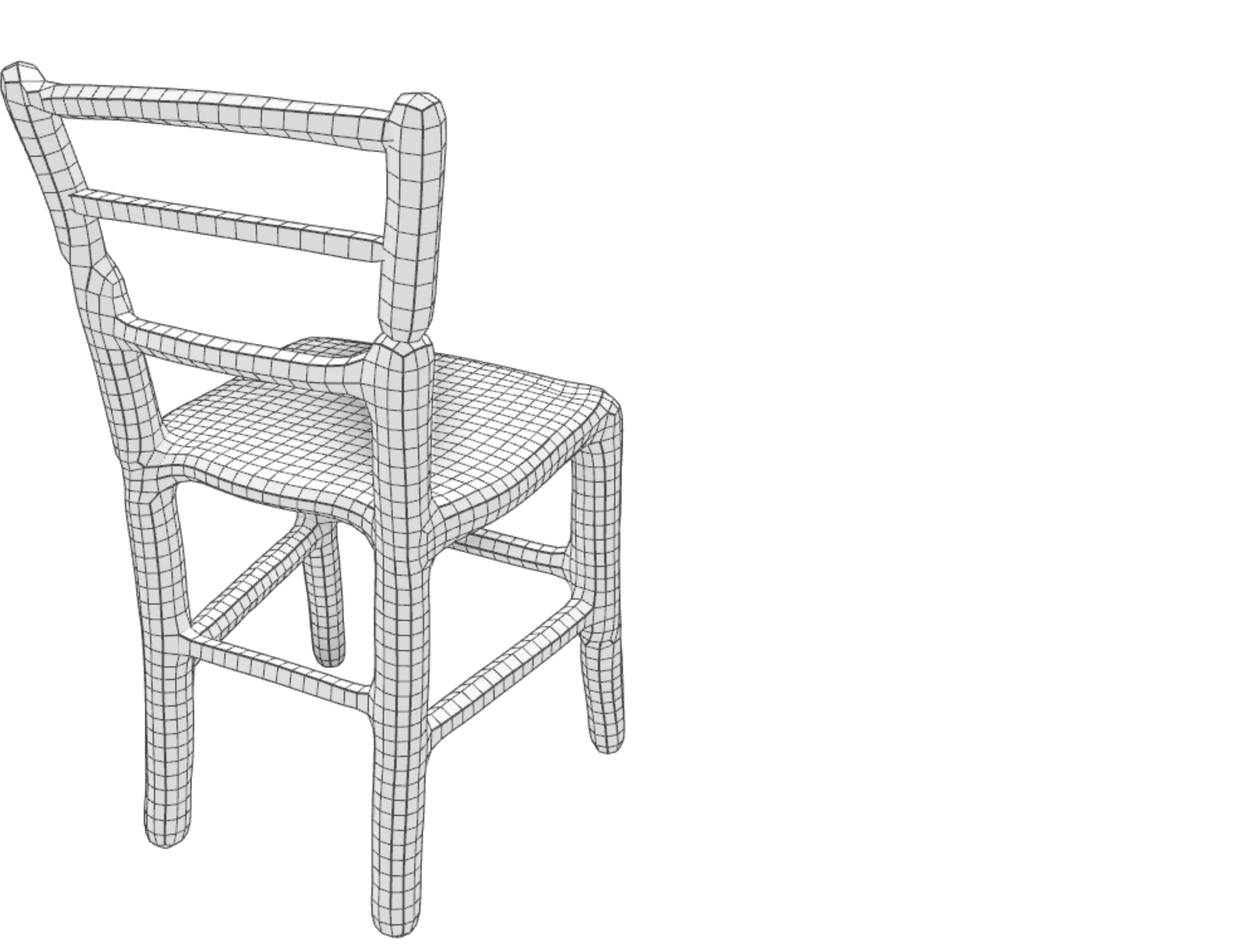} &
    \hspace*{-0.1in}
    \includegraphics[height=1.0in,trim=5.2in 1.in 0 1in, clip]{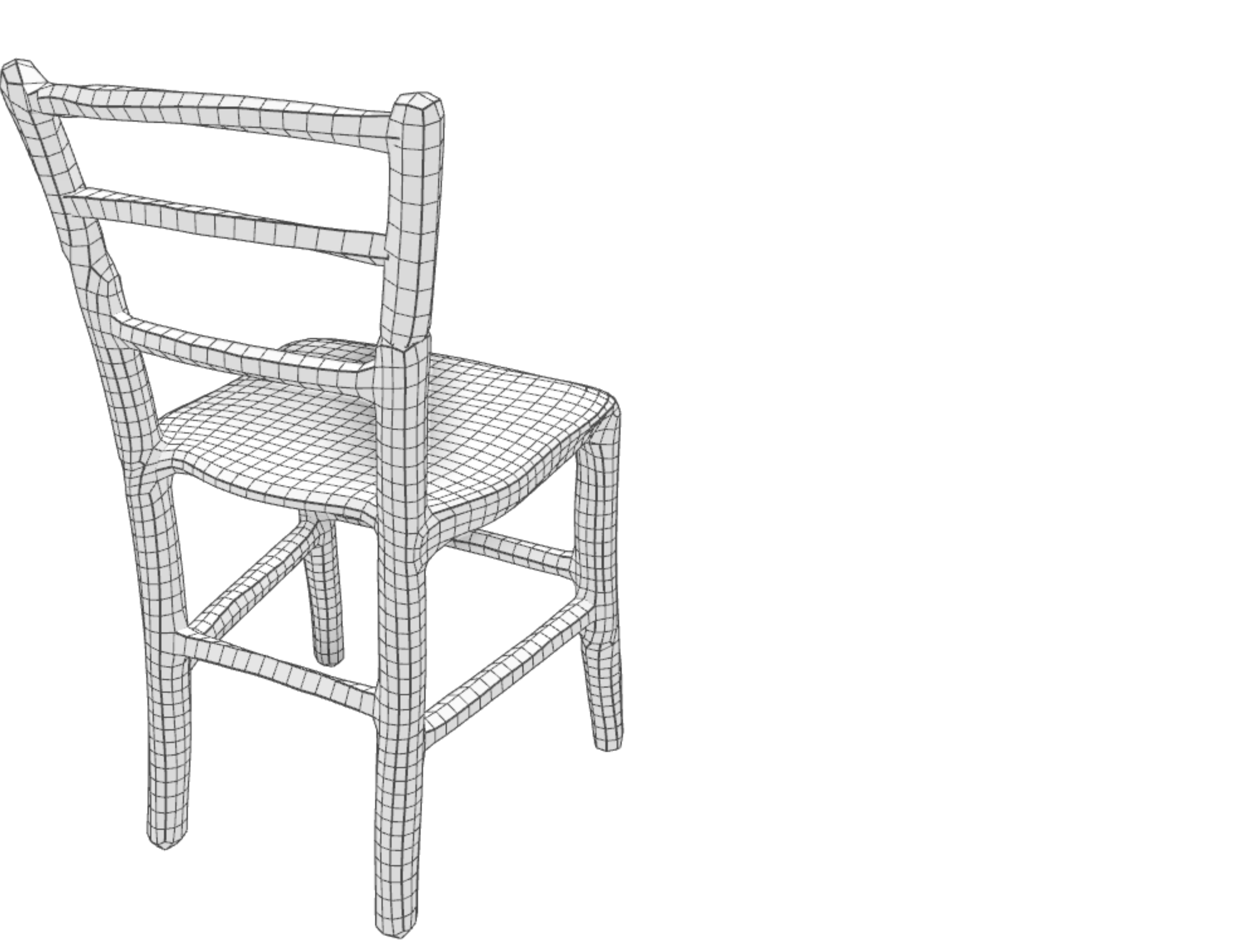}  \\
    {\footnotesize (a) Input} & {\footnotesize (b) PolyCube}  & \multicolumn{3}{c}{\footnotesize (c) Pullback and mesh quality optimization}
  \end{tabular}
  \caption{Robustness of our inversion-free pullback and mesh quality optimization.
    The three columns in (c) show the result after the pullback followed by two more steps of mesh quality optimization.
    For the \textit{homer} model (top row), we intentionally use an extremely crude PolyCube consisting of a single cuboid.
    The hexahedralization stage is capable of recovering a hex mesh that is similar to the input (except it fails to recover the highly non-convex gap between the legs).
    For the \textit{chair} model (bottom row), the PolyCube has isolated cuboids that are disconnected from the rest (circled red).
    As a result, the generated hex mesh also has isolated components and may result in intersection. 
    Note this is not the intended usage of the pipeline---the user should fix the connectivity during either the decomposition stage or the discretization stage.
    For both models, no inverted hex occurs throughout the process.
  }
  \label{fig:no_inv_stress_test}
\end{figure}

\paragraph{Inversion-free pullback stress test}
We test the robustness of our hexahedralization stage (\cref{stage:hexahedralization}) when the voxelized PolyCube from the previous stage is too coarse or has topological deficiencies (\cref{fig:no_inv_stress_test}).
In both cases, our method is capable of pulling back the voxelized PolyCube in an inversion-free manner.

\begin{figure*}
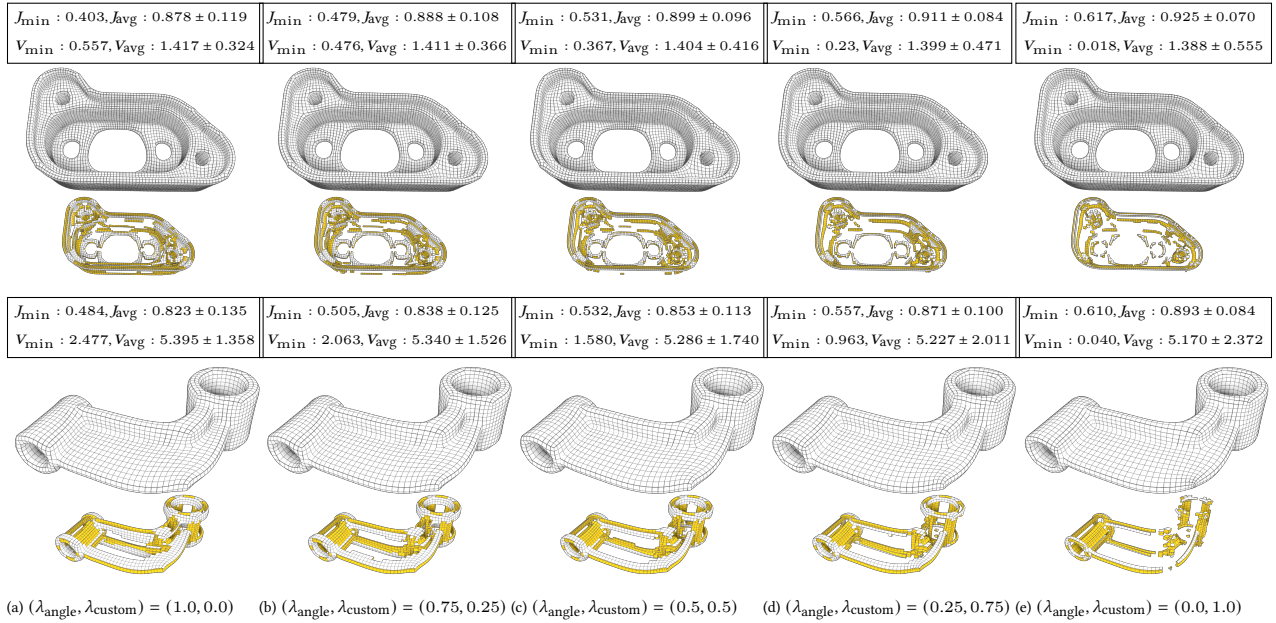

  \newcolumntype{C}[1]{>{\centering\arraybackslash}m{#1}}
  \begin{tabular}{C{3.0cm} C{3.0cm} C{3.0cm} C{3.0cm} C{3.0cm}}
    \input{annotated/lock_good_aut.tex} &
    \input{annotated/lock_ok_aut_meh_sj.tex} &
    \input{annotated/lock_mid_aut_sj.tex} &
    \input{annotated/lock_meh_aut_ok_sj.tex} &
    \input{annotated/lock_good_sj.tex} \\
    \input{annotated/hanger_good_aut.tex} &
    \input{annotated/hanger_ok_aut_meh_sj.tex} &
    \input{annotated/hanger_mid_aut_sj.tex} &
    \input{annotated/hanger_meh_aut_ok_sj.tex} &
    \input{annotated/hanger_good_sj.tex} \\
    \mbox{\scriptsize(a) $(\lambda_{\text{angle}},\lambda_\text{custom}) = (1.0, 0.0)$} & 
    \mbox{\scriptsize(b) $(\lambda_{\text{angle}},\lambda_\text{custom}) =(0.75, 0.25)$}  &
    \mbox{\scriptsize(c) $(\lambda_{\text{angle}},\lambda_\text{custom}) =(0.5, 0.5)$} & 
    \mbox{\scriptsize(d) $(\lambda_{\text{angle}},\lambda_\text{custom}) =(0.25, 0.75)$} & 
    \mbox{\scriptsize(e) $(\lambda_{\text{angle}},\lambda_\text{custom}) =(0.0, 1.0)$ }
  \end{tabular}
  \caption{
    Trade-off between angle preservation (large \textit{conformal} parameter $\lambda_{\text{angle}}$) and scaled Jacobian values (large \textit{custom} parameter $\lambda_{\text{custom}}$ with the scaled Jacobian energy).
    For each hex mesh we show the minimum scaled Jacobian $J_{\min}$, the average scaled Jacobian $J_{\avg}$, the minimum (unscaled) Jacobian $V_{\min}$, and the average (unscaled) Jacobian $V_\avg$. The standard deviation across hexes is shown after the $\pm$ sign. Both $V_{\min}$ and $V_\avg$ are multiplied by $10^5$.
    The bottom image of each row shows the hex elements with scaled Jacobian less than $0.8$. Yellow indicates an interior quad.
    As we gradually decrease the $\lambda_{\text{angle}}$ and increase $\lambda_{\text{custom}}$, we see that $J_{\min}$ and $J_\avg$ increase while $V_{\min}$ decreases, and the standard deviation of the (unscaled) Jacobian grows, indicating the hex mesh has more unevenly sized elements.
    The user can conduct such trade-off exploration interactively to decide whether they want more volume preservation or better scaled Jacobian values.
    To produce each mesh, we start from the same initialized inversion-free hex mesh and use default weights for all parameters except for $\lambda_{\text{angle}}$ and $\lambda_{\text{custom}}$.
    We then run the optimization for 1000 steps.
    We use \cref{eqn:iso_lse} to improve the worst element for both the distortion and the custom energy.
  }
  \label{fig:trade_off_aut_sj}
\end{figure*}

\begin{figure*}
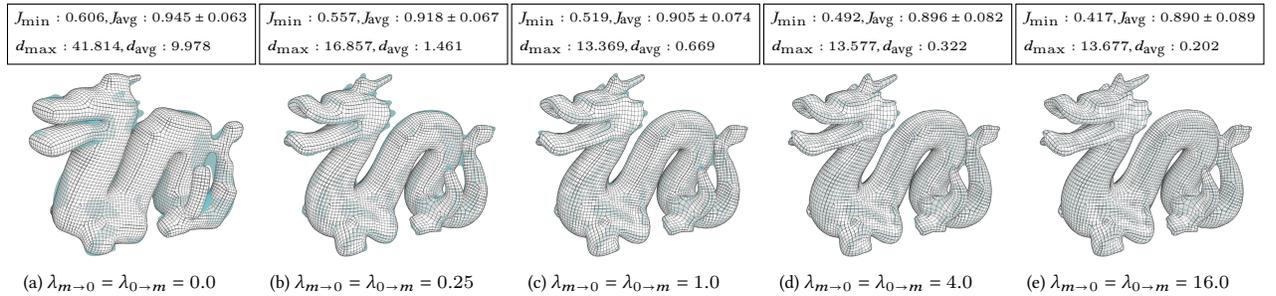

  \newcolumntype{C}[1]{>{\centering\arraybackslash}m{#1}}
  \begin{tabular}{C{3.0cm} C{3.0cm} C{3.0cm} C{3.0cm} C{3.0cm}}
    \input{annotated/dragon_0.tex} &
    \input{annotated/dragon_05.tex} &
    \input{annotated/dragon_2.tex} &
    \input{annotated/dragon_4.tex} &
    \input{annotated/dragon_6.tex} 
    \\
    {\footnotesize (a) $\lambda_{m\sto 0}=\lambda_{0\sto m}=0.0$} & 
    {\footnotesize (b) $\lambda_{m\sto 0}=\lambda_{0\sto m}=0.25$} &
    {\footnotesize (c) $\lambda_{m\sto 0}=\lambda_{0\sto m}=1.0$} &
    {\footnotesize (d) $\lambda_{m\sto 0}=\lambda_{0\sto m}=4.0$} &
    {\footnotesize (e) $\lambda_{m\sto 0}=\lambda_{0\sto m}=16.0$} 
  \end{tabular}
  \caption{
    Trade-off between hex element quality and the approximation level of the input surface (controlled by the \textit{projection} parameter $\lambda_{m\sto 0}$ and the \textit{details} parameter $\lambda_{0\sto m}$).
    We add a blue translucent shell of the input mesh surface to visualize the discrepancy.
    For each hex mesh we show the minimum scaled Jacobian $J_{\min}$, the average scaled Jacobian $J_{\avg}$, the symmetric Hausdorff distance $d_{\max}$, and the average Hausdorff distance $d_{\min}$ (computed similarly to the ones in \cref{tbl:cmp}). 
    The standard deviation across hex elements is shown after the $\pm$ sign. 
    As we increase $\lambda_{m\sto 0}$ and $\lambda_{0\sto m}$, the average Hausdorff distance becomes smaller but at the cost of worst hex quality (i.e., reduced $J_{\min}$ and $J_{\avg}$).
    The experiment setup is the same as that of \cref{fig:trade_off_aut_sj}, except the parameters being changed are $\lambda_{m\sto 0}$ and $\lambda_{0\sto m}$.
  }
  \label{fig:trade_off_detail}
\end{figure*}

\paragraph{Trade-off exploration}
Our mesh quality optimization component of the hexahedralization stage allows the user to explore and choose their preferred trade-off.
In \cref{fig:trade_off_aut_sj}, we demonstrate the exploration of the trade-off between angle preservation and scaled Jacobian values by simply changing the relative weights before optimizing.
In \cref{fig:trade_off_detail}, we show the exploration of the trade-off between hex element quality and approximation level of the input surface.
Since trade-off exploration is incorporated seamlessly within the system, the user can easily try out any combination of the parameters from \cref{tbl:hexahedralization_parameters}, change the surface parameterization mode, or put landmarks to alter surface vertex positions manually (\cref{fig:hexahedralization_untangle_fingers}) before reoptimization.

\begin{figure}
  \newcolumntype{C}[1]{>{\centering\arraybackslash}m{#1}}
  \begin{tabular}{C{0.8cm} C{1.4cm} C{1.4cm} C{1.4cm} C{1.4cm}} 
      \includegraphics[height=0.7in,trim=6.0in 2.in 0 3in, clip]{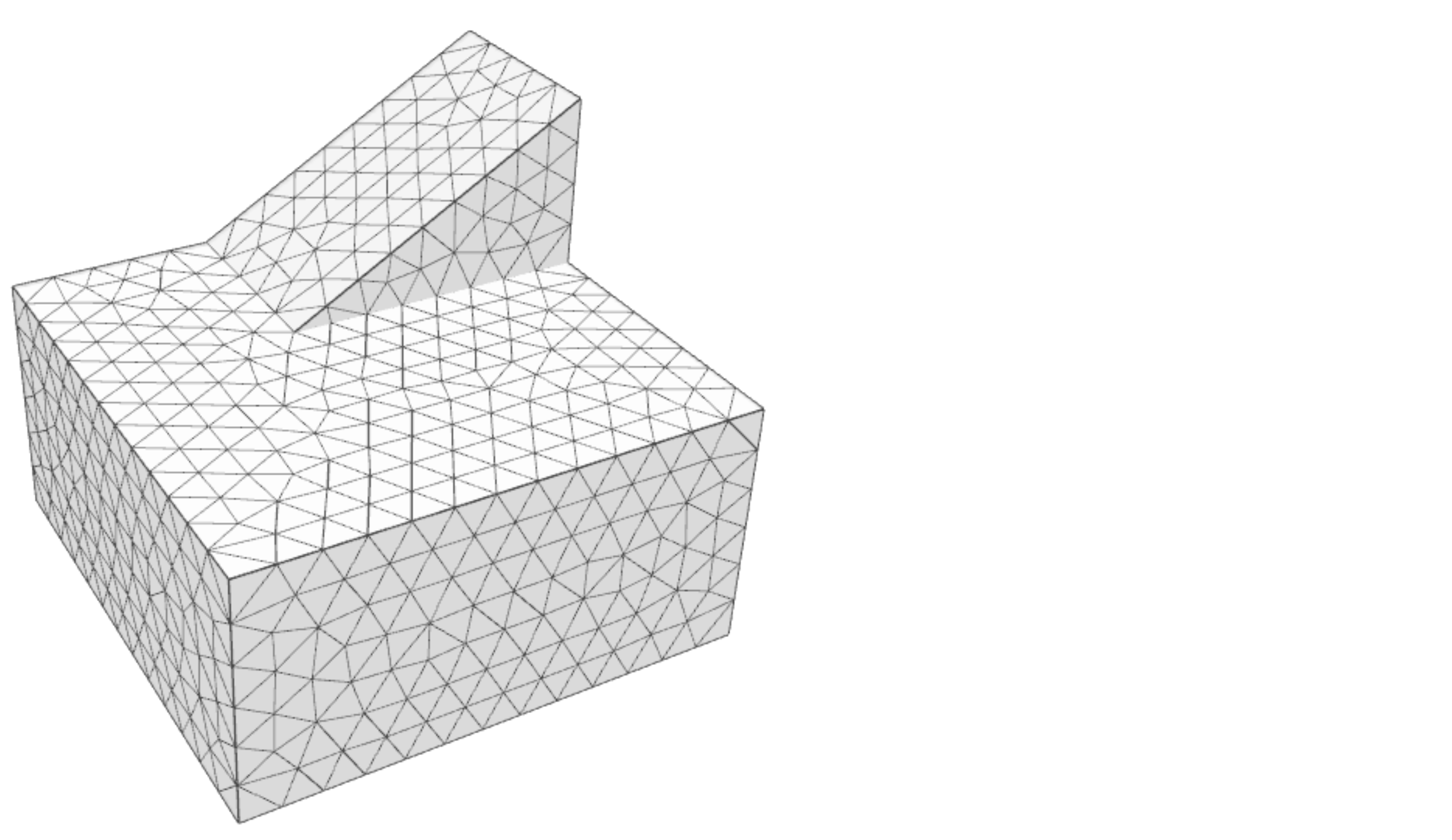} 
      &
      \includegraphics[height=0.7in,trim=6.0in 2.in 0 3in, clip]{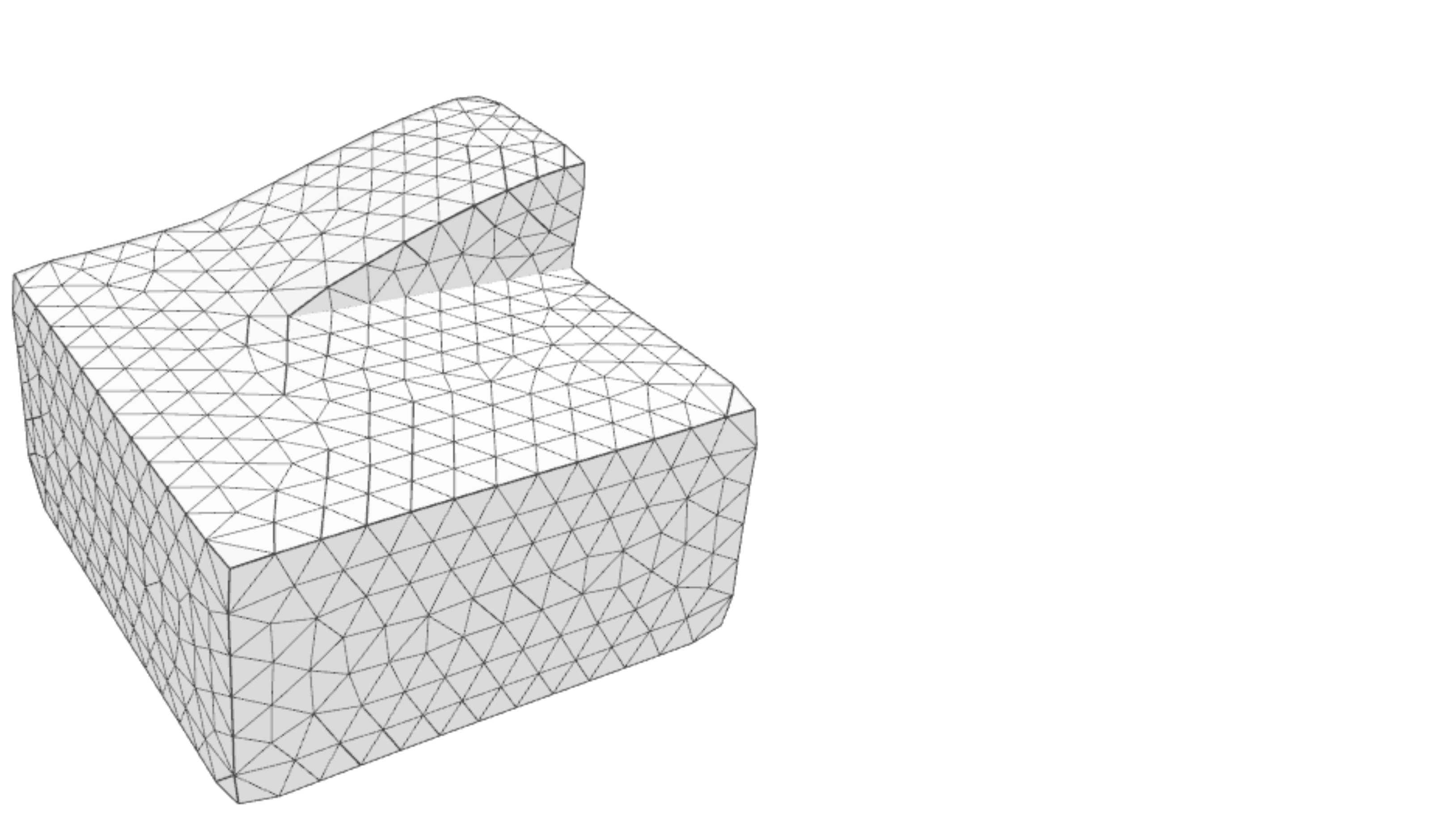} 
                                                                                    &
       \includegraphics[height=0.7in,trim=6.0in 2.in 0 3in, clip]{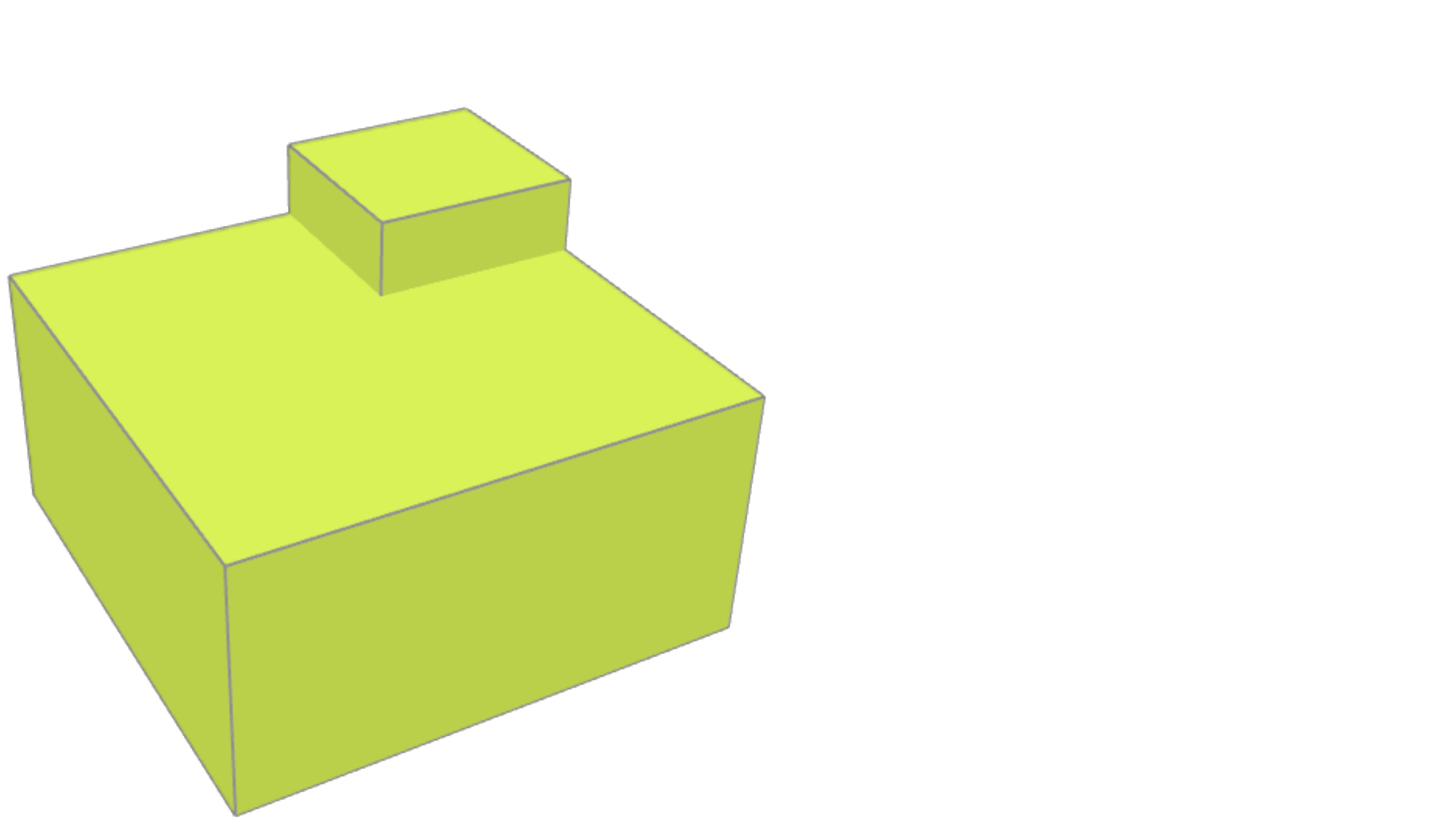} 
                                                                                       &
      \includegraphics[height=0.7in,trim=6.0in 2.in 0 3in, clip]{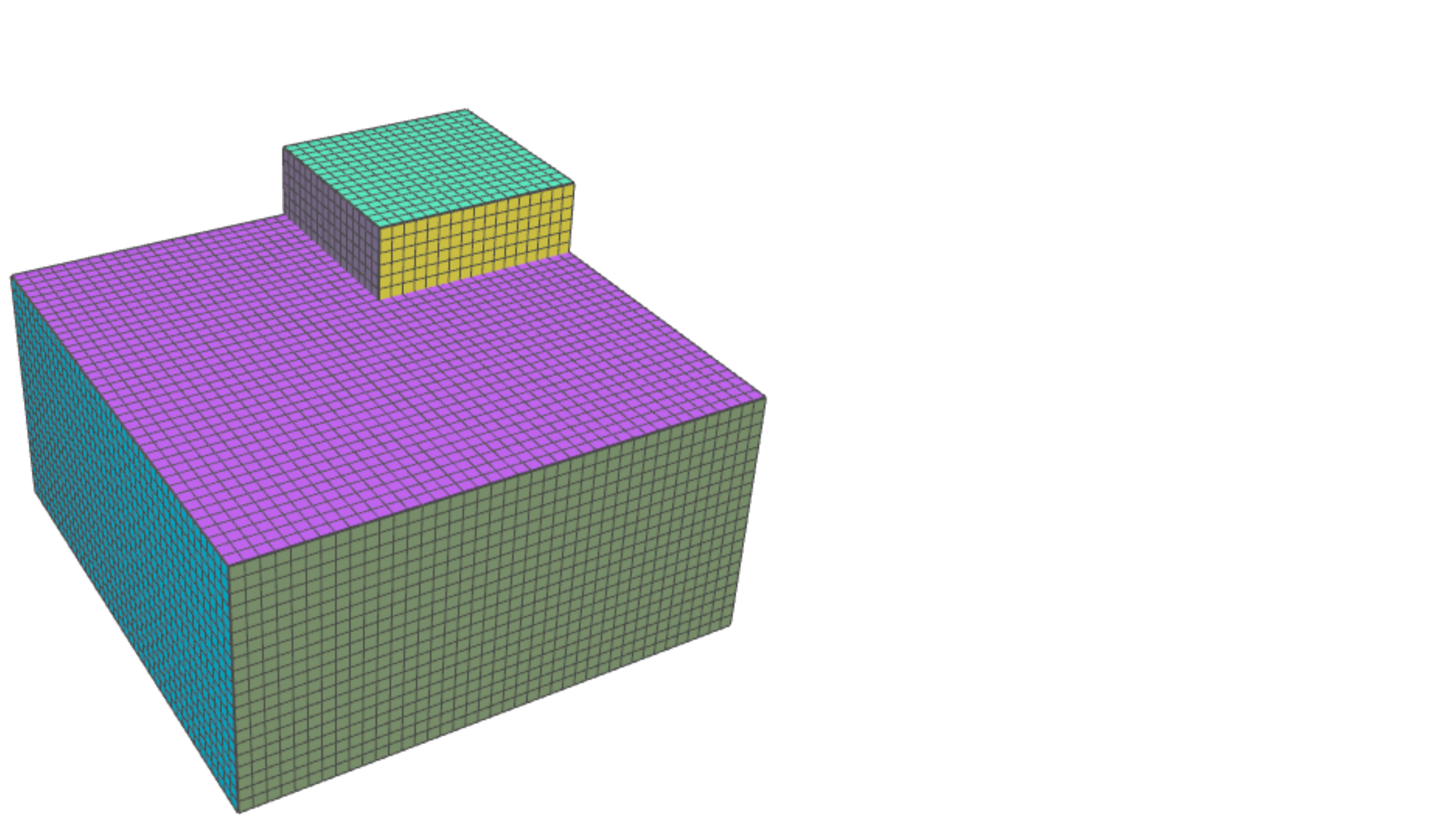} &
      \includegraphics[height=0.7in,trim=6.0in 2.in 0 3in, clip]{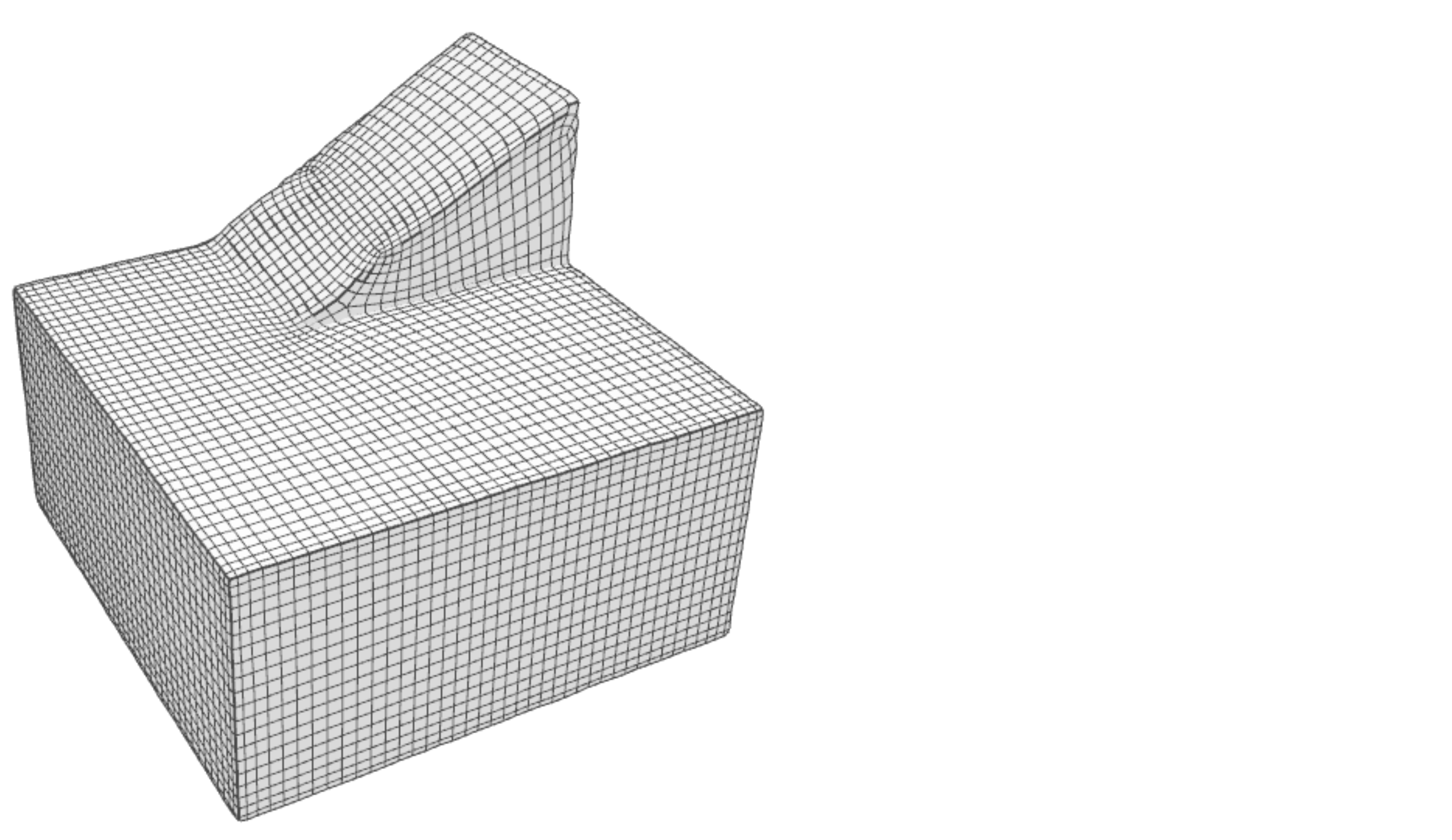}  \\
      \includegraphics[height=0.7in,trim=6.0in 2.in 0 3in, clip]{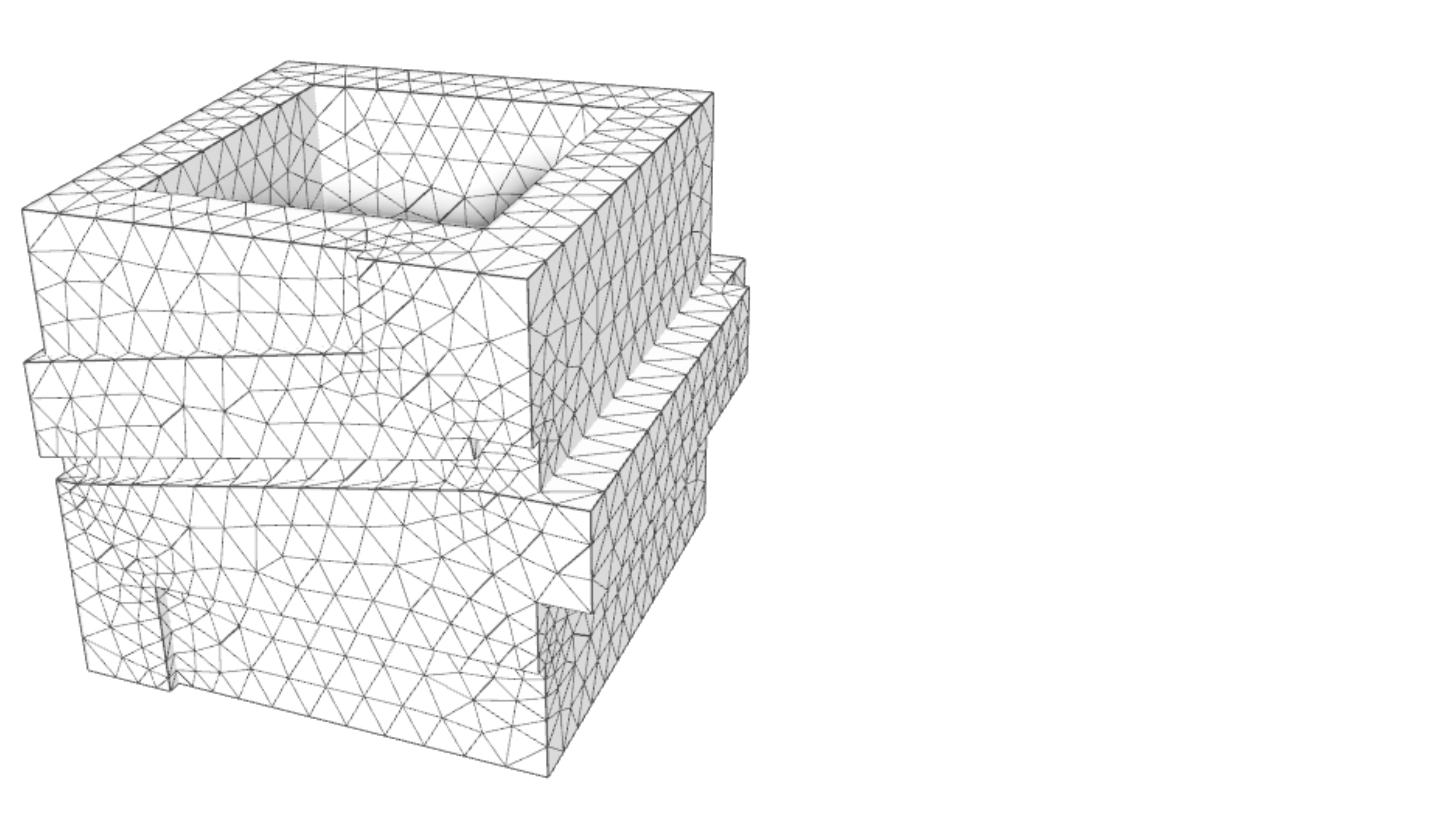} 
                                                                                          &
      \includegraphics[height=0.7in,trim=6.0in 2.in 0 3in, clip]{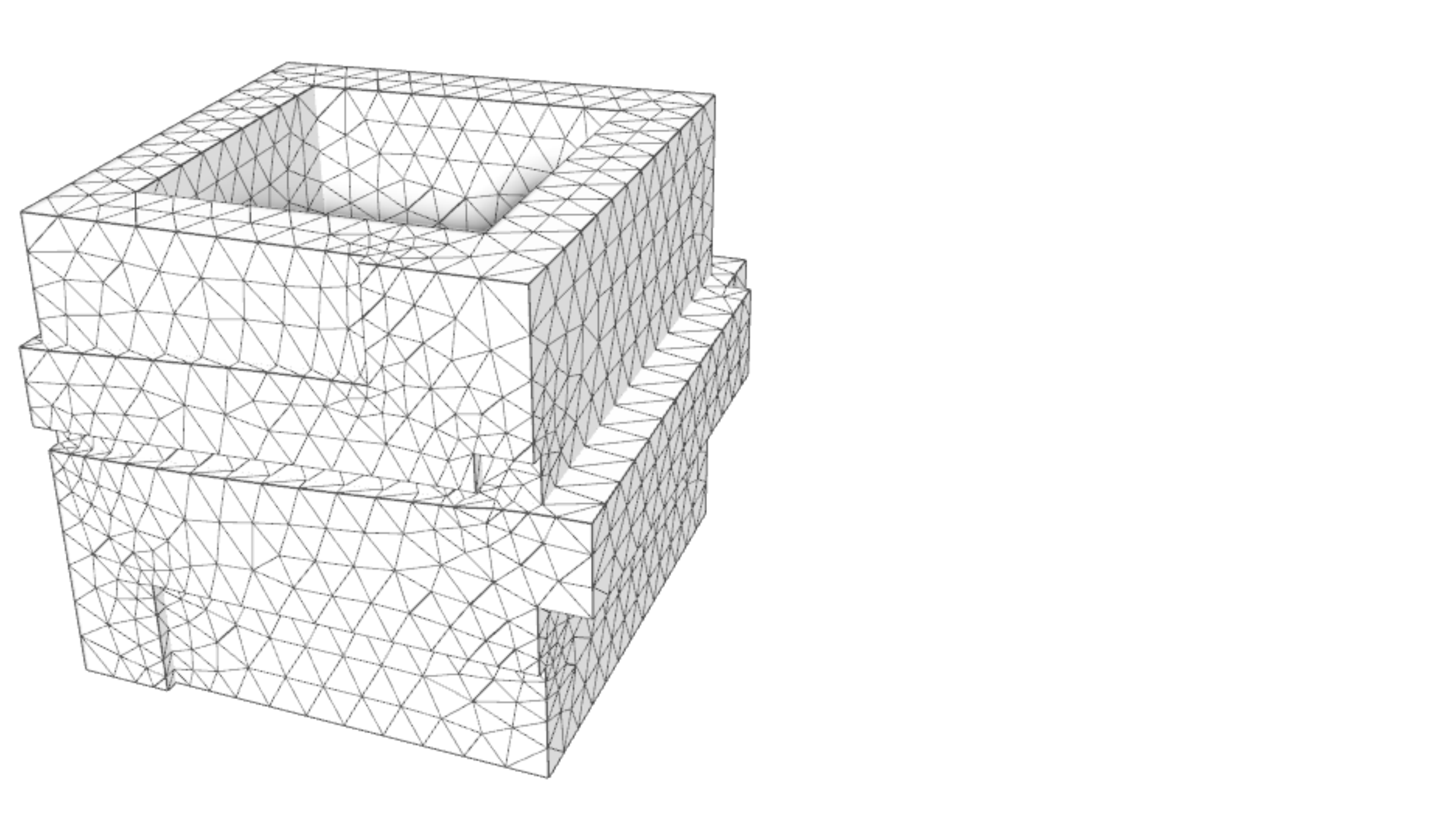} 
                                                                                    &
      \includegraphics[height=0.7in,trim=6.0in 2.in 0 3in, clip]{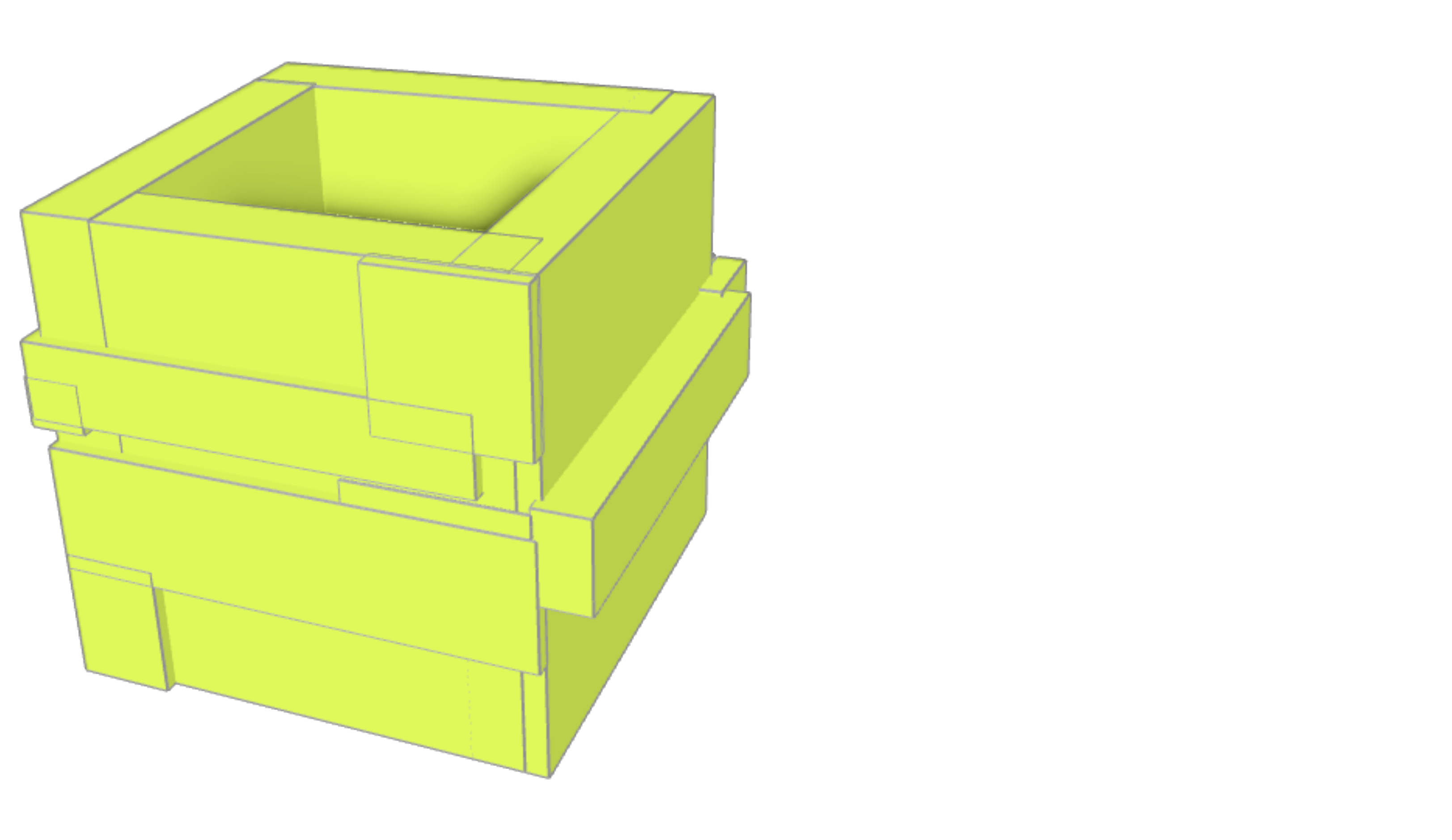} 
                                                                                    &
      \includegraphics[height=0.7in,trim=6.0in 2.in 0 3in, clip]{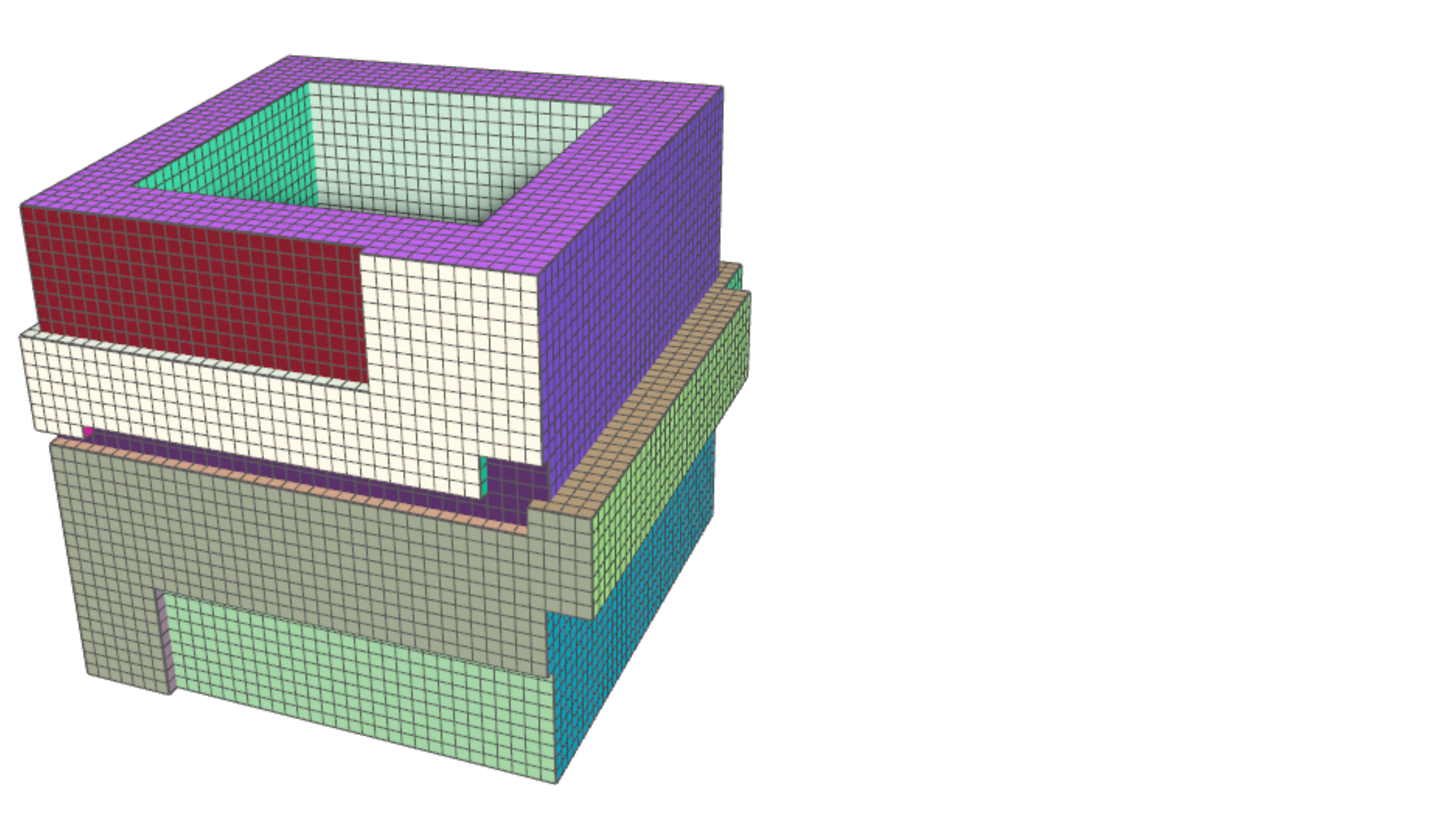} &
      \includegraphics[height=0.7in,trim=6.0in 2.in 0 3in, clip]{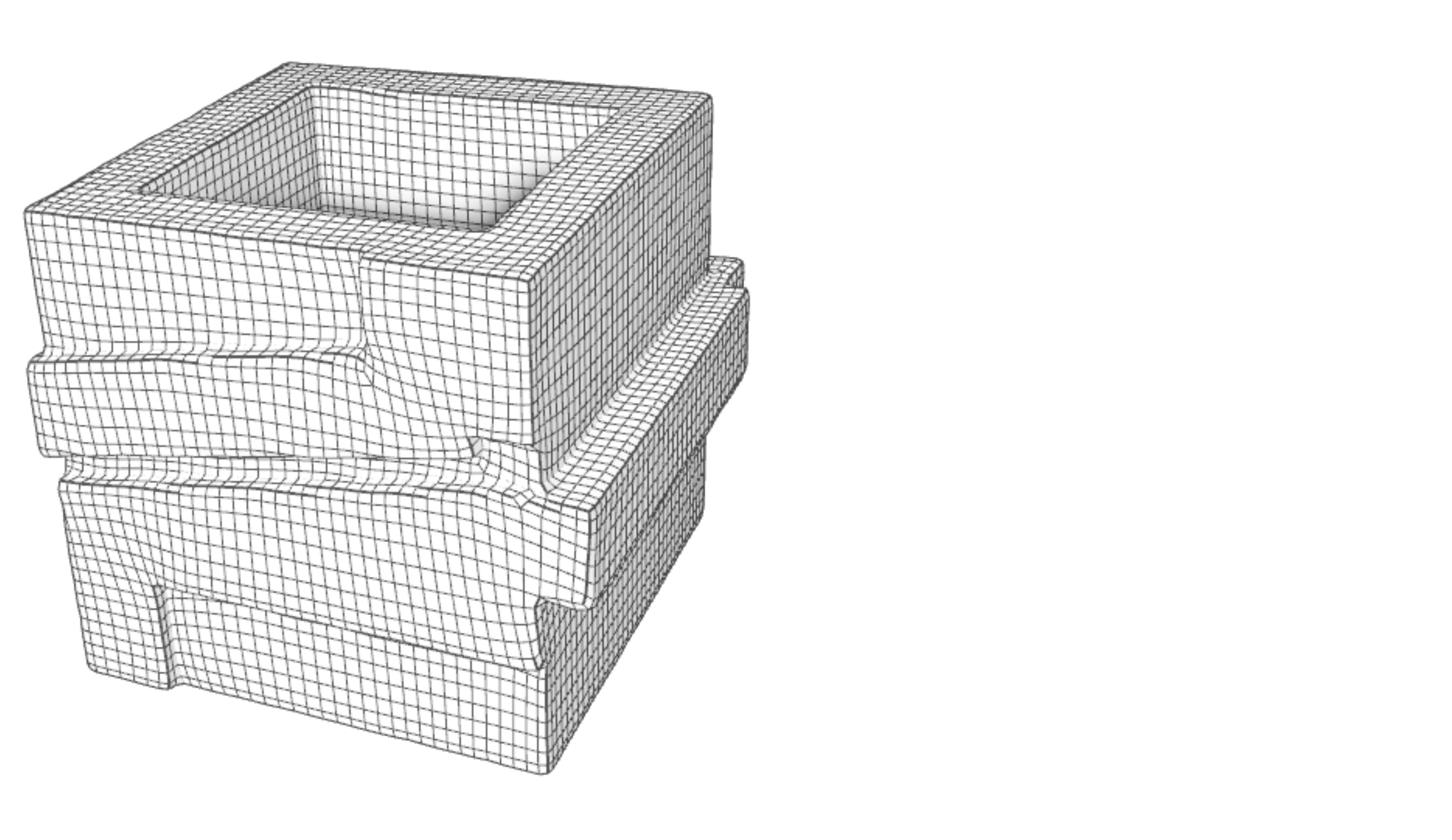}  \\
      \includegraphics[height=0.7in,trim=5.0in 1.in 0 1in, clip]{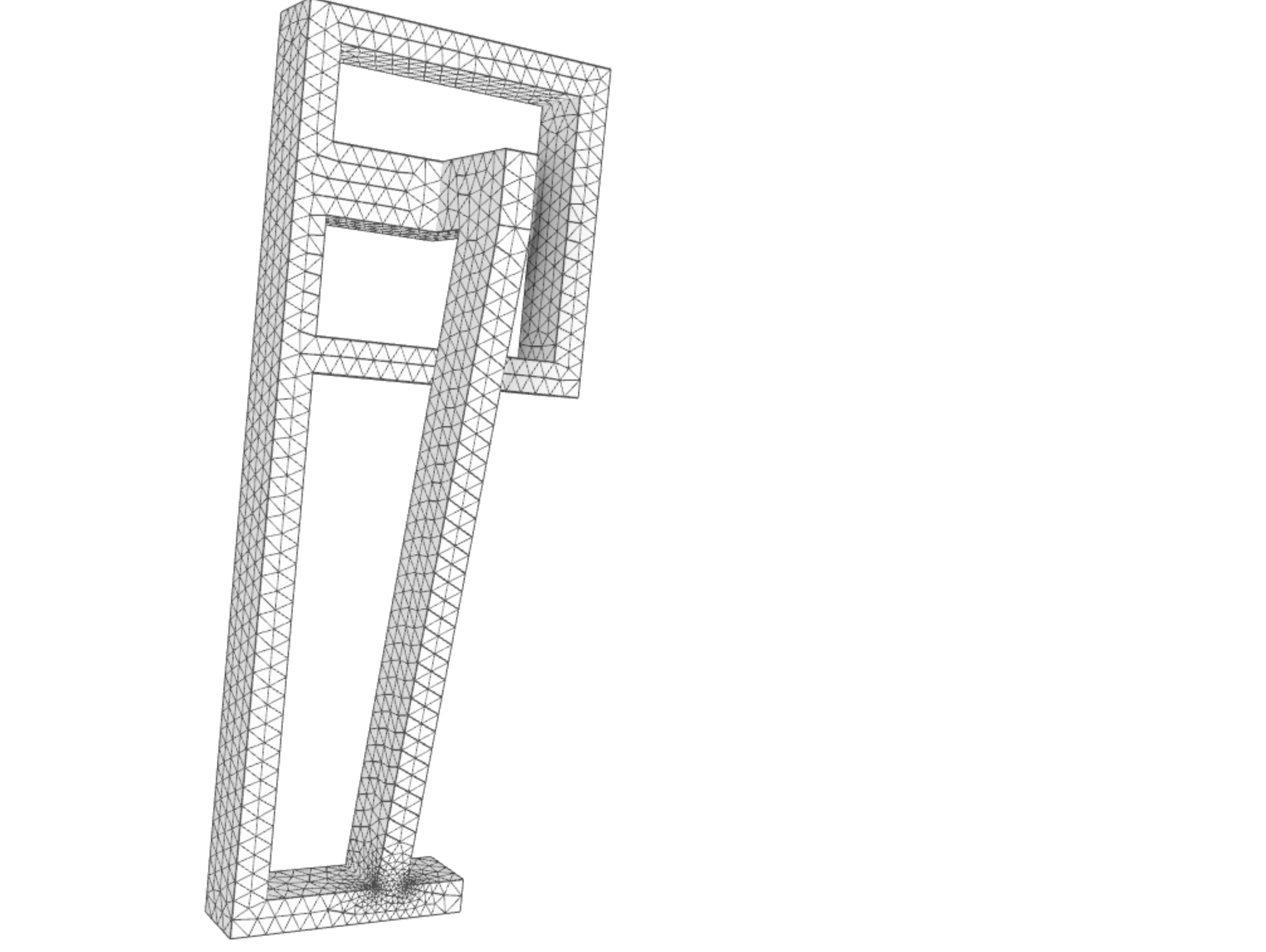} &
      \includegraphics[height=0.7in,trim=5.0in 1.in 0 1in, clip]{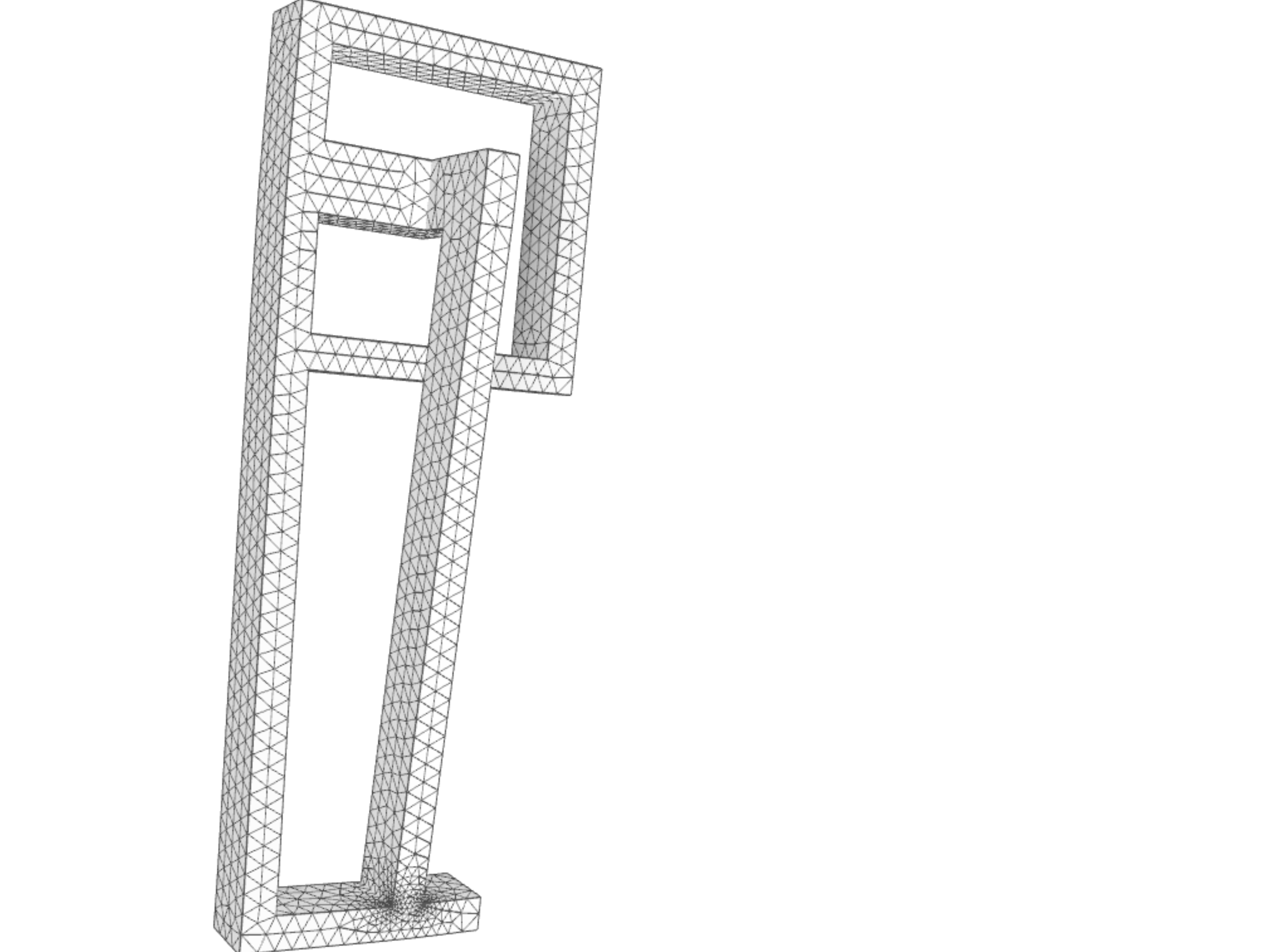} &
      \includegraphics[height=0.7in,trim=5.0in 1.in 0 1in, clip]{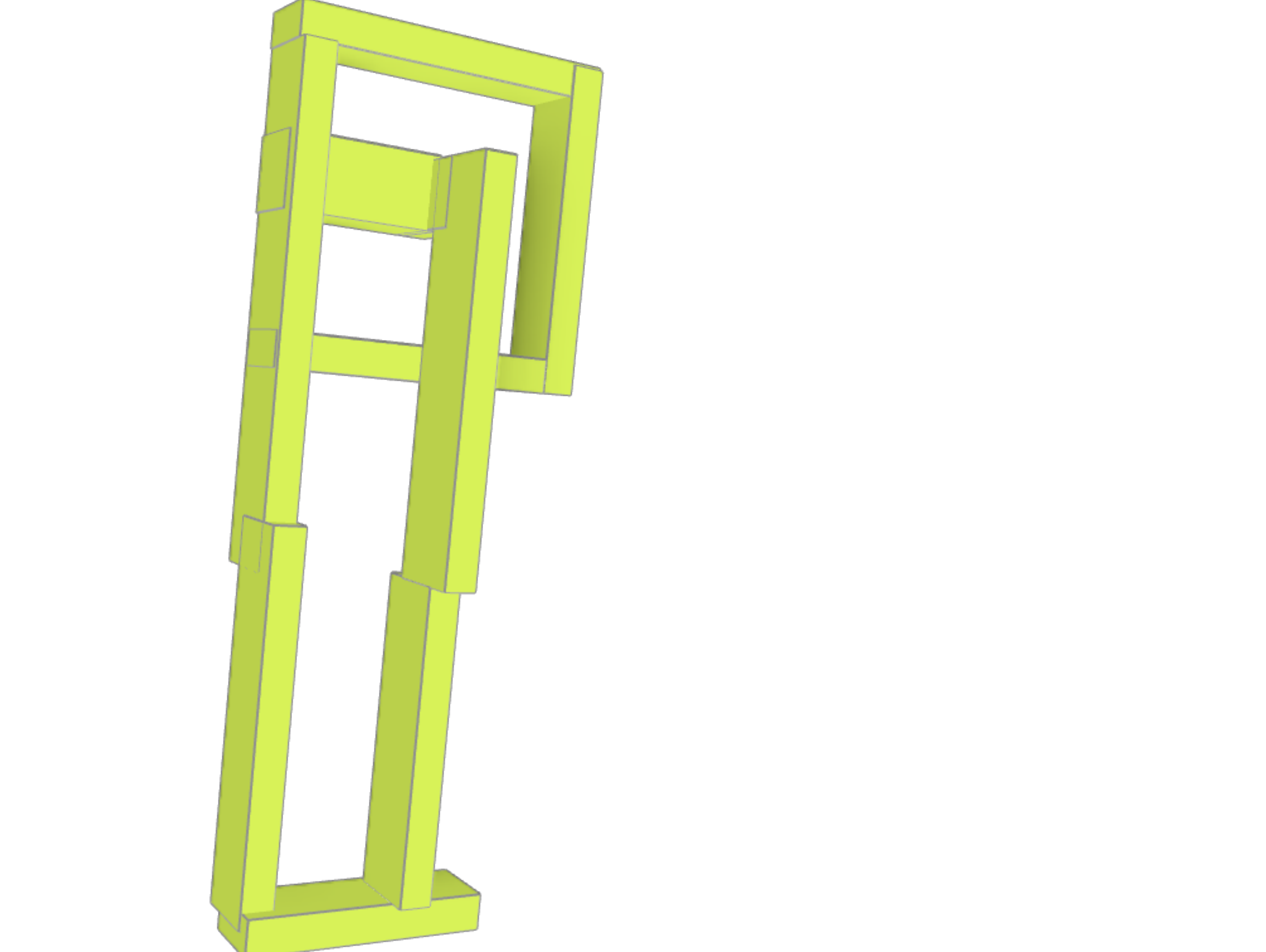} &
      \includegraphics[height=0.7in,trim=5.0in 1.in 0 1in, clip]{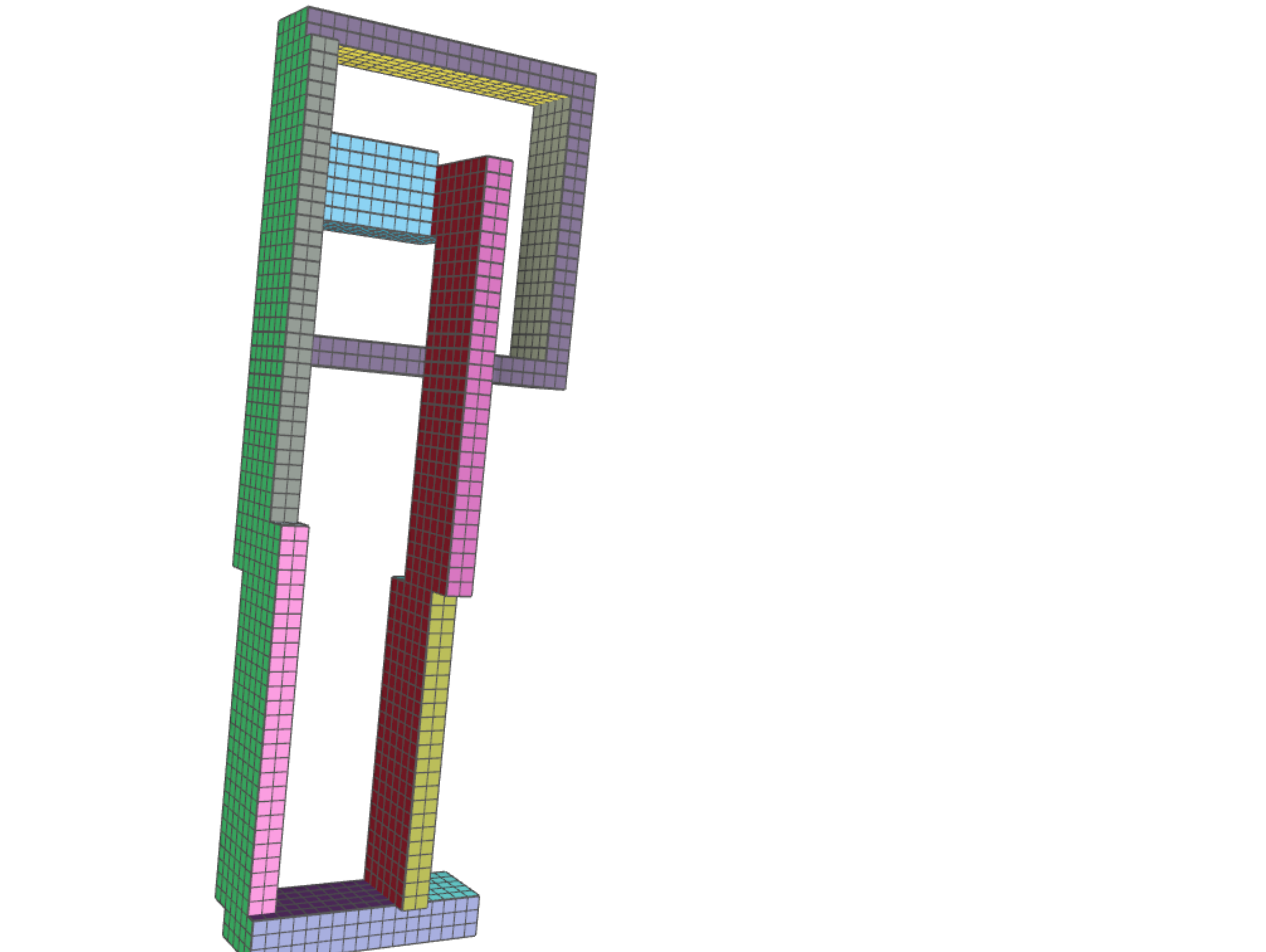} &
      \includegraphics[height=0.7in,trim=5.0in 1.in 0 1in, clip]{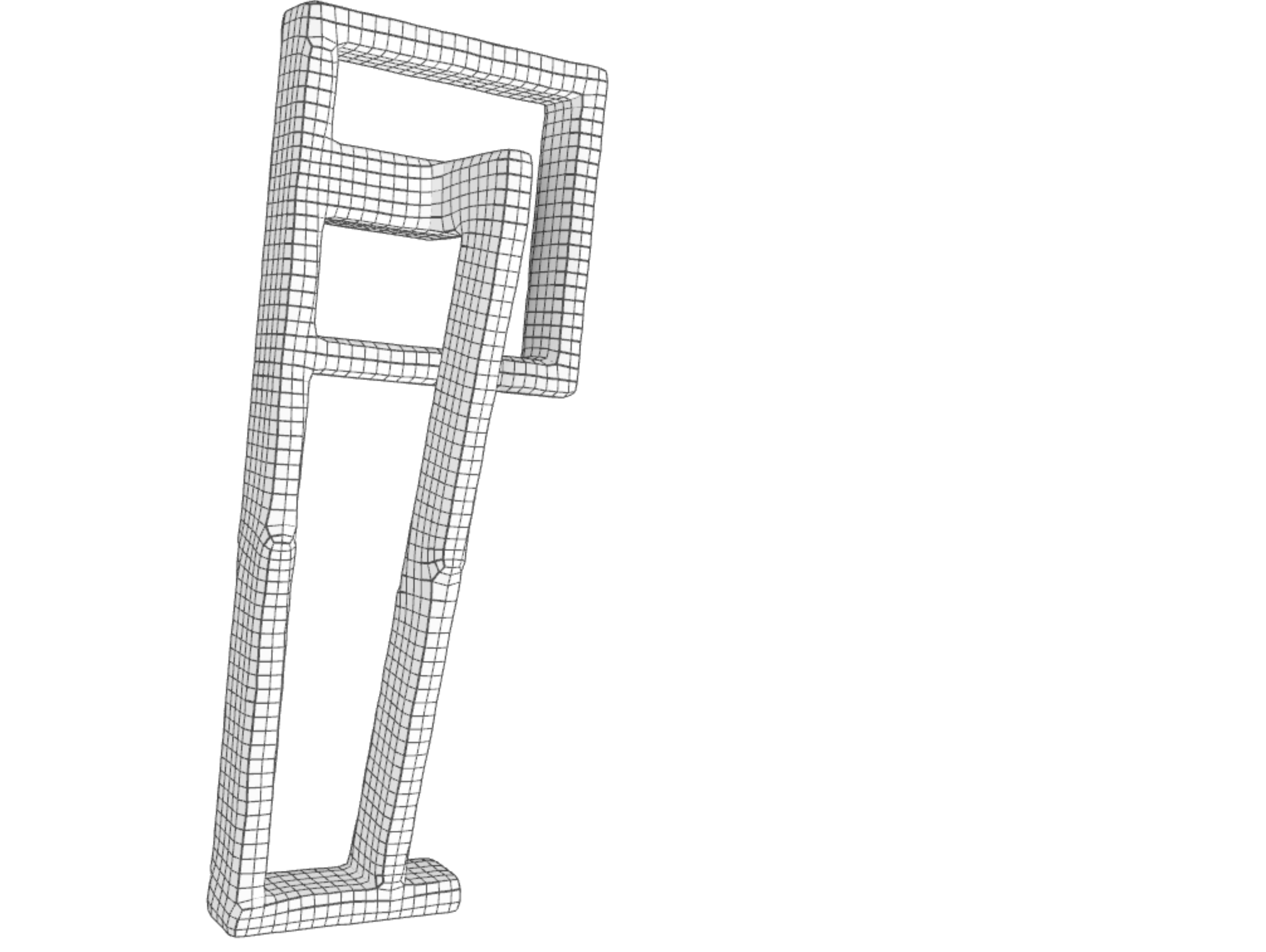}  \\
      \includegraphics[height=0.7in,trim=6.0in 3.in 0 3in, clip]{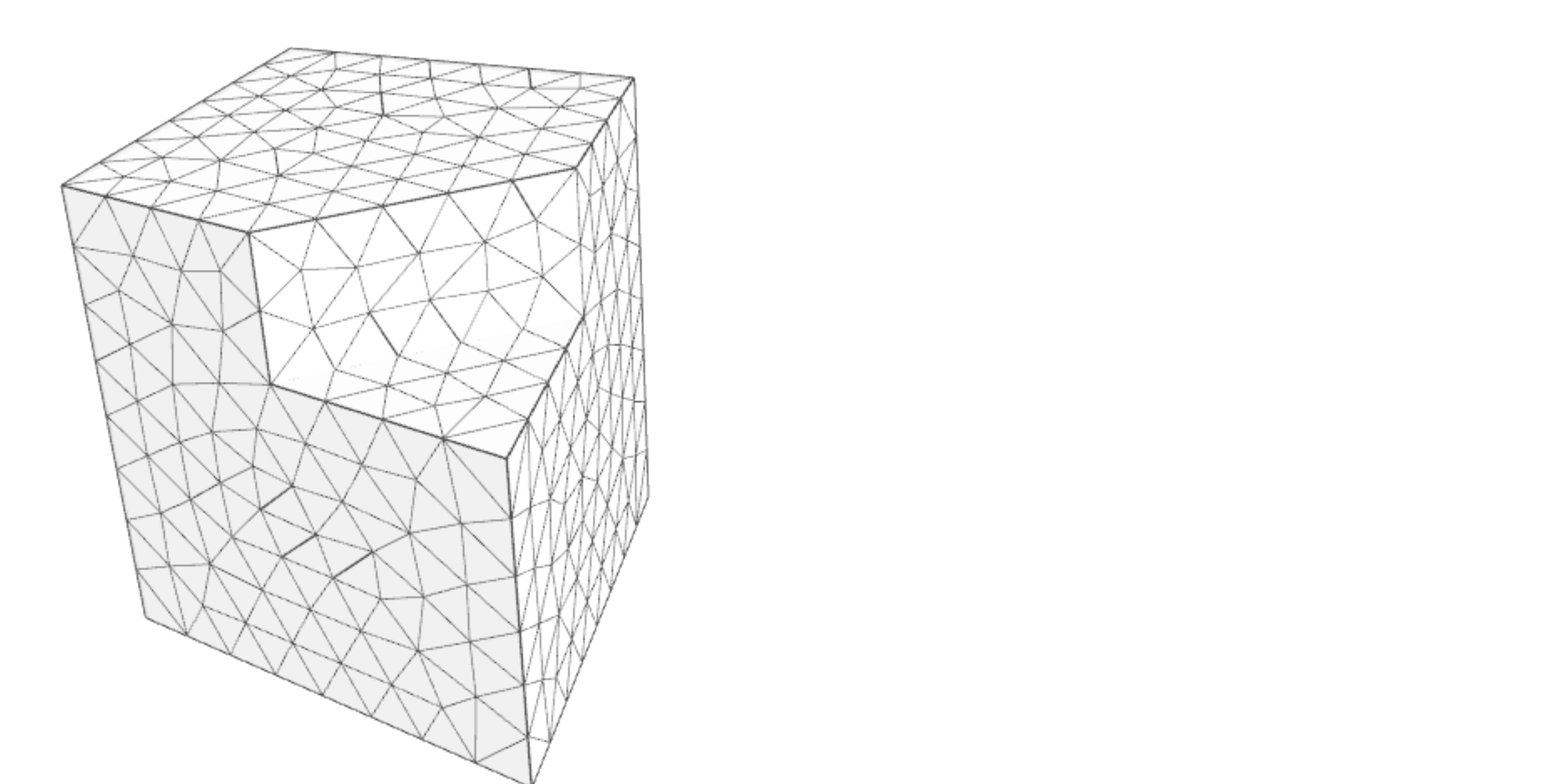} &
      \includegraphics[height=0.7in,trim=6.0in 3.in 0 3in, clip]{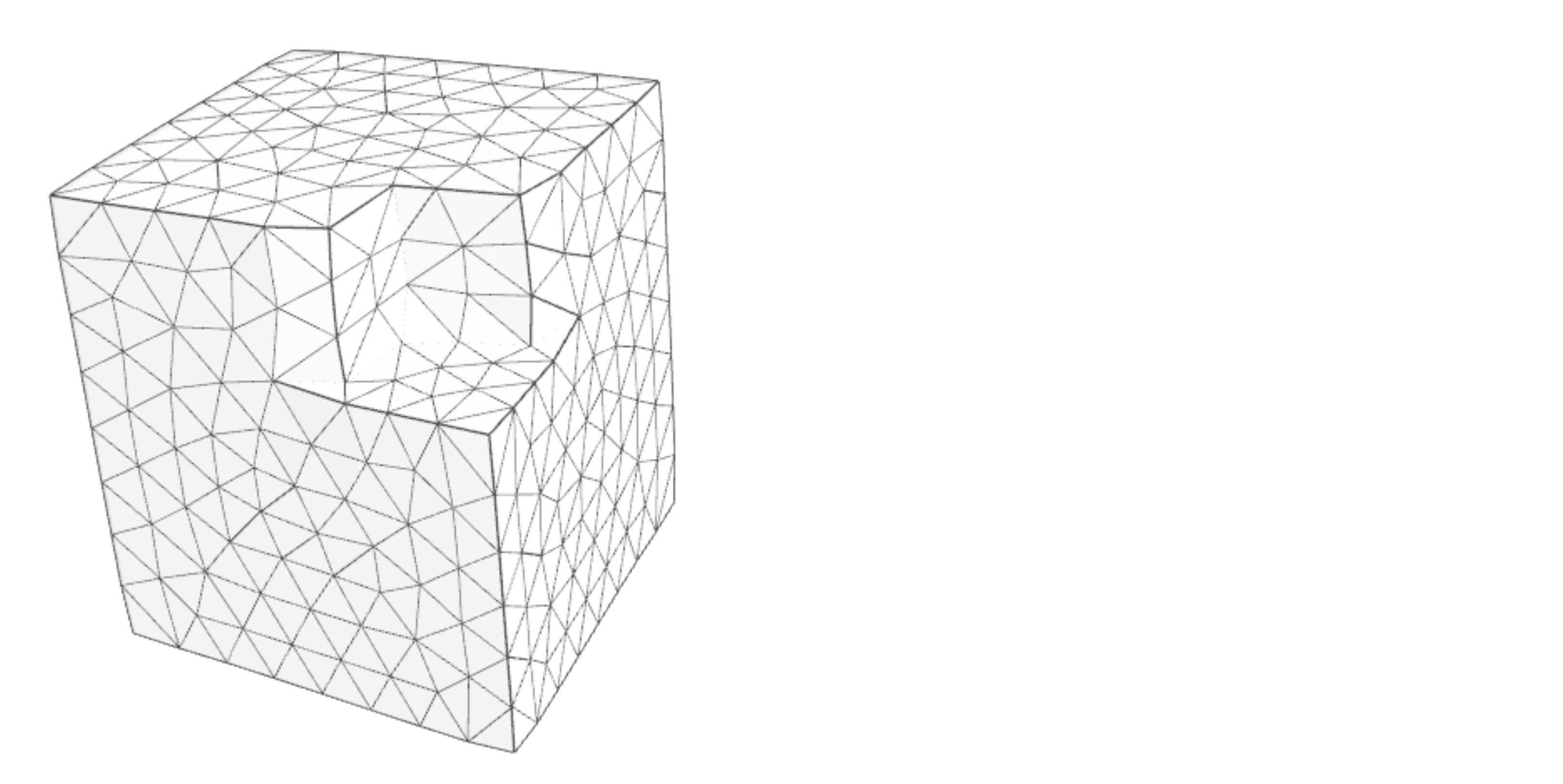} &
      \includegraphics[height=0.7in,trim=6.0in 3.in 0 3in, clip]{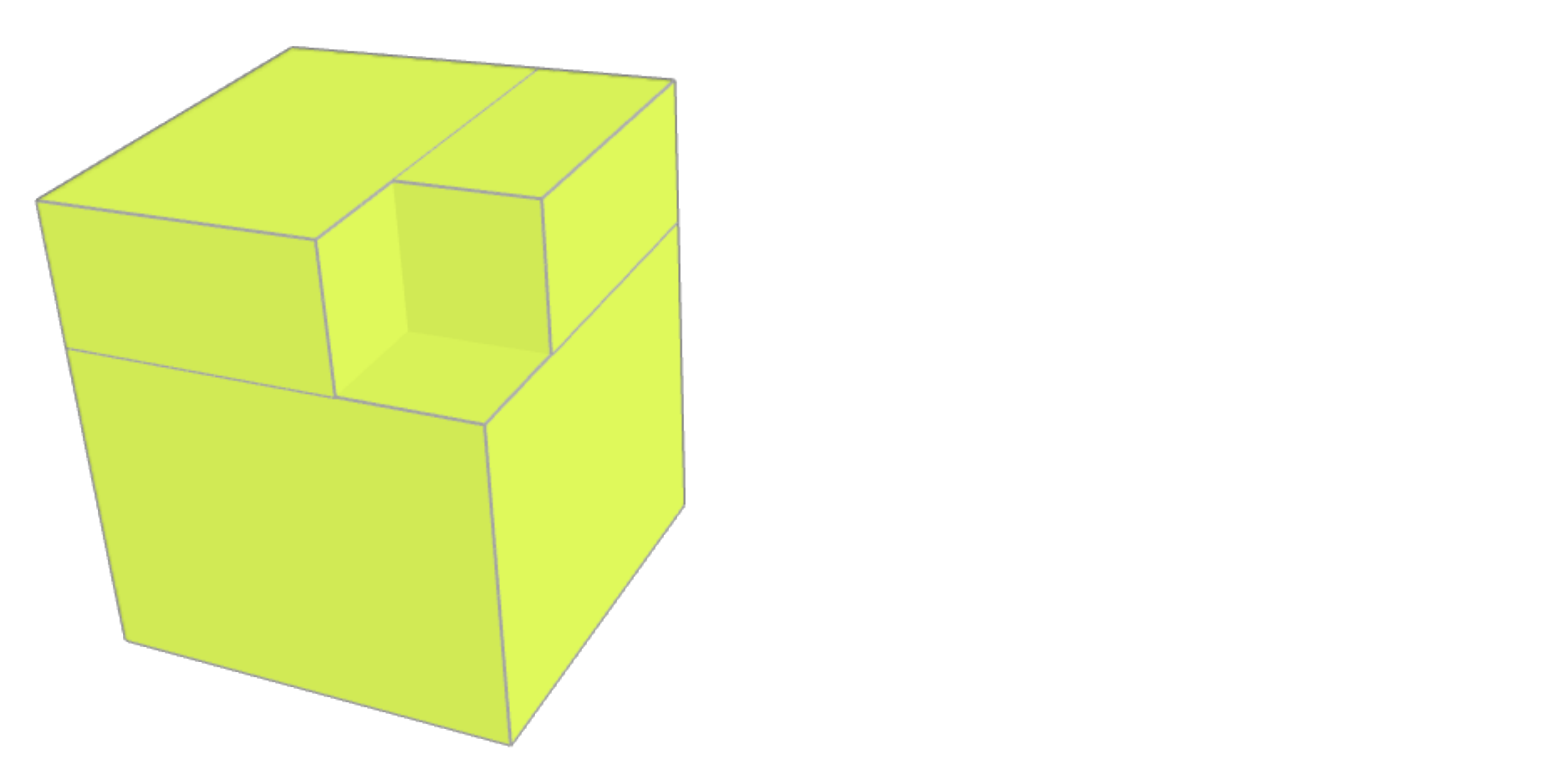} &
      \includegraphics[height=0.7in,trim=6.0in 3.in 0 3in, clip]{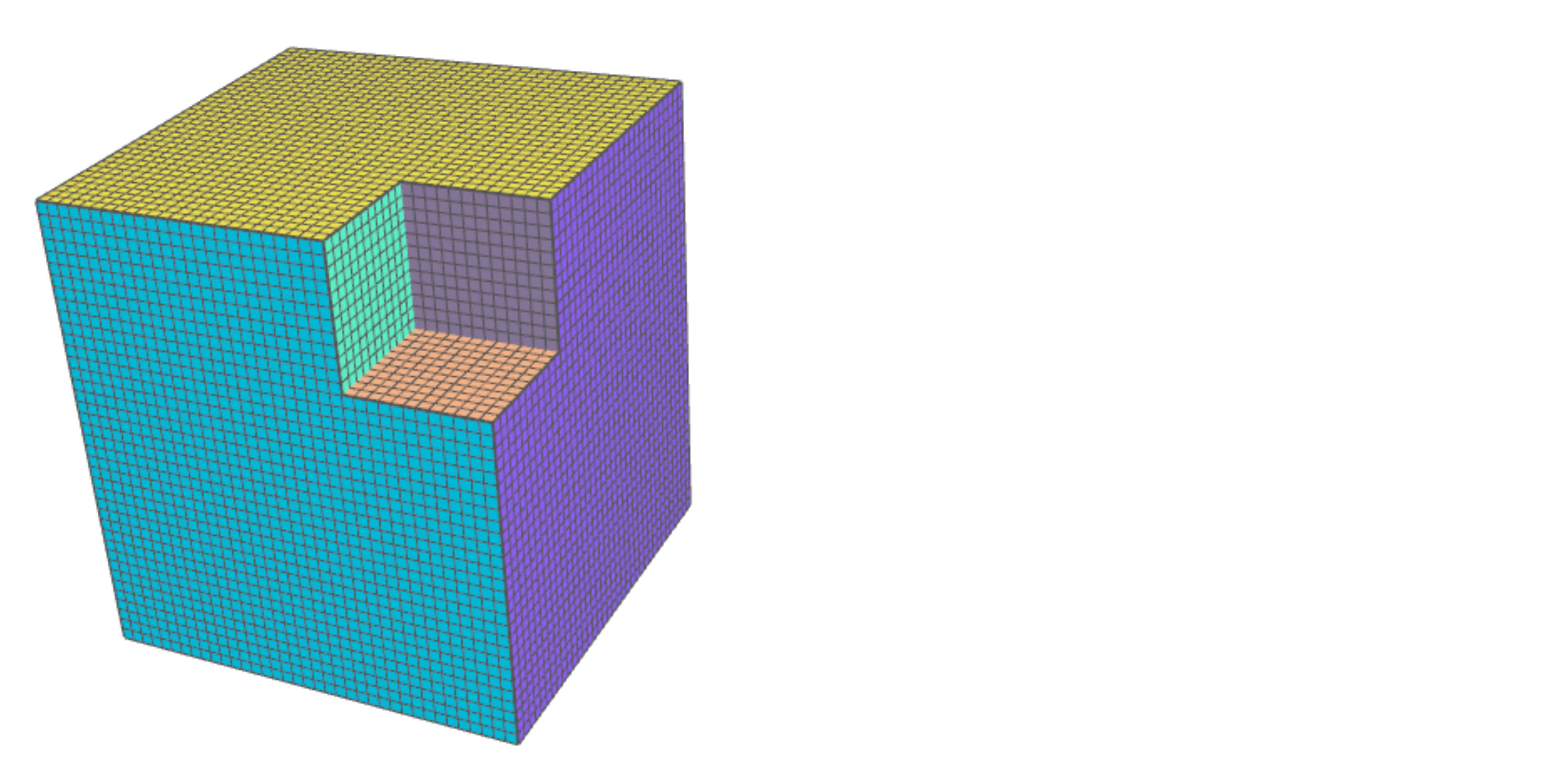} &
      \includegraphics[height=0.7in,trim=6.0in 3.in 0 3in, clip]{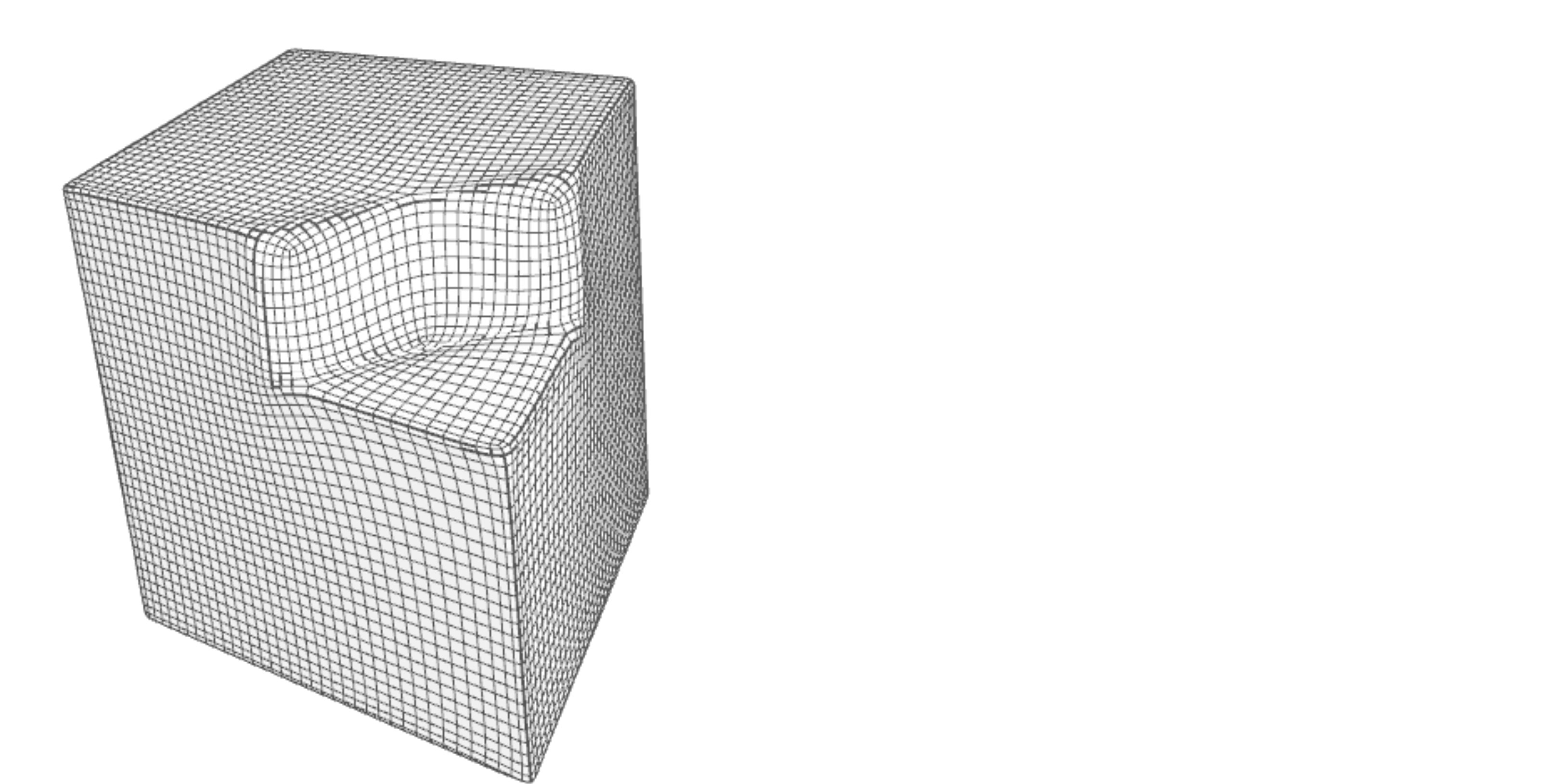}  \\
      \multicolumn{2}{c}{{\footnotesize (a) Input and deformed mesh}} & {\footnotesize (b) Cuboids} & {\footnotesize (c) PolyCube} & {\footnotesize (d) Result}
    \end{tabular}
  \caption{
    Challenging cases for deformation-based PolyCube methods and frame-field-based hexahedral meshing.
    In the first three rows the deformation fails to produce the needed stairs, even with a large \textit{cubeness} value.
    Frame-field-based hexahedral meshing approaches invariably produce un-meshable singularities on the mesh in the fourth row \cite{viertel2016analysis}.
    Our method of distance-field-based PolyCube optimization successfully meshes each case.
    A small amount of user interaction (less than 5 minutes total) is needed for the models in the middle two rows.
    \textit{Ramp} (top row): $47224$ hexes, $J_{\min} = 0.392, J_{\avg} = 0.942, d_{\max} = 3.350$.
    \textit{Ex14} (second row): $48756$ hexes, $J_{\min} = 0.472, J_{\avg} = 0.896, d_{\max} = 5.465$.
    \textit{Ex16} (third row): $9256$ hexes, $J_{\min} = 0.478, J_{\avg} = 0.872, d_{\max} = 7.244$.
    \textit{Notch7} (last row): $68704$ hexes, $J_{\min} = 0.579, J_{\avg} = 0.957, d_{\max} = 3.861$.
  }
  \label{fig:result_challenging}
\end{figure}

\paragraph{Challenging models}
Existing deformation-based PolyCube hex meshing methods rely heavily on the deformation and may fail when correct stairs cannot be created \citep{sokolov2015fixing}.
In comparison, the continuous PolyCube optimization in our method always results in a valid PolyCube and  resolves stairs correctly most of the time, while allowing user intervention if needed.
We test our method on some challenging examples from \citet{sokolov2015fixing}, and our pipeline reliably generates good hex meshes with limited user interaction (\cref{fig:result_challenging}).

\begin{figure}
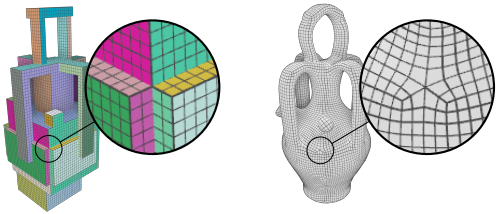

  \newcolumntype{C}[1]{>{\centering\arraybackslash}m{#1}}
  \begin{tabular}{C{3.3cm}  C{3.3cm}}
    \input{annotated/botjio_voxelized.tex} &
    \input{annotated/botjio_result.tex} 
  \end{tabular}
  \caption{An example where the PolyCube has corners incident to 6 charts (left), and the resulting hex mesh (right).
  }
  \label{fig:flex_charts}
\end{figure}

\paragraph{Larger class of PolyCubes}
Compared to previous PolyCube-based methods \citep{livesu2013polycut, fu2016efficient} where the heuristic repairs limit the PolyCube structure to have exactly three adjacent charts on every corner, our pipeline allows PolyCubes to have corners adjacent to more than three charts (\cref{fig:flex_charts}).

\paragraph{Comparison}
We test our pipeline on a number of models and quantitatively compare the obtained results with recent hex meshing methods \citep{yu2014optimizing, fang2016all, fu2016efficient, takayama2019dual, guo2020cut, livesu2019loopy}.
The statistics and timings are shown in \cref{tbl:cmp}.
For all models tested except for \textit{joint} and \textit{sculpt}, we obtain better minimum scaled Jacobian while having competing numbers in other metrics.
\rev{
The reported runtime for our method is the recorded total time for an expert user for each model.
The detailed timings for each stage vary across different models; empirically the decomposition and discretization stages take the most time.
}

\begin{table*}
  \caption{
    Comparison of hex mesh quality between our results and other recent methods.
    We report the amount of time spent in minutes (explained later), the number of vertices and hexes of the generated hex meshes, the minimum and average scaled Jacobian $J_{\min}, J_\avg$, and the maximum and mean Hausdorff distance $d_{\max}, d_\avg$ (scaled by the diagonal length of the bounding box of the input mesh).
    Both $d_{\max}$ and $d_\avg$ are computed from the point-to-mesh distances of $50k$ uniformly sampled points from the considered mesh surfaces.
    The results of \citet{yu2014optimizing, fang2016all, fu2016efficient, takayama2019dual, guo2020cut, livesu2019loopy} are obtained from their papers and \url{www.hexalab.net} \citep{bracci2019hexalab}.
    The unavailable entries are marked as `-'.
    For \citet{fu2016efficient}, the reported time is only for the PolyCube construction step and does not include the final hex meshing step.
    The reported time of our method is the recorded amount of time for an expert user to generate the corresponding hex mesh.
    Since the results for different methods are generated on different machines, the reported runtimes are not directly comparable, and we include them only for the sake of completeness.
    When optimizing for hex quality in our method, the expert prioritizes improving the worst scaled Jacobian values (large $\lambda_{\text{custom}}$) and minimizing the Hausdorff distance (large $\lambda_{m\sto 0}, \lambda_{0\sto m}$).
  }
  \begin{tabular}{c c c c c}
    \toprule
    Model & Time (m) & \# vert/\# hex & $J_{\min}$/$J_{\avg}$ & $d_{\max}$/$d_{\avg} (\times 10^{-3})$ \\
    \midrule 
    \textit{Armadillo} \citep{fu2016efficient} & $\ge 0.46$ & 87303/78376 & 0.265/$0.909 \pm 0.080$ & 4.59/- \\
    \textit{Armadillo} \citep{gao2019feature} & 320.97 & 72728/60340 & 0.159/$0.779 \pm 0.023$  & 4.9/- \\
    \textit{Armadillo} (Ours) & 21 & 21836/24709& \textbf{0.569}/$\mathbf{0.928}\pm 0.071$  & \textbf{2.806}/0.238 \\
    \hline
    \textit{Bimba} \citep{fu2016efficient} & $\ge 0.40$ & 73104/67039 & 0.361/$0.935 \pm 0.068$ & \textbf{3.69}/- \\
    \textit{Bimba} \citep{gao2019feature} & 168.90 & 63679/55035 & 0.056/$0.792 \pm 0.026$  & 4.9/- \\
    \textit{Bimba} \citep{livesu2019loopy} & - & 1973/1560 & 0.408/$0.881 \pm 0.026$  & -/- \\
    \textit{Bimba} (Ours) & 8 & 26191/28920 & \textbf{0.542}/$\mathbf{0.952}\pm 0.056$  & 4.590/0.160 \\
    \hline
    \textit{Bunny} \citep{yu2014optimizing} & $14.7$ & 47549/42543 & -0.948/$0.900 \pm 0.177$ & 18.8/- \\
    \textit{Bunny} \citep{fu2016efficient} & $\ge 0.46$ & 65603/59841 & 0.422/$0.942 \pm 0.062$ & 4.28/- \\
    \textit{Bunny} \citep{takayama2019dual} & - & 3724/2832& -0.771/$0.749 \pm 0.242$ & -/- \\
    \textit{Bunny} (Ours) & 6 & 17230/19145 & \textbf{0.601}/$\mathbf{0.944} \pm 0.060$  & \textbf{3.280}/0.276 \\
    \hline
    \textit{Buste} \citep{fu2016efficient} & $\ge 0.51$ & 86595/79294 & 0.439/$0.943 \pm 0.057$ & \textbf{3.24}/- \\
    \textit{Buste} (Ours) & 8 & 22973/25514& \textbf{0.620}/$\mathbf{0.949}\pm 0.055$  & 3.432/0.218\\
    \hline
    \textit{Dancing children} \citep{fu2016efficient} & $\ge 0.27$ & 36353/30691 & 0.251/$0.878 \pm 0.099$  & \textbf{5.11}/- \\
    \textit{Dancing children} (Ours) & 21 & 26718/30770& \textbf{0.502}/$\mathbf{0.907}\pm 0.074$  & 16.718/0.172\\
    \hline
    \textit{Dilo} \citep{gao2019feature} & 89.61 & 100110/84977 & 0.100/$0.791 \pm 0.022$  & \textbf{4.2}/- \\
    \textit{Dilo} (Ours) & 15 & 32622/36500& \textbf{0.551}/$\mathbf{0.949}\pm 0.067$  & 6.611/0.100 \\
    \hline
    \textit{Dragon} \citep{fang2016all} & 125.93 & 127360/114178 & 0.162/$0.919 \pm 0.084$  & \textbf{12.6}/0.278 \\
    \textit{Dragon} \citep{fu2016efficient} & $\ge 0.76$ & 118610/106244 & 0.265/$0.862 \pm 0.106$  & 16.1/- \\
    \textit{Dragon} (Ours) & 19 & 35845/40421& \textbf{0.391}/$\mathbf{0.927}\pm 0.073$  & 13.438/0.154 \\
    \hline
    \textit{Fandisk} \citep{takayama2019dual} & - & 2404/1774& 0.217/$0.905\pm 0.114$  & -/- \\
    \textit{Fandisk} \citep{guo2020cut} & 5.69 & 45156/39858 & 0.242/$\mathbf{0.959}\pm 0.020$  & 4.151/0.044 \\
    \textit{Fandisk} (Ours) & 7 & 32488/35841& \textbf{0.500}/$0.942\pm 0.077$  & \textbf{3.993}/0.056 \\
    \hline
    \textit{Hanger} \citep{gao2019feature} & 25.60 & 33002/26918 & 0.155 /$0.828 \pm 0.028$  & \textbf{2.3}/- \\
    \textit{Hanger} \citep{takayama2019dual} & - & 2229/1382 & 0.333 /$\mathbf{0.944} \pm 0.094$  & -/- \\
    \textit{Hanger} \citep{guo2020cut} & 8.76 & 9411/7080 & 0.412 /$0.881 \pm 0.118$  & -/- \\
    \textit{Hanger} (Ours) & 10 & 8798/10500& \textbf{0.559}/$0.870\pm 0.108$  & 3.154/0.111 \\
    \hline
    \textit{Joint} \citep{fang2016all} & 5.10 & 5181/3785 & 0.778/$\mathbf{0.984} \pm 0.032$  & 8.08/0.254 \\
    \textit{Joint} \citep{livesu2019loopy} & - & 751/456 & \textbf{0.798} /$0.949 \pm 0.045$  & -/- \\
    \textit{Joint} \citep{takayama2019dual} & - & 2711/2010 & 0.249 /$0.927 \pm 0.118$  & -/- \\
    \textit{Joint} \citep{guo2020cut} & 3.33 & 39658/33565 & 0.683/$0.980 \pm 0.032$  & -/- \\
    \textit{Joint} (Ours) & 8 & 37227/41768& 0.735/$0.976\pm 0.049$  & \textbf{3.995}/0.120 \\
    \hline
    \textit{Kitten} \citep{fang2016all} & 2.30 & 14459/11941 & 0.435/$0.922 \pm 0.077$  & 12.10/0.652 \\
    \textit{Kitten} \citep{livesu2019loopy} & - & 2126/1728 & -0.002/$0.744 \pm 0.204$  & -/- \\
    \textit{Kitten} (Ours) & 8 & 11531/12929 & \textbf{0.648}/$\mathbf{0.943}\pm 0.059$  & \textbf{4.604}/0.251 \\
    \hline
    \textit{Lock} \citep{guo2020cut} & 6.33 & 7634/5224 & 0.052 /$0.887 \pm 0.122$  & -/- \\
    \textit{Lock} (Ours) & 11 & 35502/40645 & \textbf{0.545}/$\mathbf{0.899} \pm 0.093$  & 1.996/0.060 \\
    \hline
    \textit{Rocker arm} \citep{yu2014optimizing} & 16.6 & 10078/7874 & -0.506/$0.869 \pm 0.210$  & 7.08/- \\
    \textit{Rocker arm} \citep{fang2016all} & 61.85 & 20680/17594 & 0.464/$0.936 \pm 0.071$  & 8.06/0.421\\
    \textit{Rocker arm} \citep{fu2016efficient} & $\ge 0.15$ & 27174/23421 & 0.477/$0.924 \pm 0.074$ & 5.27/- \\
    \textit{Rocker arm} \citep{takayama2019dual} & - & 2651/1858 & -0.189/$0.805 \pm 0.200$ & -/- \\
    \textit{Rocker arm} (Ours) & 14 & 40929/45789& \textbf{0.689}/$\mathbf{0.940} \pm 0.055$  & \textbf{3.696}/0.163 \\
    \hline
    \textit{Sculpt} \citep{livesu2019loopy} & - & 327/168& \textbf{0.806}/$0.918 \pm 0.044$ & -/- \\
    \textit{Sculpt} \citep{guo2020cut} & 4.35 & 25562/21695 & 0.528/$\mathbf{0.949} \pm 0.044$ & \textbf{1.927}/0.108 \\
    \textit{Sculpt} (Ours) & 13 & 18995/21488& 0.614/$0.917\pm 0.075$ & 6.321/0.202\\
    \bottomrule
  \end{tabular}
  \label{tbl:cmp}
\end{table*}

\paragraph{User Study}

Our system offers an interactive user-in-the-loop experience for producing hex meshes. Unlike previous algorithms, a key advantage of our method is the ability for the user to make an intuitive choice at each juncture, all of which contribute to the output. To evaluate this aspect of our work, we conduct a user study. Each participant accessed a build of our software preinstalled on a remote server. They were first asked to complete a detailed tutorial, which outlined the process of obtaining a hex
mesh using our tool (taking the \textit{spot} model as an example). Users were also provided with a video walk-through of the steps described in the tutorial.

Upon finishing the tutorial, users were asked to select one or more additional meshes \rev{(from the publicly available models of \citet{fu2016efficient})} to experiment with and produce a hex mesh using our software. Finally, they completed a survey about their prior experience with 3D modeling, their overall experience with our tool, and their experience with each mesh that they chose.

A total of 21 participants took part in the user study, with a self-reported average score of 2.2 for familiarity with 3D modeling tools and 2.0 familiarity with hex meshing (both on a 1-5 scale). The users spent an average of 20.1 minutes on each mesh that they worked with after the tutorial. On average, 81\% of the users were satisfied with the hex meshes that they obtained. In particular, 18 out of the 21 participants agreed that they were able to to make fine-grained adjustments to their final mesh, and 13 out of 21 preferred their meshes compared to hex meshes obtained from automatic algorithms (on \url{https://www.hexalab.net}). Users ``appreciated the speed and the interpretability of [the] optimizers and hyperparameters'' and were ``particularly impressed with the final hex mesh.'' Participants wrote that ``[the tool] allowed a lot of manual changes but the default parameters worked well which was great as an inexperienced user,'' and ``the tool provides a good balance of personal customization (i.e., what parts to capture/focus on) and computer-assisted optimization (no need to worry too much about placement and sizing of the cubes [...]).''  This highlights the fact that our method enables low-level control without adding superfluous complexity.

\begin{figure}
  \centering
  \includegraphics[width=\linewidth]{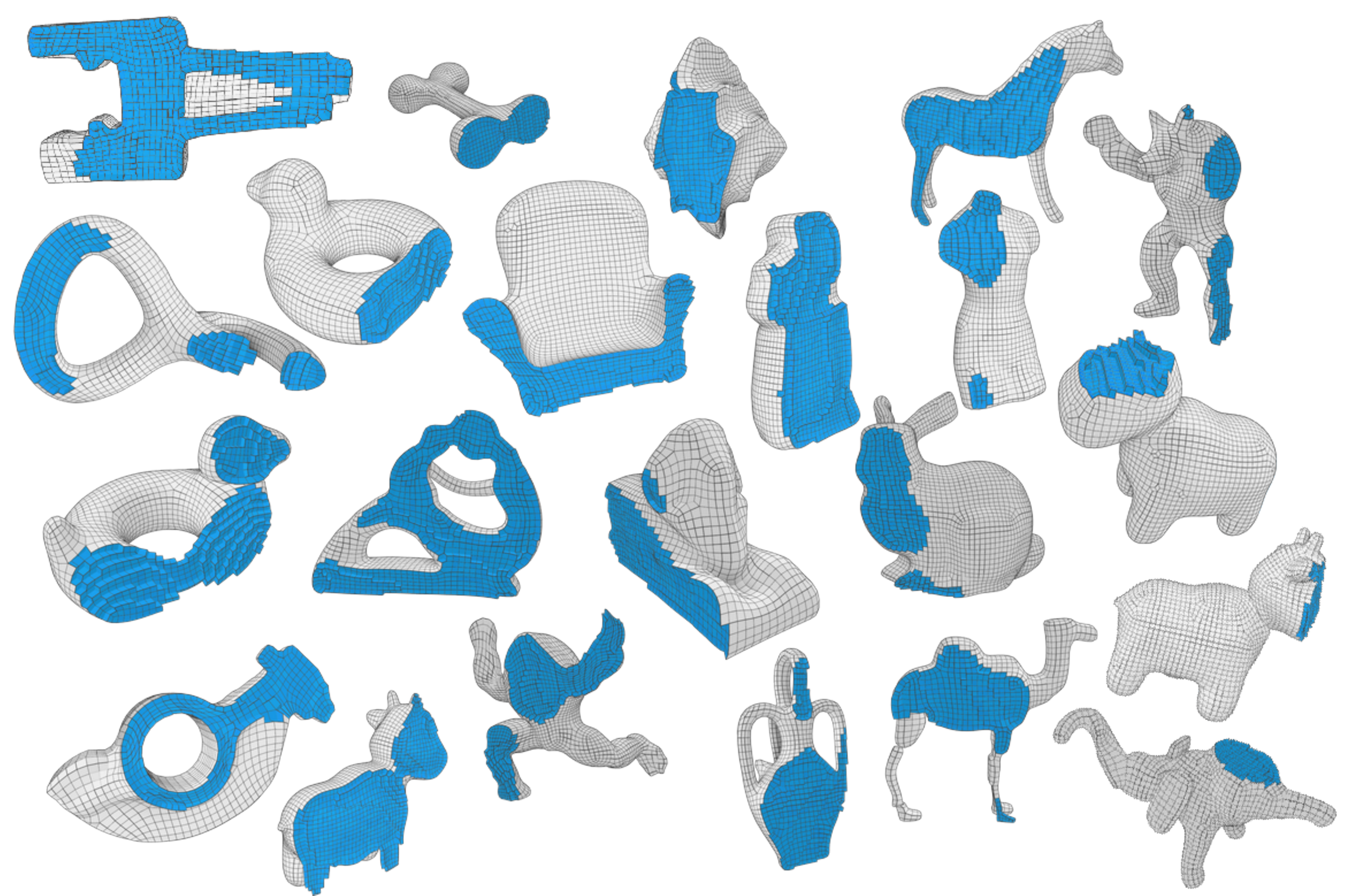}
  \caption{A sampling of hex meshes produced by novice users as part of our user study.}
\label{fig:user_study_meshes}
\end{figure}

\cref{fig:user_study_meshes} shows some hex meshes that were produced during the user study.

\section{Conclusion and Future Work}
\label{sec:conclusion_and_limitations}
The task of converting a tetrahedral mesh to a hex mesh is a highly under-constrained problem, and many trade-offs must be considered to produce a satisfactory result. Different applications and aesthetics motivate different choices for the placement and number of singularities, articulation or smoothing of sharp features, simplicity versus fidelity, and so on. Rather than proposing a one-size-fits-all algorithm like many previous works, we instead design a comprehensive interactive system, consisting of several stages. At each stage, the user can intervene, making the design choices that influence the final output. While our system has a shallow learning curve for the non-expert user due to robust default parameter choices, as demonstrated by our user study, it also allows extremely fine-grained control, yielding high-quality meshes that achieve or exceed state-of-the-art quantitative metrics.

Our system generates high-quality hex meshes on all the test models. Below we discuss some exciting directions to further extend our pipeline in future work.

\paragraph{Interior singularity support}
Like other PolyCube-based methods, the singularities on the hex meshes produced by our system are restricted to the surface (except for a layer of global padding). Allowing interior singularities could improve mesh quality (e.g., results of \textit{sculpt} from \citet{livesu2019loopy} and \citet{guo2020cut} in \cref{tbl:cmp}).
One strategy would be to augment our pipeline by employing the selective padding of \citet{cherchi2019selective} instead of global padding or by injecting interior singularities by cutting PolyCube edges open \citep{guo2020cut}.
Alternatively, we could follow \citet{fang2016all} by %
breaking all tunnel loops of the input shape and adding additional constraints during optimization to glue back the input.

\paragraph{Invariance to orientation}
Although our user-guided deformation (\cref{fig:cubeness_adjustment}) can orient coarse features of the input, the deformation still largely depends on the initial orientation of the shape. %
For instance, for the \textit{dragon} model, an ideal deformation map would straighten the body instead of creating unnecessary zig-zag patterns.
A potential solution is to use a frame field to guide the deformation, like in \cite{fang2016all}.

\paragraph{Sharp feature preservation}
For certain classes of input shapes, it may be desirable to preserve sharp features explicitly. One way to achieve this would be to integrate feature-aware PolyCube generation from \citet{guo2020cut} by allowing the user to draw feature curves on the input surface, which would then get mapped to PolyCube edges.
\rev{Then, during the hexahedralization stage, we can include another energy term that favors pairs of orthogonal faces that meet on a feature curve, such as Eq. (5) from \citet{guo2020cut}.}

\paragraph{Topological consistency}
We rely on the user to make sure the topology is correct, allowing them to repair issues during the decomposition stage (\cref{fig:cuboid_user_fix}) or the discretization stage (\cref{fig:dig_extrude_example}).
While directly incorporating topological priors into gradient-based optimization is a challenging open problem, we could detect incorrect topology and interactively alert the user to ease the process.

\rev{
\paragraph{Concavities and tunnels}
For highly complex models with concavities and tunnels, it could be strenuous to manually place cuboids in the decomposition stage while avoiding all the empty regions.
One promising future direction is to allow users to construct a PolyCube for the complement of the shape
and then to cut out this complement from the primal PolyCube. 
Our distance-field-based formulation is particularly suitable for this approach. For instance, the signed distance field of the complement can be obtained by reversing signs, and getting the signed distance field of the subtracted PolyCube is similarly straightforward.
Orthogonally, we can extend the discretization stage to allow users to make interior edits by introducing ways to hide layers of voxels.
}

\paragraph{Additional UI features}
The feedback gathered during our user study could be incorporated to further enhance the user experience.
Several users commented that the PolyCube optimization %
may create gaps (\cref{fig:sdf_illustration}(b)) or unnecessary stairs (\cref{fig:cuboid_user_fix}(d)) despite already being at a local minimum.
We could add a postprocessing step after the decomposition stage to glue nearby cuboids together and remove superfluous stairs, e.g., using the erasing-and-filling operators from \citet{yang2019computing}.

The experience of navigating between the various stages of our pipeline could be improved by propagating visual cues across the stages.
For instance, if the user notices a region that needs changing (such as an undesired singularity or topological issue) in the final stage of the pipeline only after visualizing the generated hex mesh,
it may be hard to locate the corresponding region in the decomposition or discretization stage.
Visualizing correspondences between the final hex mesh and the voxelized PolyCube would simplify this workflow.

Some users reported difficulties avoiding inversions when placing landmarks. %
One way to avoid inversions caused by arbitrarily placing and fixing surface vertices would be to to encode the new positions as a soft constraint and use an energy term similar to \cref{eqn:pullback_energy} to gradually deform the mesh so that it satisfies the new positional constraints.

\begin{acks}
The MIT Geometric Data Processing group acknowledges the generous support of Army Research Office grant W911NF2010168, of Air Force Office of Scientific Research award FA9550-19-1-031, of National Science Foundation grants IIS-1838071 and CHS-1955697, from the CSAIL Systems that Learn program, from the MIT–IBM Watson AI Laboratory, from the Toyota–CSAIL Joint Research Center, from a gift from Adobe Systems, from an MIT.nano Immersion Lab/NCSOFT Gaming Program seed grant, and from the Skoltech–MIT Next Generation Program. We would also like to thank Zoya Bylinskii for valuable advice on designing and conducting the user study.
\end{acks}

\bibliographystyle{ACM-Reference-Format}
\bibliography{hex-bibliography}

\appendix

\rev{
\section{More implementation details}
\label{sec:app_impl_details}

\subsection{Heuristics in \cref{stage:decomposition}}
\label{sec:app_heuristics}
For the two heuristics in the \textit{Add} operation of \cref{stage:decomposition}, we create a $32\times 32\times 32$ uniform grid over the shape and then compute the signed distances of all grid points to both the PolyCube and the deformed mesh.
For the distance-based heuristic, we can then detect an uncovered point of the deformed mesh that is furthest away from the PolyCube. 
For the volume-based heuristic, to find the largest cuboid inside the deformed mesh and outside the PolyCube, we extend the algorithm for finding the largest rectangle in a histogram to 3D with time complexity $O(n^4)$ with $n=32$, similar to \citet{yang2019computing}.
For the \textit{Subtract} operation, the implementation is similar.

\subsection{Custom CUDA functions}
\label{sec:app_cuda}
We implement the following CUDA functions to enable our pipeline to run at interactive speed.
Each functions takes $(P, M)$ as the input, where $P$ is a list of query points and $M$ is a representation of a triangular or tetrahedral mesh, and returns a quantity for each query point.
In our case, $P$ can contain tens of thousands of points, and $M$ can have up to $10^5$ cells.
In our CUDA code, we use a thread block of dimension $128$ for each query point and evenly divide the mesh triangles or tetrahedrons among the 128 threads to process the query point in a thread block.
The results from all threads are collected using parallel reduction.

Below are details of the implementations for each CUDA device function for a pair of point and triangle/tetrahedron.
\paragraph{Point-triangle projection.}
Let $p \in \RR^3$ be the point we want to project onto a triangle with vertices $v_0, v_1, v_2 \in \RR^3$.
Denote $e_1 = v_1 - v_0$ and $e_2 = v_2 - v_0$.
Consider first projecting $p$ onto the plane spanned by $v_0, v_1, v_2$.
This can be done by solving the following least-squares problem:
\begin{equation}
  d(p, \{v_i\}) \defeq \min_{w_1,w_2\in\RR} \norm{p - (v_0 + w_1e_1 + w_2e_2)}^2.
  \label{eqn:cuda_tri_proj_opt}
\end{equation}
The closed-form solution is given by, for $i=1,2$,
\begin{equation}
  w_i = \frac{(p - v_0) \cdot \left(e_i - \frac{(e_{2-i} \cdot e_{i})e_{2-i}}{\norm{e_{2-i}}^2}\right)}{
  \norm{e_i}^2 - \frac{(e_{2-i}\cdot e_i)^2}{\norm{e_{2-i}}^2}
  }.
  \label{eqn:cuda_w}
\end{equation}
Let $w_0 = 1 - w_1 - w_2$. 
Note $(w_0,w_1,w_2)$ is the barycentric coordinate of $p$ in this triangle.
If $w_i \ge 0$ for $i\in\{0,1,2\}$, then the projection of $p$ lands inside the triangle.
Otherwise, we project $p$ onto line segment $(p_0,p_1), (p_1,p_2), (p_2, p_0)$ and choose the projection with the smallest distance.

When backpropagating through the proximity energy in \cref{eqn:prox_energy}, we need to compute the gradient of the projected point $q$ with respect to both $p$ and $v_0,v_1,v_2$:
\begin{itemize}
  \item
    If $w_i \ge 0$ for $i \in \{0,1,2\}$, i.e., $q$ lands inside the triangle, then we have $q = \sum_{i=0}^2 w_i v_i$,
    so $\frac{\partial q_k}{\partial p} = \sum_{i=0}^2\frac{\partial w_i}{\partial p} (v_i)_k$, where $k \in \{x,y,z\}$, and each $\frac{\partial w_i}{\partial p}$ is a constant deduced from \cref{eqn:cuda_w} that can be precomputed for each triangle ahead of time.
    The case for when $q$ is outside the triangle can be handled similarly. 
  \item
    To compute the gradient of $q$ with respect to $v_i$'s, direct calculation from \cref{eqn:cuda_w} can be cumbersome. 
    This is only needed for \cref{eqn:prox_energy_dd}, so we just need to differentiate $d \defeq d(p, \{v_i\})$ from \cref{eqn:cuda_tri_proj_opt} with respect to the $v_i$'s.
    By the Envelope Theorem, we have $\frac{\partial d}{\partial v_i} = w_i(q-p)$ for each $i$.
\end{itemize}

\paragraph{Point-tetrahedron projection.}
To compute the pullback energy from \cref{eqn:pullback_energy}, we need to project a point onto the input tetrahedral mesh.
Consider projecting a point $p$ to a single tetrahedron with vertices $\{v_i\}_{i=0}^3$.
Similar to \cref{eqn:cuda_tri_proj_opt}, we can find the barycentric coordinates of the projection by solving a least-squares problem but with three variables rather than two:
\begin{equation}
  d(p, \{v_i\}) \defeq \min_{w_1,w_2,w_3\in\RR} \norm{p - (v_0 + w_1e_1 + w_2e_2 + w_3e_3)}^2.
  \label{eqn:cuda_tet_proj_opt}
\end{equation}
Taking the derivative and setting it equal to zero gives a linear system of the form $Aw=b$, where $A$ is invertible and does not depend on $p$.
We can precompute the inverse of $A$ for each tetrahedron.
If any $w_i < 0$, then we project $p$ onto the four triangular faces of the tetrahedron in the same way as in the last paragraph.
We do not need gradient information for point-tetrahedron projection.

\paragraph{Point-in-mesh inclusion test.}
To compute the signed distance field of the deformed tet mesh in \cref{eqn:df_positive}, in addition to computing point-triangle projection, we need to determine the sign for each query point, i.e., whether it is inside the mesh or not.
Our implementation relies on the fact that we know the volume mesh on which we want to test: we check if the query point is inside any tetrahedron, which can be done by checking if the point is on the correct side of each face of the tetrahedron.
We found this approach robust even with single-precision floating-point computation.
}

\end{document}